\documentclass[amsmath,amssymb,aps,prl,reprint,superscriptaddress]{revtex4-1}
\usepackage{graphicx}
\usepackage{xcolor}
\usepackage[colorlinks=true,urlcolor=blue,citecolor=blue,linkcolor=blue]{hyperref}

\usepackage{physics}
\usepackage{siunitx}
\usepackage{xspace}
\usepackage{bm}
\usepackage{amssymb,ulem,amsmath}
\usepackage{verbatim}
\usepackage{comment}

\newcommand{\bk}{\textbf{k}}
\newcommand{\bq}{\textbf{q}}

\definecolor{AJ-color}{named}{blue}
\definecolor{PT-color}{rgb}{0.97,0.57,0.11}
\definecolor{GB-color}{RGB}{128,0,128}

\definecolor{AJ-color2}{named}{blue}
\definecolor{PT-color2}{rgb}{0.87,0.47,0.01}
\definecolor{GB-color2}{RGB}{128,0,128}

\begin{document}

\onecolumngrid
This document includes two papers: a letter "\textbf{Quantum geometry and flat band Bose-Einstein condensation}" and a longer, more detailed article "\textbf{Excitations of a  Bose-Einstein condensate and the quantum geometry of a flat band}". The former presents the main results of the work, whereas the latter provides the details of the calculations, considers physical quantities not studied in the letter, and provides a substantially extended discussion of the subject.
\title{Quantum geometry and flat band Bose-Einstein condensation}

\newcommand{\affiliationAalto}{Department of Applied Physics, Aalto University, P.O.Box 15100, 00076 Aalto, Finland}
\newcommand{\affiliationAarhus}{Center for Complex Quantum Systems, Department of Physics and Astronomy, Aarhus University, Ny Munkegade 120, DK-8000 Aarhus C, Denmark}
\newcommand{\affiliationChina}{Shenzhen Institute for Quantum Science and Engineering and Department of Physics, Southern University of Science and Technology, Shenzhen 518055, China}

\author{Aleksi Julku}
\affiliation{\affiliationAalto}
\affiliation{\affiliationAarhus}
\author{Georg M.\ Bruun}
\affiliation{\affiliationAarhus}
\affiliation{\affiliationChina}
\author{P\"aivi T\"orm\"a}
\affiliation{\affiliationAalto}


\date{\today}

\begin{abstract}

We study the properties of a weakly interacting Bose-Einstein condensate (BEC) in a flat band lattice system by using multiband Bogoliubov theory, and discover fundamental connections to the underlying quantum geometry. In a flat band, the speed of sound and the quantum depletion of the condensate are dictated by the quantum geometry, and a finite quantum distance between the condensed and other states guarantees stability of the BEC. Our results reveal that a suitable quantum geometry allows one to reach the strong quantum correlation regime even with weak interactions.


        
\end{abstract}


\maketitle


\textit{Introduction ---} Geometric and topological properties of Bloch wave functions in periodic lattice systems~\cite{Resta:2011,Provost:1980} - i.e.\ the quantum geometry -- are important to describe a range of physical phenomena. Tremendous progress in the understanding of the physical relevance of concepts such as the quantum metric~\cite{Souza2000,Neupert:2013,Gu:2010}, Berry curvature, Chern number, and other topological
invariants has been made~\cite{Klitzing1980,Thouless1982,Haldane1988,Kane2005,Hasan2010,Qi2011,Bernevig2013}, and experimental techniques to probe quantum geometry have been  developed~\cite{Gianfrate:2019,Tan2019}. Quantum geometric phenomena are especially striking in systems that feature dispersionless (flat) Bloch bands, where the kinetic energy is quenched and quantum states are strongly localized~\cite{Leykam:2018}. Due to a vanishing kinetic energy, the transport properties of a flat band are determined by the overlap between Bloch states, that is, by the quantum geometry~\cite{Liang:2017b}. Indeed,   previous studies have shown that the superfluid density of flat band systems is determined by the  Chern number, quantum metric or Berry curvature~\cite{Peotta2015,Liang2017,julku2016} despite the fact that effective mass of the electrons in a flat band is infinite. Recently it has been proposed~\cite{Hu:2019,Julku:2020,xie:2019,classen:2019} that the observed  superconductivity in twisted bilayer graphene~\cite{Cao:2018,yankowitz:2019} stems from quantum geometric properties of quasi-flat Bloch bands.


Geometric properties of quantum states are widely studied in fermionic systems but less is known about their role in bosonic systems where particles can undergo Bose-Einstein condensation (BEC). While bosonic flat band geometries have been studied experimentally~\cite{Taie:2015,Vicencio2015,Mukherjee2015,kajiwara2016,Harder2020a,Baboux2016,Whittaker2018,Scafirimuto2021,Harder2020b}, and quantum geometry is experimentally accessible in bosonic systems~\cite{Gianfrate:2019}, understanding how the quantum geometry affects the physical properties of a BEC is still lacking. In this letter and in our more detailed joint work of Ref.~\cite{julku2021}, by using multiband Bogoliubov formalism, we theoretically unravel fundamental connections between a weakly-interacting BEC taking place in a multiband lattice system and the quantum geometric properties of the underlying Bloch states. Our focus is on the systems where the condensation takes place within a flat band. We show how the quantum geometry crucially determines the stability and excitation properties of a flat band BEC. 

A fundamental question on flat band BEC relates to the stability of the condensate: can the bosons coherently condense to a single flat Bloch band when all the other flat band states have the same energy? As a first guess, one could think that the interaction effects renormalize the energy dispersion so that the lowest excitation band is not flat anymore, ensuring the stability of a BEC. We, however, show that one can realize a stable BEC even in the limit of \textit{vanishing interaction strength} $U$. Intriguingly, a non-zero quantum distance $D(\bq)$ (defined below with $\bq$ being the quasi-momentum and $0 \leq D(\bq) \leq 1$) between the flat band states prevents the scenario where all the particles escape the condensate even if such excitations do not in the limit of $U\rightarrow 0$ cost any extra energy. This mechanism guarantees a stable flat band BEC. Because some of the non-condensed Bloch states can overlap with the condensed state (i.e. $D(\bq) < 1$), in the limit of $U\rightarrow 0$ there can exists finite quantum depletion, i.e.~finite density of non-condensed bosons $n_{\textrm{ex}}$. This is in stark contrast to conventional dispersive-band BEC where $\lim_{U\rightarrow 0}n_{\textrm{ex}} =0$~\cite{Fetter1971,Pitaevskii2003}. 
We also find that the quantum geometric origin of a stable BEC is  manifested by the speed of sound $c_s$ which turns out to be determined by the quantum metric~\cite{Resta:2011,Provost:1980} at the condensed state, i.e. the second derivative of the quantum distance.

Importantly, we show that $\lim_{U\rightarrow 0} n_{\textrm{ex}}$ is determined by the quantum geometry only and not by the total density $n_{\textrm{tot}}$. Therefore, by decreasing the condensation density, one can increase the relative depletion of the condensate, $n_{\textrm{ex}}/n_{\textrm{tot}}$, even in the $U\rightarrow 0$ limit. In this way, the importance of quantum fluctuations and correlations can be significantly enhanced. We demonstrate this in our joint work~\cite{julku2021} where we calculate the density-density correlation function to show that the quantum geometry can provide access to a regime dominated by interaction effects even with infinitesimally small $U$. This is highly relevant in systems where  interactions are inherently small such as photon and polariton condensates.

In this letter, we consider a two-dimensional kagome lattice geometry that supports a flat band. In our joint work of Ref.~\cite{julku2021}, we provide the details of the calculations for a generic flat band system and furthermore show the results for density-density correlations and superfluid density. These two works together thus establish fundamental connections between quantum geometry and various physical properties of weakly interacting flat band condensates.

\begin{figure}
  \centering
    \includegraphics[width=1.0\columnwidth]{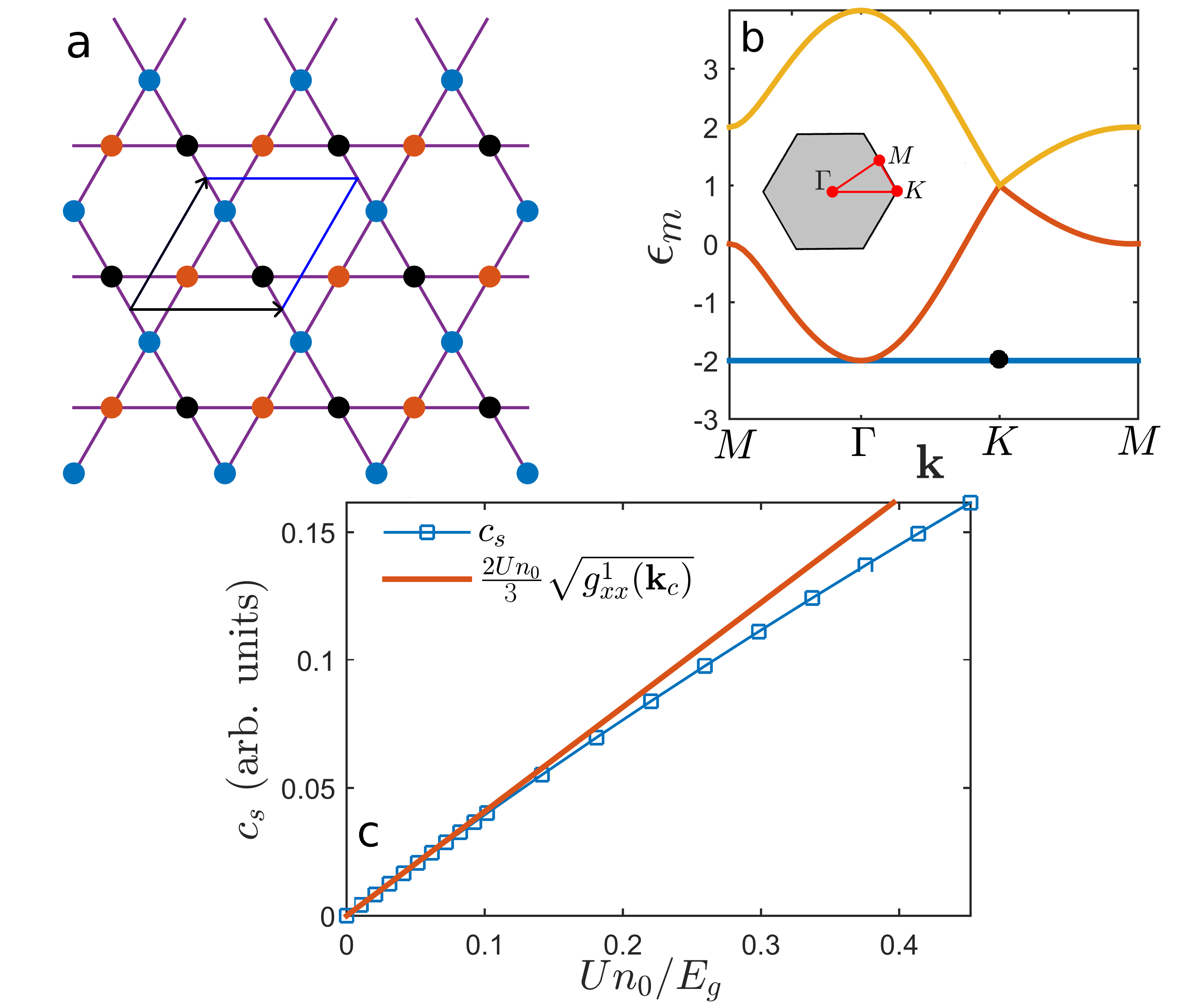}
    \caption{(a) Kagome lattice geometry. The unit cell is shown as a blue parallelogram and black arros are the basis vectors $\textbf{a}_1$ and $\textbf{a}_2$. Purple lines depict NN hopping terms of strength $t$. (b) Bloch bands of the kagome lattice with $t=1$ along the path connecting the high-symmetry points shown in the inset. The lowest band is strictly flat. The black dot marks the Dirac point $\bk=[4\pi/3,0]$ in which BEC can take place. (c) Speed of sound $c_s$ for the kagome flat band BEC as a function of $U$. Total density was chosen to be $n_{\textrm{tot}} = 3$, i.e.\ one particle per lattice site. We also show the weak-coupling result of Eq.~\eqref{sos_kag} as a solid line. The energy scale $E_g = 3t$ is the energy gap from the flat band to the dispersive bands at $\bk_c$. }
   \label{Fig:1:PRL}
\end{figure}

\textit{Kagome flat band model} ---  We consider a Bose-Hubbard Hamiltonian $H = H_0 + H_{int}$ in kagome lattice [see Fig.~\ref{Fig:1:PRL}(a)] whose one-particle Hamiltonian in momentum-space reads $H_0 = \sum_\bk\big( c^\dag_{\bk\alpha} \mathcal{H}_{\alpha\beta}(\bk)c_{\bk\beta} -\mu c^\dag_{\bk\alpha} c_{\bk\alpha}\big)$ with summations over repeated indices assumed. Here, $c_{\bk\alpha}$ annihilates a boson of momentum $\bk$ in the $\alpha$th sublattice and $\mu$ is the chemical potential. For kagome lattice there exists three sublattices and the hopping matrix $\mathcal{H}(\bk)$ is 
\begin{align}
\mathcal{H}(\bk) = 2t\begin{bmatrix}
	0 &  \cos(k_1/2) & \cos(k_2/2) \\
	\cos(k_1/2) & 0 & \cos( k_3/2) \\
	\cos(k_2/2) & \cos(k_3/2) & 0
\end{bmatrix},
\end{align}
where $k_i = \bk \cdot \textbf{a}_i$ for $i=\{1,2\}$ and $k_3 = k_1 - k_2$. Here $\textbf{a}_i$ are the basis vectors [Fig.~\ref{Fig:1:PRL}(a)] and $t>0$ is the nearest-neighbour hopping. One can diagonalize  $\mathcal{H}(\bk)$ as $\mathcal{H}(\bk) |u_n(\bk) \rangle = \epsilon_n(\bk)|u_{n}(\bk) \rangle$, where $\epsilon_n(\bk)$ ($|u_{n\bk} \rangle$) are the eigenenergies (Bloch states) and $n$ is the band index so that $\epsilon_1(\bk) \leq \epsilon_2(\bk) \leq \epsilon_3(\bk)$. The lowest Bloch band is strictly flat, i.e. $\epsilon_1(\bk) = -2t$, see Fig.~\ref{Fig:1:PRL}(b).   

The interaction Hamiltonian is $H_{int} = \frac{U}{2N} \sum_{\alpha\bk,\bk',\bq} c^\dag_{\bk\alpha} c_{\bk-\bq\alpha}c^\dag_{\bk'\alpha} c_{\bk'+\bq\alpha}$, where $N$ is the number of unit cells and $U>0$ describes the repulsive on-site interaction. Because the lowest band is flat, it is the interaction term that determines the momentum $\bk_c$ and Bloch state $|\phi_0 \rangle \equiv| u_1(\bk_c)\rangle$  in which the BEC takes place~\cite{you:2012}. Via a mean-field analysis~\cite{Pitaevskii2003,you:2012} it is shown that for kagome lattice the condensation takes place in one of the Dirac points, e.g. in $\bk_c = [4\pi/3,0]$ [black dot in Fig.~\ref{Fig:1:PRL}(b)] with $|\phi_0 \rangle =[-1,-1,1]^T$. For this Bloch state the particle density is distributed uniformly among all three sublattices so that the repulsive Hubbard interaction is minimized~\cite{you:2012}.

To analyze the stability and excitation properties of BEC, we utilize the multiband Bogoliubov approximation (details are provided in Ref.~\cite{julku2021}) where the bosonic operators for the condensate are treated as complex numbers, i.e. we write $c_{\bk_c\alpha} = \sqrt{Nn_0}\langle \alpha | \phi_0  \rangle$, where $n_0$ is the number of condensed bosons per unit cell and $\langle \alpha | \phi_0 \rangle$ is the projection of $| \phi_0 \rangle$ to the $\alpha$th sublattice. In the Bogoliubov theory, one considers only the interaction terms that are quadratic in fluctuations $c_{\bk\alpha}$ and $c^\dag_{\bk\alpha}$ with $\bk\neq\bk_c$. The total Hamiltonian is then $H = E_c + H_B$, where $E_c$ is a constant giving the ground energy of the condensate, and the Bogoliubov Hamiltonian $H_B$ describes the fluctuations of the condensate:
\begin{align}
\label{ham}
H_B = \frac{1}{2}\sum_{\bk}{}^{'} \Psi^\dag_\bk  \mathcal{H}_B(\bk)  \Psi_{\bk},
\end{align}
where $\mathcal{H}_B(\bk)$ is a $6\times 6$ matrix given by 
\begin{align}\label{BogMatrix}
&\mathcal{H}_B(\bk) = \begin{bmatrix}
	\mathcal{H}(\bk) -\mu_{\textrm{eff}} &  \Delta \\
	\Delta^* & \mathcal{H}^*(2\bk_c - \bk) -\mu_{\textrm{eff}},
\end{bmatrix}, \nonumber \\
& \Psi_\bk = [c_{\bk 1}, c_{\bk 2},c_{\bk 3}, c^\dag_{2\bk_c -\bk 1},c^\dag_{2\bk_c -\bk 2},c^\dag_{2\bk_c -\bk 3}]^T, \nonumber \\
&[\Delta]_{\alpha\beta} = \delta_{\alpha,\beta} U n_0/3, \nonumber \\
&[\mu_{\textrm{eff}}]_{\alpha\beta} = (\epsilon_0 - \frac{Un_0}{3})\delta_{\alpha,\beta}.
\end{align}
The primed sum in Eq.~\eqref{ham} includes the momenta for non-condensed states only, i.e. $\bk \neq \bk_c$ and $2\bk_c-\bk \neq \bk_c$.


The excitation energies of the BEC can be accessed by diagonalizing $L(\bk) \equiv \sigma_z \mathcal{H}_B(\bk)$, where $\sigma_z$ is the Pauli matrix acting in the particle-hole space~\cite{castin:book}. We then obtain Bogoliubov bands of the energies $E_3(\bk) \geq E_2(\bk) \geq E_1(\bk) \geq 0 \geq -E_1(2\bk_c-\bk) \geq -E_2(2\bk_c -\bk) \geq -E_3(2\bk_c-\bk)$. Positive (negative) energies describe quasi-particle (-hole) excitations and the corresponding quasi-particle (-hole) states are labelled as $|\psi^+_m (\bk) \rangle$ ($|\psi^-_m (\bk) \rangle$). The lowest quasi-particle energy band becomes gapless at $\bk_c$, i.e. $E_1(\bk \rightarrow \bk_c) = 0$, which corresponds to the Goldstone mode emerging from the spontaneous gauge $U(1)$ symmetry breaking of the complex phase of the BEC wavefunction~\cite{Fetter1971,Pitaevskii2003}.

\textit{Speed of sound of kagome flat band BEC}--- As the speed of sound $c_s$ for a BEC is given by the slope of the gapless Goldstone mode $E_1(\bk)$ at $\bk_c$, we write $\bk = \bk_c + \bq$, where $\bq << 1$. We then unitary transform $L(\bk)$ to the Bloch band basis and discard the dispersive bands of freedom to obtain the $2\times2$ matrix $L_p(\bk)$ projected to the flat band space:
\begin{align}
\label{projectedL}
L_p(\bk) = \frac{Un_0}{3}\begin{bmatrix} 1 & \alpha(\bq) \\
- \alpha^*(\bq)  & -1
\end{bmatrix} 
\end{align}
for $\bq \rightarrow 0$. Here, $\alpha(\bq) \equiv \langle u_1(\bk_c + \bq)| u_1(\bk_c - \bq) \rangle$. Diagonalizing \eqref{projectedL}, we find the Goldstone mode as $E_1(\bk_c+\bq) = \frac{Un_0}{3}D(\bq)$, where $D(\bq) = \sqrt{1- |\alpha(\bq)|^2}$ is the Hilbert-Schmidt quantum distance~\cite{Berry1989} which for fermionic flat band systems was recently shown to dictate the spread of the Landau levels~\cite{Rhim2020}. By definition, $0 \leq D(\bq) \leq 1$. We can immediately see that non-zero $D(\bq)$ is required to have finite speed of sound $c_s$.

By Taylor expanding the Bloch states up to second order in $\bq$, one finds for $c_s$
\begin{align}
\label{sos_kag}
c_s = \frac{2Un_0}{3}\sqrt{ g^1(\bk_c)},    
\end{align}
where the quantity inside the square root is called \textit{quantum metric} and defined as~\cite{Resta:2011}
\begin{align}
g^n_{\mu\nu}(\bk) = \textrm{Re}\Big[\bra{\partial_\mu u_n(\bk)} \Big(1 - | u_n(\bk) \rangle \langle u_n(\bk)  |\Big) | \partial_\nu u_n(\bk)\rangle \Big].  
\end{align}
with the notation $\partial_\mu = \frac{\partial}{\partial k_\mu}$. In case of kagome lattice we have $g^1_{xx}(\bk_c) = g^1_{yy}(\bk_c) \equiv g^1(\bk_c)$ and $g^1_{xy}(\bk_c) = g^1_{yx}(\bk_c) =0$. For anisotropic systems (for derivation see~\cite{julku2021}), $c_s(\theta_\bq) = \frac{2Un_0}{M}\sqrt{ \hat{\textbf{e}}^T_\bq g^1(\bk_c) \hat{\textbf{e}}_\bq}$, where $ \hat{\textbf{e}}_\bq = \bq/|\bq|$, $\tan \theta_\bq = q_y/q_x$, $[g^1]_{\mu\nu} = g^1_{\mu\nu}$, and $M$ the number of orbitals. 

A remarkable consequence of Eq.~\eqref{sos_kag} is that a \textit{finite quantum metric of the condensed state} guarantees finite $c_s$ -- and thus possibility for superfluidity -- even if the condensation takes place within a strictly flat band. Conversely, by measuring the speed of sound of  a flat band condensate, one can extract the quantum metric at the condensation point $\bk_c$. This should be compared to fermionic systems, where flat band superfluidity is guaranteed by finite Chern numbers or \textit{integrals} of the quantum metric over the first Brillouin zone (BZ)~\cite{Peotta2015,Liang2017,julku:2016}. Moreover, in Ref.~\cite{Torma2018} it was shown that for a fermionic two-body problem, the effective mass $m_{\textrm{eff}}^{C}$ of the Cooper pairs within a flat band is inversely proportional to the the quantum metric integrated over the whole BZ. Via the usual dependence of $c_s \propto 1/\sqrt{m_{\textrm{eff}}^{C}}$, one could anticipate a similar relationship between $c_s$ and quantum geometry. However, the result presented here is different: only the quantum metric of the condensed Bloch state is needed, not an integral over the whole BZ. Furthermore, in Ref.~\cite{Iskin2020} the speed of sound was analyzed for spin-orbit coupled Fermi gases: the Goldstone mode was shown to depend on the momentum-space integrals in which the quantum metric is convoluted with other non-geometric terms. Thus, the significance of quantum geometry was obscured due to the presence of more prominent non-geometric contributions. In contrast to this, we have shown that the quantum geometry plays a dominant role for determining the speed of sound in a flat band BEC.

In Fig.~\ref{Fig:1:PRL}(c) we plot $c_s$ for the kagome flat band condensate as a function of $U$ by numerically extracting the speed of sound from the full Bogoliubov Hamiltonian~\eqref{ham}. Moreover, we also plot the weak-coupling result of Eq.~\eqref{sos_kag}. The agreement at small $U$ is excellent.

Note that we find linear Goldstone modes for flat band condensates. In contrast, in Refs.~\cite{ozawa2012,barnett2012} the sound mode for spin-orbit (SO) coupled BEC is quadratic in the direction of dispersionless one-dimensional flat band. This is due to the inter-sublattice interaction term, induced by the SO coupling, and thus does not contradict our results as we only consider intra-sublattice interaction. 

\textit{Excitation density} --- An important question related to the stability of flat band BEC is how the excitation density $n_{ex}$ behaves,  in particular when $U\rightarrow 0$. For the usual dispersive band BEC, one has $\lim_{U\rightarrow 0}n_{ex}= 0$~\cite{Fetter1971}. However, for a strictly flat band, the $U\rightarrow 0$ limit of Eq.~\eqref{projectedL} implies that the Goldstone modes becomes flat. One could then conclude that the condensate becomes unstable as exciting particles out of the condensate does not cost energy. We now show that this is not the case as the quantum distance ensures the stability of a flat band BEC in the non-interacting limit.

The expression for $n_{\textrm{ex}}$ reads~\cite{julku2021}:
\begin{align}
\label{nex}
n_{ex} &= \frac{1}{N}\sum_{\bk m}{}^{'} \langle c^\dag_{\bk m} c_{\bk m} \rangle = \frac{1}{2N} \sum_{\bk m}{}^{'}[-1 + \langle \psi^-_m(\bk) |  \psi^-_m(\bk)  \rangle]\nonumber \\
&\equiv \frac{1}{N}\sum_{\bk}{}^{'}n_{ex}(\bk),
\end{align}
where $c^\dag_{\bk m}$ creates a boson in the Bloch band $m$ with momentum $\bk$. We again consider the projected $L_p(\bk)$ of Eq.~\eqref{projectedL} and neglect the higher bands as we are considering the $U\rightarrow 0$ limit. By diagonalizing Eq.~\eqref{projectedL}, one obtains
\begin{align}
\label{nex_ana}
\lim_{U\rightarrow 0}\langle c^\dag_{\bk 1} c_{\bk 1} \rangle =  \frac{1-D(\bq)}{2D(\bq)}, 
\end{align}
where $\bq = \bk - \bk_c$. 
Equation \eqref{nex_ana} provides a remarkable link between  the density of non-condensed bosons, $n_{\textrm{ex}}$, and the quantum distance $D(\bq)$. We see that $n_{ex}(\bk)$ diverges for $D(\bq) = 0$, implying the breakdown of the Bogoliubov theory. This is intuitively easy to understand as $D(\bq)=0$ indicates the perfect overlap between the condensed state $|\phi_0\rangle$ and other flat band condensates, i.e. $\langle u_1(\bk_c + \bq) | \phi_0 \rangle = 1$~\cite{julku2021}. On the other hand, finite $D(\bq)$ sets the limit for the excitation density, allowing a stable flat band BEC at arbitrarily small interaction values. The Eqs.~\ref{sos_kag} and \ref{nex_ana} are valid for any flat band with real Bloch functions $\ket{u_1}$; the relation to quantum geometry is similar also for arbitrary wavefunctions although the formula are slightly more complicated~\cite{julku2021}.

\begin{figure}
  \centering
    \includegraphics[width=1.0\columnwidth]{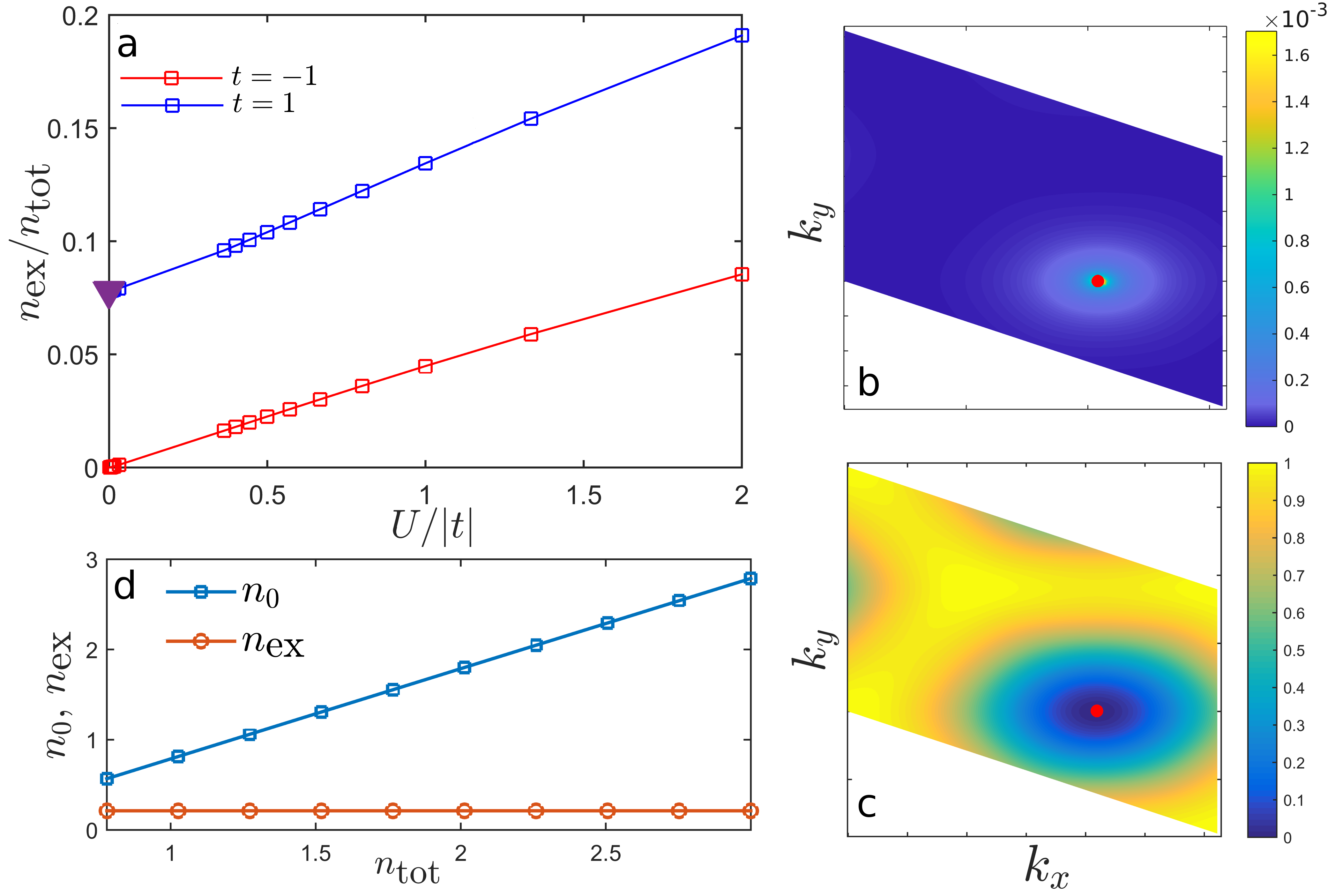}
    \caption{(a) Excitation fraction $n_{\textrm{ex}}/n_{\textrm{tot}}$ at $n_{\textrm{tot}}=3$ as a function of $U$ for the flat band BEC ($t=1$) and dispersive band BEC ($t=-1$). Purple triangle depicts the analytical result of Eq.~\eqref{nex_ana} integrated over the first BZ. (b) Momentum dependence of $n_{\textrm{ex}}(\bk)$ at  $Un_0/E_g= 5.13\times 10^{-4}$. (c) Quantum distance $D(\bq)$ as a function of $\bk = \bk_c + \bq$. In (b) and (c) the red dot depicts the momentum $\bk_c = [4\pi/3,0]$ of the flat band BEC. (d) Densities $n_0$ and $n_{\textrm{ex}}$ as a function of $n_{\textrm{tot}}$ for the flat band condensation at $U = |t|/1800$. Excitation density $n_{\textrm{ex}}$ remains constant, as it is determined by the quantum distance. }
   \label{Fig:2:PRL}
\end{figure}

In Fig.~\ref{Fig:2:PRL}(a) we present $n_{\textrm{ex}}/n_{\textrm{tot}}$, where $n_{\textrm{tot}}$ is the total density, for the kagome lattice as a function of  $U$. In addition to the flat band BEC, we also provide the result for dispersive band BEC. Condensation to one of the dispersive bands of the kagome lattice can be achieved by changing the sign of the NN hopping term, i.e. $t <0$. This choice flips the Bloch band structure such that the dispersive band is the lowest band for which the condensation takes place at $\bk_c=0$. From Fig.~\ref{Fig:2:PRL}(a) we see that $\lim_{U\rightarrow 0}n_{\textrm{ex}}=0$ for the dispersive band BEC, as expected. However, for the flat band BEC, the non-interacting asymptote of $n_{\textrm{ex}}$ is given by Eq.~\eqref{nex_ana} integrated over the first BZ. This clearly illustrates that the quantum distance  determines the excitation density and protects the stability of flat band BEC in the weak-coupling limit.

In Figs.~\ref{Fig:2:PRL}(b)-(c) we show $n_{\textrm{ex}}(\bk)$ for small $U$ and $D(\bq = \bk-\bk_c)$, respectively, as a function of momentum $\bk$ across the first BZ. We see that indeed the quantum distance is imprinted to the momentum distribution of excitation density. Importantly, $n_{ex}(\bk)$ is the Fourier transform of the following first-order spatial coherence function: $\tilde{g}^{(1)}(j) \equiv \frac{1}{N}\sum_{i\alpha} \langle \delta c^\dag_{i+j \alpha} \delta c_{i\alpha} \rangle$, where $\delta c_{i\alpha}$ annihilates a non-condensed boson in the $i$th unit cell and $\alpha$th sublattice.  Thus, first order coherence is fundamentally determined by quantum geometry, and measuring it provides a direct access to the quantum distance. 

Surprisingly, the number of atoms excited out of the condensate for a vanishing interaction strength, $\lim_{U\rightarrow 0}n_{\textrm{ex}}$ given by  Eq.~\eqref{nex_ana}, does not depend on the total density $n_{\textrm{tot}}$ but is solely determined by the quantum geometry of the flat band. This implies that by decreasing $n_{\textrm{tot}}$, the excitation fraction $n_{\textrm{ex}}/n_{\textrm{tot}}$ of the flat band BEC and the role of the interactions can be made large even at $U\rightarrow 0$. We demonstrate this in Fig.~\ref{Fig:2:PRL}(d) by presenting $n_0$ and $n_{\textrm{ex}}$ as a function of $n_{\textrm{tot}}$ for small $U$. We see that $n_{\textrm{ex}}$ remains constant, consistent with Eq.~\eqref{nex_ana}, whereas $n_0$ decreases with decreasing $n_{\textrm{tot}}$, implying that \textit{at the low density regime, the condensate depletion and interaction effects can be made significant, even at the non-interacting limit of $U\rightarrow 0$}.  The validity of the Bogoliubov theory for large $n_{\textrm{ex}}/n_{\textrm{tot}}$ ratios is addressed in Ref.~\cite{julku2021}.

\textit{Discussion}---By using Bogoliubov theory, we have studied fundamental connections between the excitations of a BEC and quantum geometry of the Bloch states. The properties of the flat band BEC are dictated by the underlying  quantum geometry and are strikingly different from the dispersive band case. The speed of sound $c_s$ is proportional to the quantum metric of the condensed state, and the excitation density $n_{\textrm{ex}}$ does not vanish with interactions as in case of a dispersive band BEC. In contrast, it obtains a finite value given by the quantum distance between the Bloch states. These results have a common origin; the quantum metric is the small momentum limit of the quantum distance, meaning that long-wavelength physical quantities such as $c_s$ and low energy excitations depend on the quantum metric, while those that involve higher momenta, e.g. $n_{\textrm{ex}}$, are governed by the quantum distance. While the quantum distance has been previously connected to Landau level spreading in non-interacting flat band models~\cite{Rhim2020}, our results are among the first to unravel the deep connections between the quantum distance and relevant physical quantities in an \textit{interacting many-body quantum system}.

Our predictions should be readily observable. The linear dependence of the speed of sound in a flat band BEC on the interaction strength is in stark contrast to the usual quadratic dependence of a dispersive band BEC and can be detected by tuning the interaction for example in experimental ultracold gas settings~\cite{Torma_book,bloch:2008}. Furthermore, as the excitation fraction is the Fourier transform of the first order coherence, measurement of the latter gives access to the quantum geometry effects. In addition to ultracold systems~\cite{Leung2020,Taie:2015}, flat band condensates can be also created in polaritonic platforms~\cite{Harder2020a,Baboux2016,Whittaker2018,Scafirimuto2021,Harder2020b} which therefore could be used to study quatum geometric effects discussed here.

Enhancing interaction effects has been a key motivation for studying flat band systems. The present work, alongside the accompanying study of Ref.~\cite{julku2021}, shows that indeed this promise is realized in the context of BEC. Even more importantly, we show that these effects are controlled by the  non-trivial quantum geometry. Therefore, bosons in a flat band provide a highly promising platform to explore beyond mean-field physics and effects of the quantum geometry, as well as to realize strong correlations even in the weak interaction limit. This is particularly important for photon and polariton systems where effective interactions in general are small. The results presented here are thus relevant for efforts of realizing strongly correlated photons, important for both fundamental research and opto-electronic components. In the future, it would be interesting to explore how quantum geometry affects the spatial and temporal dependence of the first and second order correlation functions, the physics of the strong interaction limit~\cite{Caleffi2020}, and driven-dissipative BECs.


\begin{acknowledgments}
\textit{Acknowledgements}
We thank Long Liang and Menderes Iskin for useful discussions. A.J.~and P.T.~acknowledge support by the Academy of Finland under project numbers 303351, 307419 and 327293. A.J.~acknowledges financial support from the Jenny and Antti Wihuri Foundation. This research was supported in part by the National Science Foundation under Grant No.~PHY-1748958. 
\end{acknowledgments}


\begin{thebibliography}{50}%
\makeatletter
\providecommand \@ifxundefined [1]{%
 \@ifx{#1\undefined}
}%
\providecommand \@ifnum [1]{%
 \ifnum #1\expandafter \@firstoftwo
 \else \expandafter \@secondoftwo
 \fi
}%
\providecommand \@ifx [1]{%
 \ifx #1\expandafter \@firstoftwo
 \else \expandafter \@secondoftwo
 \fi
}%
\providecommand \natexlab [1]{#1}%
\providecommand \enquote  [1]{``#1''}%
\providecommand \bibnamefont  [1]{#1}%
\providecommand \bibfnamefont [1]{#1}%
\providecommand \citenamefont [1]{#1}%
\providecommand \href@noop [0]{\@secondoftwo}%
\providecommand \href [0]{\begingroup \@sanitize@url \@href}%
\providecommand \@href[1]{\@@startlink{#1}\@@href}%
\providecommand \@@href[1]{\endgroup#1\@@endlink}%
\providecommand \@sanitize@url [0]{\catcode `\\12\catcode `\$12\catcode
  `\&12\catcode `\#12\catcode `\^12\catcode `\_12\catcode `\%12\relax}%
\providecommand \@@startlink[1]{}%
\providecommand \@@endlink[0]{}%
\providecommand \url  [0]{\begingroup\@sanitize@url \@url }%
\providecommand \@url [1]{\endgroup\@href {#1}{\urlprefix }}%
\providecommand \urlprefix  [0]{URL }%
\providecommand \Eprint [0]{\href }%
\providecommand \doibase [0]{http://dx.doi.org/}%
\providecommand \selectlanguage [0]{\@gobble}%
\providecommand \bibinfo  [0]{\@secondoftwo}%
\providecommand \bibfield  [0]{\@secondoftwo}%
\providecommand \translation [1]{[#1]}%
\providecommand \BibitemOpen [0]{}%
\providecommand \bibitemStop [0]{}%
\providecommand \bibitemNoStop [0]{.\EOS\space}%
\providecommand \EOS [0]{\spacefactor3000\relax}%
\providecommand \BibitemShut  [1]{\csname bibitem#1\endcsname}%
\let\auto@bib@innerbib\@empty
\bibitem [{\citenamefont {Resta}(2011)}]{Resta:2011}%
  \BibitemOpen
  \bibfield  {author} {\bibinfo {author} {\bibfnamefont {R.}~\bibnamefont
  {Resta}},\ }\href {\doibase 10.1140/epjb/e2010-10874-4} {\bibfield  {journal}
  {\bibinfo  {journal} {Eur. Phys. J. B}\ }\textbf {\bibinfo {volume} {79}},\
  \bibinfo {pages} {121} (\bibinfo {year} {2011})}\BibitemShut {NoStop}%
\bibitem [{\citenamefont {Provost}\ and\ \citenamefont
  {Vallee}(1980)}]{Provost:1980}%
  \BibitemOpen
  \bibfield  {author} {\bibinfo {author} {\bibfnamefont {J.~P.}\ \bibnamefont
  {Provost}}\ and\ \bibinfo {author} {\bibfnamefont {G.}~\bibnamefont
  {Vallee}},\ }\href {\doibase 10.1007/BF02193559} {\bibfield  {journal}
  {\bibinfo  {journal} {Commun. Math. Phys.}\ }\textbf {\bibinfo {volume}
  {76}},\ \bibinfo {pages} {289} (\bibinfo {year} {1980})}\BibitemShut
  {NoStop}%
\bibitem [{\citenamefont {Souza}\ \emph {et~al.}(2000)\citenamefont {Souza},
  \citenamefont {Wilkens},\ and\ \citenamefont {Martin}}]{Souza2000}%
  \BibitemOpen
  \bibfield  {author} {\bibinfo {author} {\bibfnamefont {I.}~\bibnamefont
  {Souza}}, \bibinfo {author} {\bibfnamefont {T.}~\bibnamefont {Wilkens}}, \
  and\ \bibinfo {author} {\bibfnamefont {R.~M.}\ \bibnamefont {Martin}},\
  }\href {\doibase 10.1103/PhysRevB.62.1666} {\bibfield  {journal} {\bibinfo
  {journal} {Phys. Rev. B}\ }\textbf {\bibinfo {volume} {62}},\ \bibinfo
  {pages} {1666} (\bibinfo {year} {2000})}\BibitemShut {NoStop}%
\bibitem [{\citenamefont {Neupert}\ \emph {et~al.}(2013)\citenamefont
  {Neupert}, \citenamefont {Chamon},\ and\ \citenamefont
  {Mudry}}]{Neupert:2013}%
  \BibitemOpen
  \bibfield  {author} {\bibinfo {author} {\bibfnamefont {T.}~\bibnamefont
  {Neupert}}, \bibinfo {author} {\bibfnamefont {C.}~\bibnamefont {Chamon}}, \
  and\ \bibinfo {author} {\bibfnamefont {C.}~\bibnamefont {Mudry}},\ }\href
  {\doibase 10.1103/PhysRevB.87.245103} {\bibfield  {journal} {\bibinfo
  {journal} {Phys. Rev. B}\ }\textbf {\bibinfo {volume} {87}},\ \bibinfo
  {pages} {245103} (\bibinfo {year} {2013})}\BibitemShut {NoStop}%
\bibitem [{\citenamefont {Gu}(2010)}]{Gu:2010}%
  \BibitemOpen
  \bibfield  {author} {\bibinfo {author} {\bibfnamefont {S.-J.}\ \bibnamefont
  {Gu}},\ }\href {\doibase 10.1142/S0217979210056335} {\bibfield  {journal}
  {\bibinfo  {journal} {Int. J. Mod. Phys. A}\ }\textbf {\bibinfo {volume}
  {24}},\ \bibinfo {pages} {4371} (\bibinfo {year} {2010})}\BibitemShut
  {NoStop}%
\bibitem [{\citenamefont {Klitzing}\ \emph {et~al.}(1980)\citenamefont
  {Klitzing}, \citenamefont {Dorda},\ and\ \citenamefont
  {Pepper}}]{Klitzing1980}%
  \BibitemOpen
  \bibfield  {author} {\bibinfo {author} {\bibfnamefont {K.~v.}\ \bibnamefont
  {Klitzing}}, \bibinfo {author} {\bibfnamefont {G.}~\bibnamefont {Dorda}}, \
  and\ \bibinfo {author} {\bibfnamefont {M.}~\bibnamefont {Pepper}},\ }\href
  {\doibase 10.1103/PhysRevLett.45.494} {\bibfield  {journal} {\bibinfo
  {journal} {Phys. Rev. Lett.}\ }\textbf {\bibinfo {volume} {45}},\ \bibinfo
  {pages} {494} (\bibinfo {year} {1980})}\BibitemShut {NoStop}%
\bibitem [{\citenamefont {Thouless}\ \emph {et~al.}(1982)\citenamefont
  {Thouless}, \citenamefont {Kohmoto}, \citenamefont {Nightingale},\ and\
  \citenamefont {Den~Nijs}}]{Thouless1982}%
  \BibitemOpen
  \bibfield  {author} {\bibinfo {author} {\bibfnamefont {D.~J.}\ \bibnamefont
  {Thouless}}, \bibinfo {author} {\bibfnamefont {M.}~\bibnamefont {Kohmoto}},
  \bibinfo {author} {\bibfnamefont {M.~P.}\ \bibnamefont {Nightingale}}, \ and\
  \bibinfo {author} {\bibfnamefont {M.}~\bibnamefont {Den~Nijs}},\ }\href
  {\doibase 10.1103/PhysRevLett.49.405} {\bibfield  {journal} {\bibinfo
  {journal} {Phys. Rev. Lett.}\ }\textbf {\bibinfo {volume} {49}},\ \bibinfo
  {pages} {405} (\bibinfo {year} {1982})}\BibitemShut {NoStop}%
\bibitem [{\citenamefont {Haldane}(1988)}]{Haldane1988}%
  \BibitemOpen
  \bibfield  {author} {\bibinfo {author} {\bibfnamefont {F.~D.~M.}\
  \bibnamefont {Haldane}},\ }\href {\doibase 10.1103/PhysRevLett.61.2015}
  {\bibfield  {journal} {\bibinfo  {journal} {Phys. Rev. Lett.}\ }\textbf
  {\bibinfo {volume} {61}},\ \bibinfo {pages} {2015} (\bibinfo {year}
  {1988})}\BibitemShut {NoStop}%
\bibitem [{\citenamefont {Kane}\ and\ \citenamefont {Mele}(2005)}]{Kane2005}%
  \BibitemOpen
  \bibfield  {author} {\bibinfo {author} {\bibfnamefont {C.~L.}\ \bibnamefont
  {Kane}}\ and\ \bibinfo {author} {\bibfnamefont {E.~J.}\ \bibnamefont
  {Mele}},\ }\href {\doibase 10.1103/PhysRevLett.95.226801} {\bibfield
  {journal} {\bibinfo  {journal} {Phys. Rev. Lett.}\ }\textbf {\bibinfo
  {volume} {95}},\ \bibinfo {pages} {226801} (\bibinfo {year}
  {2005})}\BibitemShut {NoStop}%
\bibitem [{\citenamefont {Hasan}\ and\ \citenamefont {Kane}(2010)}]{Hasan2010}%
  \BibitemOpen
  \bibfield  {author} {\bibinfo {author} {\bibfnamefont {M.~Z.}\ \bibnamefont
  {Hasan}}\ and\ \bibinfo {author} {\bibfnamefont {C.~L.}\ \bibnamefont
  {Kane}},\ }\href {\doibase 10.1103/RevModPhys.82.3045} {\bibfield  {journal}
  {\bibinfo  {journal} {Rev. Mod. Phys.}\ }\textbf {\bibinfo {volume} {82}},\
  \bibinfo {pages} {3045} (\bibinfo {year} {2010})}\BibitemShut {NoStop}%
\bibitem [{\citenamefont {Qi}\ and\ \citenamefont {Zhang}(2011)}]{Qi2011}%
  \BibitemOpen
  \bibfield  {author} {\bibinfo {author} {\bibfnamefont {X.-L.}\ \bibnamefont
  {Qi}}\ and\ \bibinfo {author} {\bibfnamefont {S.-C.}\ \bibnamefont {Zhang}},\
  }\href {\doibase 10.1103/RevModPhys.83.1057} {\bibfield  {journal} {\bibinfo
  {journal} {Rev. Mod. Phys.}\ }\textbf {\bibinfo {volume} {83}},\ \bibinfo
  {pages} {1057} (\bibinfo {year} {2011})}\BibitemShut {NoStop}%
\bibitem [{\citenamefont {Bernevig}\ and\ \citenamefont
  {Hughes}(2013)}]{Bernevig2013}%
  \BibitemOpen
  \bibfield  {author} {\bibinfo {author} {\bibfnamefont {B.~A.}\ \bibnamefont
  {Bernevig}}\ and\ \bibinfo {author} {\bibfnamefont {T.~L.}\ \bibnamefont
  {Hughes}},\ }\href@noop {} {\emph {\bibinfo {title} {{"Topological Insulators
  and Topological Superconductors"}}}}\ (\bibinfo  {publisher} {Princeton
  University Press},\ \bibinfo {year} {2013})\ p.\ \bibinfo {pages}
  {247}\BibitemShut {NoStop}%
\bibitem [{\citenamefont {{Gianfrate}}\ \emph {et~al.}(2020)\citenamefont
  {{Gianfrate}}, \citenamefont {{Bleu}}, \citenamefont {{Dominici}},
  \citenamefont {{Ardizzone}}, \citenamefont {{De Giorgi}}, \citenamefont
  {{Ballarini}}, \citenamefont {{Lerario}}, \citenamefont {{West}},
  \citenamefont {{Pfeiffer}}, \citenamefont {{Solnyshkov}}, \citenamefont
  {{Sanvitto}},\ and\ \citenamefont {{Malpuech}}}]{Gianfrate:2019}%
  \BibitemOpen
  \bibfield  {author} {\bibinfo {author} {\bibfnamefont {A.}~\bibnamefont
  {{Gianfrate}}}, \bibinfo {author} {\bibfnamefont {O.}~\bibnamefont {{Bleu}}},
  \bibinfo {author} {\bibfnamefont {L.}~\bibnamefont {{Dominici}}}, \bibinfo
  {author} {\bibfnamefont {V.}~\bibnamefont {{Ardizzone}}}, \bibinfo {author}
  {\bibfnamefont {M.}~\bibnamefont {{De Giorgi}}}, \bibinfo {author}
  {\bibfnamefont {D.}~\bibnamefont {{Ballarini}}}, \bibinfo {author}
  {\bibfnamefont {G.}~\bibnamefont {{Lerario}}}, \bibinfo {author}
  {\bibfnamefont {K.}~\bibnamefont {{West}}}, \bibinfo {author} {\bibfnamefont
  {L.~N.}\ \bibnamefont {{Pfeiffer}}}, \bibinfo {author} {\bibfnamefont
  {D.~D.}\ \bibnamefont {{Solnyshkov}}}, \bibinfo {author} {\bibfnamefont
  {D.}~\bibnamefont {{Sanvitto}}}, \ and\ \bibinfo {author} {\bibfnamefont
  {G.}~\bibnamefont {{Malpuech}}},\ }\href {\doibase
  https://doi.org/10.1038/s41586-020-1989-2} {\bibfield  {journal} {\bibinfo
  {journal} {Nature}\ }\textbf {\bibinfo {volume} {578}},\ \bibinfo {pages}
  {381} (\bibinfo {year} {2020})}\BibitemShut {NoStop}%
\bibitem [{\citenamefont {Tan}\ \emph {et~al.}(2019)\citenamefont {Tan},
  \citenamefont {Zhang}, \citenamefont {Yang}, \citenamefont {Chu},
  \citenamefont {Zhu}, \citenamefont {Li}, \citenamefont {Yang}, \citenamefont
  {Song}, \citenamefont {Han}, \citenamefont {Li}, \citenamefont {Dong},
  \citenamefont {Yu}, \citenamefont {Yan}, \citenamefont {Zhu},\ and\
  \citenamefont {Yu}}]{Tan2019}%
  \BibitemOpen
  \bibfield  {author} {\bibinfo {author} {\bibfnamefont {X.}~\bibnamefont
  {Tan}}, \bibinfo {author} {\bibfnamefont {D.-W.}\ \bibnamefont {Zhang}},
  \bibinfo {author} {\bibfnamefont {Z.}~\bibnamefont {Yang}}, \bibinfo {author}
  {\bibfnamefont {J.}~\bibnamefont {Chu}}, \bibinfo {author} {\bibfnamefont
  {Y.-Q.}\ \bibnamefont {Zhu}}, \bibinfo {author} {\bibfnamefont
  {D.}~\bibnamefont {Li}}, \bibinfo {author} {\bibfnamefont {X.}~\bibnamefont
  {Yang}}, \bibinfo {author} {\bibfnamefont {S.}~\bibnamefont {Song}}, \bibinfo
  {author} {\bibfnamefont {Z.}~\bibnamefont {Han}}, \bibinfo {author}
  {\bibfnamefont {Z.}~\bibnamefont {Li}}, \bibinfo {author} {\bibfnamefont
  {Y.}~\bibnamefont {Dong}}, \bibinfo {author} {\bibfnamefont {H.-F.}\
  \bibnamefont {Yu}}, \bibinfo {author} {\bibfnamefont {H.}~\bibnamefont
  {Yan}}, \bibinfo {author} {\bibfnamefont {S.-L.}\ \bibnamefont {Zhu}}, \ and\
  \bibinfo {author} {\bibfnamefont {Y.}~\bibnamefont {Yu}},\ }\href {\doibase
  10.1103/PhysRevLett.122.210401} {\bibfield  {journal} {\bibinfo  {journal}
  {Phys. Rev. Lett.}\ }\textbf {\bibinfo {volume} {122}},\ \bibinfo {pages}
  {210401} (\bibinfo {year} {2019})}\BibitemShut {NoStop}%
\bibitem [{\citenamefont {Leykam}\ \emph {et~al.}(2018)\citenamefont {Leykam},
  \citenamefont {Andreanov},\ and\ \citenamefont {Flach}}]{Leykam:2018}%
  \BibitemOpen
  \bibfield  {author} {\bibinfo {author} {\bibfnamefont {D.}~\bibnamefont
  {Leykam}}, \bibinfo {author} {\bibfnamefont {A.}~\bibnamefont {Andreanov}}, \
  and\ \bibinfo {author} {\bibfnamefont {S.}~\bibnamefont {Flach}},\ }\href
  {\doibase 10.1080/23746149.2018.1473052} {\bibfield  {journal} {\bibinfo
  {journal} {Advances in Physics: X}\ }\textbf {\bibinfo {volume} {3}},\
  \bibinfo {pages} {1473052} (\bibinfo {year} {2018})}\BibitemShut {NoStop}%
\bibitem [{\citenamefont {Liang}\ \emph
  {et~al.}(2017{\natexlab{a}})\citenamefont {Liang}, \citenamefont {Peotta},
  \citenamefont {Harju},\ and\ \citenamefont {T\"orm\"a}}]{Liang:2017b}%
  \BibitemOpen
  \bibfield  {author} {\bibinfo {author} {\bibfnamefont {L.}~\bibnamefont
  {Liang}}, \bibinfo {author} {\bibfnamefont {S.}~\bibnamefont {Peotta}},
  \bibinfo {author} {\bibfnamefont {A.}~\bibnamefont {Harju}}, \ and\ \bibinfo
  {author} {\bibfnamefont {P.}~\bibnamefont {T\"orm\"a}},\ }\href@noop {}
  {\bibfield  {journal} {\bibinfo  {journal} {Phys. Rev. B}\ } (\bibinfo {year}
  {2017}{\natexlab{a}})}\BibitemShut {NoStop}%
\bibitem [{\citenamefont {Peotta}\ and\ \citenamefont
  {T{\"o}rm{\"a}}(2015)}]{Peotta2015}%
  \BibitemOpen
  \bibfield  {author} {\bibinfo {author} {\bibfnamefont {S.}~\bibnamefont
  {Peotta}}\ and\ \bibinfo {author} {\bibfnamefont {P.}~\bibnamefont
  {T{\"o}rm{\"a}}},\ }\href {http://dx.doi.org/10.1038/ncomms9944} {\bibfield
  {journal} {\bibinfo  {journal} {Nat. Commun.}\ }\textbf {\bibinfo {volume}
  {6}},\ \bibinfo {pages} {8944} (\bibinfo {year} {2015})}\BibitemShut
  {NoStop}%
\bibitem [{\citenamefont {Liang}\ \emph
  {et~al.}(2017{\natexlab{b}})\citenamefont {Liang}, \citenamefont {Vanhala},
  \citenamefont {Peotta}, \citenamefont {Siro}, \citenamefont {Harju},\ and\
  \citenamefont {T\"orm\"a}}]{Liang2017}%
  \BibitemOpen
  \bibfield  {author} {\bibinfo {author} {\bibfnamefont {L.}~\bibnamefont
  {Liang}}, \bibinfo {author} {\bibfnamefont {T.~I.}\ \bibnamefont {Vanhala}},
  \bibinfo {author} {\bibfnamefont {S.}~\bibnamefont {Peotta}}, \bibinfo
  {author} {\bibfnamefont {T.}~\bibnamefont {Siro}}, \bibinfo {author}
  {\bibfnamefont {A.}~\bibnamefont {Harju}}, \ and\ \bibinfo {author}
  {\bibfnamefont {P.}~\bibnamefont {T\"orm\"a}},\ }\href {\doibase
  10.1103/PhysRevB.95.024515} {\bibfield  {journal} {\bibinfo  {journal} {Phys.
  Rev. B}\ }\textbf {\bibinfo {volume} {95}},\ \bibinfo {pages} {024515}
  (\bibinfo {year} {2017}{\natexlab{b}})}\BibitemShut {NoStop}%
\bibitem [{\citenamefont {Julku}\ \emph
  {et~al.}(2016{\natexlab{a}})\citenamefont {Julku}, \citenamefont {Peotta},
  \citenamefont {Vanhala}, \citenamefont {Kim},\ and\ \citenamefont
  {T\"orm\"a}}]{julku2016}%
  \BibitemOpen
  \bibfield  {author} {\bibinfo {author} {\bibfnamefont {A.}~\bibnamefont
  {Julku}}, \bibinfo {author} {\bibfnamefont {S.}~\bibnamefont {Peotta}},
  \bibinfo {author} {\bibfnamefont {T.~I.}\ \bibnamefont {Vanhala}}, \bibinfo
  {author} {\bibfnamefont {D.-H.}\ \bibnamefont {Kim}}, \ and\ \bibinfo
  {author} {\bibfnamefont {P.}~\bibnamefont {T\"orm\"a}},\ }\href {\doibase
  10.1103/PhysRevLett.117.045303} {\bibfield  {journal} {\bibinfo  {journal}
  {Phys. Rev. Lett.}\ }\textbf {\bibinfo {volume} {117}},\ \bibinfo {pages}
  {045303} (\bibinfo {year} {2016}{\natexlab{a}})}\BibitemShut {NoStop}%
\bibitem [{\citenamefont {{Hu}}\ \emph {et~al.}(2019)\citenamefont {{Hu}},
  \citenamefont {{Hyart}}, \citenamefont {{Pikulin}},\ and\ \citenamefont
  {{Rossi}}}]{Hu:2019}%
  \BibitemOpen
  \bibfield  {author} {\bibinfo {author} {\bibfnamefont {X.}~\bibnamefont
  {{Hu}}}, \bibinfo {author} {\bibfnamefont {T.}~\bibnamefont {{Hyart}}},
  \bibinfo {author} {\bibfnamefont {D.}~\bibnamefont {{Pikulin}}}, \ and\
  \bibinfo {author} {\bibfnamefont {E.}~\bibnamefont {{Rossi}}},\ }\href
  {\doibase https://doi.org/10.1103/PhysRevLett.123.237002} {\bibfield
  {journal} {\bibinfo  {journal} {Phys. Rev. Lett.}\ }\textbf {\bibinfo
  {volume} {123}},\ \bibinfo {pages} {237002} (\bibinfo {year}
  {2019})}\BibitemShut {NoStop}%
\bibitem [{\citenamefont {Julku}\ \emph {et~al.}(2020)\citenamefont {Julku},
  \citenamefont {Peltonen}, \citenamefont {Liang}, \citenamefont {Heikkil\"a},\
  and\ \citenamefont {T\"orm\"a}}]{Julku:2020}%
  \BibitemOpen
  \bibfield  {author} {\bibinfo {author} {\bibfnamefont {A.}~\bibnamefont
  {Julku}}, \bibinfo {author} {\bibfnamefont {T.~J.}\ \bibnamefont {Peltonen}},
  \bibinfo {author} {\bibfnamefont {L.}~\bibnamefont {Liang}}, \bibinfo
  {author} {\bibfnamefont {T.~T.}\ \bibnamefont {Heikkil\"a}}, \ and\ \bibinfo
  {author} {\bibfnamefont {P.}~\bibnamefont {T\"orm\"a}},\ }\href {\doibase
  10.1103/PhysRevB.101.060505} {\bibfield  {journal} {\bibinfo  {journal}
  {Phys. Rev. B}\ }\textbf {\bibinfo {volume} {101}},\ \bibinfo {pages}
  {060505} (\bibinfo {year} {2020})}\BibitemShut {NoStop}%
\bibitem [{\citenamefont {Xie}\ \emph {et~al.}(2020)\citenamefont {Xie},
  \citenamefont {Song}, \citenamefont {Lian},\ and\ \citenamefont
  {Bernevig}}]{xie:2019}%
  \BibitemOpen
  \bibfield  {author} {\bibinfo {author} {\bibfnamefont {F.}~\bibnamefont
  {Xie}}, \bibinfo {author} {\bibfnamefont {Z.}~\bibnamefont {Song}}, \bibinfo
  {author} {\bibfnamefont {B.}~\bibnamefont {Lian}}, \ and\ \bibinfo {author}
  {\bibfnamefont {B.}~\bibnamefont {Bernevig}},\ }\href {\doibase
  10.1103/PhysRevLett.124.167002} {\bibfield  {journal} {\bibinfo  {journal}
  {Phys. Rev. Lett}\ }\textbf {\bibinfo {volume} {124}},\ \bibinfo {pages}
  {167002} (\bibinfo {year} {2020})}\BibitemShut {NoStop}%
\bibitem [{\citenamefont {Classen}\ \emph {et~al.}(2019)\citenamefont
  {Classen}, \citenamefont {Honerkamp},\ and\ \citenamefont
  {Scherer}}]{classen:2019}%
  \BibitemOpen
  \bibfield  {author} {\bibinfo {author} {\bibfnamefont {L.}~\bibnamefont
  {Classen}}, \bibinfo {author} {\bibfnamefont {C.}~\bibnamefont {Honerkamp}},
  \ and\ \bibinfo {author} {\bibfnamefont {M.~M.}\ \bibnamefont {Scherer}},\
  }\href {\doibase 10.1103/PhysRevB.99.195120} {\bibfield  {journal} {\bibinfo
  {journal} {Phys. Rev. B}\ }\textbf {\bibinfo {volume} {99}},\ \bibinfo
  {pages} {195120} (\bibinfo {year} {2019})}\BibitemShut {NoStop}%
\bibitem [{\citenamefont {Cao}\ \emph {et~al.}(2018)\citenamefont {Cao},
  \citenamefont {Fatemi}, \citenamefont {Fang}, \citenamefont {Watanabe},
  \citenamefont {Taniguchi}, \citenamefont {Kaxiras},\ and\ \citenamefont
  {Jarillo-Herrero}}]{Cao:2018}%
  \BibitemOpen
  \bibfield  {author} {\bibinfo {author} {\bibfnamefont {Y.}~\bibnamefont
  {Cao}}, \bibinfo {author} {\bibfnamefont {V.}~\bibnamefont {Fatemi}},
  \bibinfo {author} {\bibfnamefont {S.}~\bibnamefont {Fang}}, \bibinfo {author}
  {\bibfnamefont {K.}~\bibnamefont {Watanabe}}, \bibinfo {author}
  {\bibfnamefont {T.}~\bibnamefont {Taniguchi}}, \bibinfo {author}
  {\bibfnamefont {E.}~\bibnamefont {Kaxiras}}, \ and\ \bibinfo {author}
  {\bibfnamefont {P.}~\bibnamefont {Jarillo-Herrero}},\ }\href@noop {}
  {\bibfield  {journal} {\bibinfo  {journal} {Nature}\ } (\bibinfo {year}
  {2018})}\BibitemShut {NoStop}%
\bibitem [{\citenamefont {Yankowitz}\ \emph {et~al.}(2019)\citenamefont
  {Yankowitz}, \citenamefont {Chen}, \citenamefont {Polshyn}, \citenamefont
  {Zhang}, \citenamefont {Watanabe}, \citenamefont {Taniguchi}, \citenamefont
  {Graf}, \citenamefont {Young},\ and\ \citenamefont {Dean}}]{yankowitz:2019}%
  \BibitemOpen
  \bibfield  {author} {\bibinfo {author} {\bibfnamefont {M.}~\bibnamefont
  {Yankowitz}}, \bibinfo {author} {\bibfnamefont {S.}~\bibnamefont {Chen}},
  \bibinfo {author} {\bibfnamefont {H.}~\bibnamefont {Polshyn}}, \bibinfo
  {author} {\bibfnamefont {Y.}~\bibnamefont {Zhang}}, \bibinfo {author}
  {\bibfnamefont {K.}~\bibnamefont {Watanabe}}, \bibinfo {author}
  {\bibfnamefont {T.}~\bibnamefont {Taniguchi}}, \bibinfo {author}
  {\bibfnamefont {D.}~\bibnamefont {Graf}}, \bibinfo {author} {\bibfnamefont
  {A.~F.}\ \bibnamefont {Young}}, \ and\ \bibinfo {author} {\bibfnamefont
  {C.~R.}\ \bibnamefont {Dean}},\ }\href {\doibase 10.1126/science.aav1910}
  {\bibfield  {journal} {\bibinfo  {journal} {Science}\ }\textbf {\bibinfo
  {volume} {363}},\ \bibinfo {pages} {1059} (\bibinfo {year}
  {2019})}\BibitemShut {NoStop}%
\bibitem [{\citenamefont {Taie}\ \emph {et~al.}(2015)\citenamefont {Taie},
  \citenamefont {Ozawa}, \citenamefont {Ichinose}, \citenamefont {Nishio},
  \citenamefont {Nakajima},\ and\ \citenamefont {Takahashi}}]{Taie:2015}%
  \BibitemOpen
  \bibfield  {author} {\bibinfo {author} {\bibfnamefont {S.}~\bibnamefont
  {Taie}}, \bibinfo {author} {\bibfnamefont {H.}~\bibnamefont {Ozawa}},
  \bibinfo {author} {\bibfnamefont {T.}~\bibnamefont {Ichinose}}, \bibinfo
  {author} {\bibfnamefont {T.}~\bibnamefont {Nishio}}, \bibinfo {author}
  {\bibfnamefont {S.}~\bibnamefont {Nakajima}}, \ and\ \bibinfo {author}
  {\bibfnamefont {Y.}~\bibnamefont {Takahashi}},\ }\href {\doibase
  10.1126/sciadv.1500854} {\bibfield  {journal} {\bibinfo  {journal} {Science
  Adv.}\ }\textbf {\bibinfo {volume} {1}} (\bibinfo {year} {2015}),\
  10.1126/sciadv.1500854}\BibitemShut {NoStop}%
\bibitem [{\citenamefont {Vicencio}\ \emph {et~al.}(2015)\citenamefont
  {Vicencio}, \citenamefont {Cantillano}, \citenamefont {Morales-Inostroza},
  \citenamefont {Real}, \citenamefont {Mej{\'{i}}a-Cort{\'{e}}s}, \citenamefont
  {Weimann}, \citenamefont {Szameit},\ and\ \citenamefont
  {Molina}}]{Vicencio2015}%
  \BibitemOpen
  \bibfield  {author} {\bibinfo {author} {\bibfnamefont {R.~A.}\ \bibnamefont
  {Vicencio}}, \bibinfo {author} {\bibfnamefont {C.}~\bibnamefont
  {Cantillano}}, \bibinfo {author} {\bibfnamefont {L.}~\bibnamefont
  {Morales-Inostroza}}, \bibinfo {author} {\bibfnamefont {B.}~\bibnamefont
  {Real}}, \bibinfo {author} {\bibfnamefont {C.}~\bibnamefont
  {Mej{\'{i}}a-Cort{\'{e}}s}}, \bibinfo {author} {\bibfnamefont
  {S.}~\bibnamefont {Weimann}}, \bibinfo {author} {\bibfnamefont
  {A.}~\bibnamefont {Szameit}}, \ and\ \bibinfo {author} {\bibfnamefont
  {M.~I.}\ \bibnamefont {Molina}},\ }\href {\doibase
  10.1103/PhysRevLett.114.245503} {\bibfield  {journal} {\bibinfo  {journal}
  {Phys. Rev. Lett.}\ }\textbf {\bibinfo {volume} {114}},\ \bibinfo {pages}
  {245503} (\bibinfo {year} {2015})}\BibitemShut {NoStop}%
\bibitem [{\citenamefont {Mukherjee}\ \emph {et~al.}(2015)\citenamefont
  {Mukherjee}, \citenamefont {Spracklen}, \citenamefont {Choudhury},
  \citenamefont {Goldman}, \citenamefont {{\"{O}}hberg}, \citenamefont
  {Andersson},\ and\ \citenamefont {Thomson}}]{Mukherjee2015}%
  \BibitemOpen
  \bibfield  {author} {\bibinfo {author} {\bibfnamefont {S.}~\bibnamefont
  {Mukherjee}}, \bibinfo {author} {\bibfnamefont {A.}~\bibnamefont
  {Spracklen}}, \bibinfo {author} {\bibfnamefont {D.}~\bibnamefont
  {Choudhury}}, \bibinfo {author} {\bibfnamefont {N.}~\bibnamefont {Goldman}},
  \bibinfo {author} {\bibfnamefont {P.}~\bibnamefont {{\"{O}}hberg}}, \bibinfo
  {author} {\bibfnamefont {E.}~\bibnamefont {Andersson}}, \ and\ \bibinfo
  {author} {\bibfnamefont {R.~R.}\ \bibnamefont {Thomson}},\ }\href {\doibase
  10.1103/PhysRevLett.114.245504} {\bibfield  {journal} {\bibinfo  {journal}
  {Phys. Rev. Lett.}\ }\textbf {\bibinfo {volume} {114}},\ \bibinfo {pages}
  {245504} (\bibinfo {year} {2015})}\BibitemShut {NoStop}%
\bibitem [{\citenamefont {Kajiwara}\ \emph {et~al.}(2016)\citenamefont
  {Kajiwara}, \citenamefont {Urade}, \citenamefont {Nakata}, \citenamefont
  {Nakanishi},\ and\ \citenamefont {Kitano}}]{kajiwara2016}%
  \BibitemOpen
  \bibfield  {author} {\bibinfo {author} {\bibfnamefont {S.}~\bibnamefont
  {Kajiwara}}, \bibinfo {author} {\bibfnamefont {Y.}~\bibnamefont {Urade}},
  \bibinfo {author} {\bibfnamefont {Y.}~\bibnamefont {Nakata}}, \bibinfo
  {author} {\bibfnamefont {T.}~\bibnamefont {Nakanishi}}, \ and\ \bibinfo
  {author} {\bibfnamefont {M.}~\bibnamefont {Kitano}},\ }\href {\doibase
  10.1103/PhysRevB.93.075126} {\bibfield  {journal} {\bibinfo  {journal} {Phys.
  Rev. B}\ }\textbf {\bibinfo {volume} {93}},\ \bibinfo {pages} {075126}
  (\bibinfo {year} {2016})}\BibitemShut {NoStop}%
\bibitem [{\citenamefont {Harder}\ \emph
  {et~al.}(2020{\natexlab{a}})\citenamefont {Harder}, \citenamefont {Egorov},
  \citenamefont {Beierlein}, \citenamefont {Gagel}, \citenamefont {Michl},
  \citenamefont {Emmerling}, \citenamefont {Schneider}, \citenamefont
  {Peschel}, \citenamefont {H\"ofling},\ and\ \citenamefont
  {Klembt}}]{Harder2020a}%
  \BibitemOpen
  \bibfield  {author} {\bibinfo {author} {\bibfnamefont {T.~H.}\ \bibnamefont
  {Harder}}, \bibinfo {author} {\bibfnamefont {O.~A.}\ \bibnamefont {Egorov}},
  \bibinfo {author} {\bibfnamefont {J.}~\bibnamefont {Beierlein}}, \bibinfo
  {author} {\bibfnamefont {P.}~\bibnamefont {Gagel}}, \bibinfo {author}
  {\bibfnamefont {J.}~\bibnamefont {Michl}}, \bibinfo {author} {\bibfnamefont
  {M.}~\bibnamefont {Emmerling}}, \bibinfo {author} {\bibfnamefont
  {C.}~\bibnamefont {Schneider}}, \bibinfo {author} {\bibfnamefont
  {U.}~\bibnamefont {Peschel}}, \bibinfo {author} {\bibfnamefont
  {S.}~\bibnamefont {H\"ofling}}, \ and\ \bibinfo {author} {\bibfnamefont
  {S.}~\bibnamefont {Klembt}},\ }\href {\doibase 10.1103/PhysRevB.102.121302}
  {\bibfield  {journal} {\bibinfo  {journal} {Phys. Rev. B}\ }\textbf {\bibinfo
  {volume} {102}},\ \bibinfo {pages} {121302} (\bibinfo {year}
  {2020}{\natexlab{a}})}\BibitemShut {NoStop}%
\bibitem [{\citenamefont {Baboux}\ \emph {et~al.}(2016)\citenamefont {Baboux},
  \citenamefont {Ge}, \citenamefont {Jacqmin}, \citenamefont {Biondi},
  \citenamefont {Galopin}, \citenamefont {Lema{\^{i}}tre}, \citenamefont {{Le
  Gratiet}}, \citenamefont {Sagnes}, \citenamefont {Schmidt}, \citenamefont
  {T{\"{u}}reci}, \citenamefont {Amo},\ and\ \citenamefont
  {Bloch}}]{Baboux2016}%
  \BibitemOpen
  \bibfield  {author} {\bibinfo {author} {\bibfnamefont {F.}~\bibnamefont
  {Baboux}}, \bibinfo {author} {\bibfnamefont {L.}~\bibnamefont {Ge}}, \bibinfo
  {author} {\bibfnamefont {T.}~\bibnamefont {Jacqmin}}, \bibinfo {author}
  {\bibfnamefont {M.}~\bibnamefont {Biondi}}, \bibinfo {author} {\bibfnamefont
  {E.}~\bibnamefont {Galopin}}, \bibinfo {author} {\bibfnamefont
  {A.}~\bibnamefont {Lema{\^{i}}tre}}, \bibinfo {author} {\bibfnamefont
  {L.}~\bibnamefont {{Le Gratiet}}}, \bibinfo {author} {\bibfnamefont
  {I.}~\bibnamefont {Sagnes}}, \bibinfo {author} {\bibfnamefont
  {S.}~\bibnamefont {Schmidt}}, \bibinfo {author} {\bibfnamefont {H.~E.}\
  \bibnamefont {T{\"{u}}reci}}, \bibinfo {author} {\bibfnamefont
  {A.}~\bibnamefont {Amo}}, \ and\ \bibinfo {author} {\bibfnamefont
  {J.}~\bibnamefont {Bloch}},\ }\href {\doibase 10.1103/PhysRevLett.116.066402}
  {\bibfield  {journal} {\bibinfo  {journal} {Phys. Rev. Lett.}\ }\textbf
  {\bibinfo {volume} {116}},\ \bibinfo {pages} {066402} (\bibinfo {year}
  {2016})}\BibitemShut {NoStop}%
\bibitem [{\citenamefont {Whittaker}\ \emph {et~al.}(2018)\citenamefont
  {Whittaker}, \citenamefont {Cancellieri}, \citenamefont {Walker},
  \citenamefont {Gulevich}, \citenamefont {Schomerus}, \citenamefont
  {Vaitiekus}, \citenamefont {Royall}, \citenamefont {Whittaker}, \citenamefont
  {Clarke}, \citenamefont {Iorsh}, \citenamefont {Shelykh}, \citenamefont
  {Skolnick},\ and\ \citenamefont {Krizhanovskii}}]{Whittaker2018}%
  \BibitemOpen
  \bibfield  {author} {\bibinfo {author} {\bibfnamefont {C.~E.}\ \bibnamefont
  {Whittaker}}, \bibinfo {author} {\bibfnamefont {E.}~\bibnamefont
  {Cancellieri}}, \bibinfo {author} {\bibfnamefont {P.~M.}\ \bibnamefont
  {Walker}}, \bibinfo {author} {\bibfnamefont {D.~R.}\ \bibnamefont
  {Gulevich}}, \bibinfo {author} {\bibfnamefont {H.}~\bibnamefont {Schomerus}},
  \bibinfo {author} {\bibfnamefont {D.}~\bibnamefont {Vaitiekus}}, \bibinfo
  {author} {\bibfnamefont {B.}~\bibnamefont {Royall}}, \bibinfo {author}
  {\bibfnamefont {D.~M.}\ \bibnamefont {Whittaker}}, \bibinfo {author}
  {\bibfnamefont {E.}~\bibnamefont {Clarke}}, \bibinfo {author} {\bibfnamefont
  {I.~V.}\ \bibnamefont {Iorsh}}, \bibinfo {author} {\bibfnamefont {I.~A.}\
  \bibnamefont {Shelykh}}, \bibinfo {author} {\bibfnamefont {M.~S.}\
  \bibnamefont {Skolnick}}, \ and\ \bibinfo {author} {\bibfnamefont {D.~N.}\
  \bibnamefont {Krizhanovskii}},\ }\href {\doibase
  10.1103/PhysRevLett.120.097401} {\bibfield  {journal} {\bibinfo  {journal}
  {Phys. Rev. Lett.}\ }\textbf {\bibinfo {volume} {120}},\ \bibinfo {pages}
  {097401} (\bibinfo {year} {2018})}\BibitemShut {NoStop}%
\bibitem [{\citenamefont {Scafirimuto}\ \emph {et~al.}(2021)\citenamefont
  {Scafirimuto}, \citenamefont {Urbonas}, \citenamefont {Becker}, \citenamefont
  {Scherf}, \citenamefont {Mahrt},\ and\ \citenamefont
  {St\"oferle}}]{Scafirimuto2021}%
  \BibitemOpen
  \bibfield  {author} {\bibinfo {author} {\bibfnamefont {F.}~\bibnamefont
  {Scafirimuto}}, \bibinfo {author} {\bibfnamefont {D.}~\bibnamefont
  {Urbonas}}, \bibinfo {author} {\bibfnamefont {M.~A.}\ \bibnamefont {Becker}},
  \bibinfo {author} {\bibfnamefont {U.}~\bibnamefont {Scherf}}, \bibinfo
  {author} {\bibfnamefont {R.~F.}\ \bibnamefont {Mahrt}}, \ and\ \bibinfo
  {author} {\bibfnamefont {T.}~\bibnamefont {St\"oferle}},\ }\href {\doibase
  10.1038/s42005-021-00548-w} {\bibfield  {journal} {\bibinfo  {journal}
  {Communications Physics}\ }\textbf {\bibinfo {volume} {4}},\ \bibinfo {pages}
  {39} (\bibinfo {year} {2021})}\BibitemShut {NoStop}%
\bibitem [{\citenamefont {Harder}\ \emph
  {et~al.}(2020{\natexlab{b}})\citenamefont {Harder}, \citenamefont {Egorov},
  \citenamefont {Krause}, \citenamefont {Beierlein}, \citenamefont {Gagel},
  \citenamefont {Emmerling}, \citenamefont {Schneider}, \citenamefont
  {Peschel}, \citenamefont {H\"ofling},\ and\ \citenamefont
  {Klembt}}]{Harder2020b}%
  \BibitemOpen
  \bibfield  {author} {\bibinfo {author} {\bibfnamefont {T.~H.}\ \bibnamefont
  {Harder}}, \bibinfo {author} {\bibfnamefont {O.~A.}\ \bibnamefont {Egorov}},
  \bibinfo {author} {\bibfnamefont {C.}~\bibnamefont {Krause}}, \bibinfo
  {author} {\bibfnamefont {J.}~\bibnamefont {Beierlein}}, \bibinfo {author}
  {\bibfnamefont {P.}~\bibnamefont {Gagel}}, \bibinfo {author} {\bibfnamefont
  {M.}~\bibnamefont {Emmerling}}, \bibinfo {author} {\bibfnamefont
  {C.}~\bibnamefont {Schneider}}, \bibinfo {author} {\bibfnamefont
  {U.}~\bibnamefont {Peschel}}, \bibinfo {author} {\bibfnamefont
  {S.}~\bibnamefont {H\"ofling}}, \ and\ \bibinfo {author} {\bibfnamefont
  {S.}~\bibnamefont {Klembt}},\ }\href {https://arxiv.org/abs/2011.10766}
  {\bibfield  {journal} {\bibinfo  {journal} {arXiv:2011.10766}\ } (\bibinfo
  {year} {2020}{\natexlab{b}})}\BibitemShut {NoStop}%
\bibitem [{jul()}]{julku2021}%
  \BibitemOpen
  \href@noop {} {}\bibinfo {note} {A. Julku, G. M. Bruun and P. T\"orm\"a,
  Excitations of a Bose-Einstein condensate and the quantum geometry of a flat
  band, arXiv:2104.14257 (2021)}\BibitemShut {NoStop}%
\bibitem [{\citenamefont {Fetter}\ and\ \citenamefont
  {Walecka}(1971)}]{Fetter1971}%
  \BibitemOpen
  \bibfield  {author} {\bibinfo {author} {\bibfnamefont {A.}~\bibnamefont
  {Fetter}}\ and\ \bibinfo {author} {\bibfnamefont {J.}~\bibnamefont
  {Walecka}},\ }\href@noop {} {\emph {\bibinfo {title} {Quantum Theory of
  Many-Particle Systems}}},\ Dover Books on Physics Series\ (\bibinfo
  {publisher} {Dover Publications},\ \bibinfo {year} {1971})\BibitemShut
  {NoStop}%
\bibitem [{\citenamefont {Pitaevskii}\ and\ \citenamefont
  {Stringari}(2003)}]{Pitaevskii2003}%
  \BibitemOpen
  \bibfield  {author} {\bibinfo {author} {\bibfnamefont {L.}~\bibnamefont
  {Pitaevskii}}\ and\ \bibinfo {author} {\bibfnamefont {S.}~\bibnamefont
  {Stringari}},\ }\href@noop {} {\emph {\bibinfo {title} {Bose-Einstein
  Condensation}}}\ (\bibinfo  {publisher} {Oxford University Press},\ \bibinfo
  {year} {2003})\BibitemShut {NoStop}%
\bibitem [{\citenamefont {You}\ \emph {et~al.}(2012)\citenamefont {You},
  \citenamefont {Chen}, \citenamefont {Sun},\ and\ \citenamefont
  {Zhai}}]{you:2012}%
  \BibitemOpen
  \bibfield  {author} {\bibinfo {author} {\bibfnamefont {Y.-Z.}\ \bibnamefont
  {You}}, \bibinfo {author} {\bibfnamefont {Z.}~\bibnamefont {Chen}}, \bibinfo
  {author} {\bibfnamefont {X.-Q.}\ \bibnamefont {Sun}}, \ and\ \bibinfo
  {author} {\bibfnamefont {H.}~\bibnamefont {Zhai}},\ }\href {\doibase
  10.1103/PhysRevLett.109.265302} {\bibfield  {journal} {\bibinfo  {journal}
  {Phys. Rev. Lett.}\ }\textbf {\bibinfo {volume} {109}},\ \bibinfo {pages}
  {265302} (\bibinfo {year} {2012})}\BibitemShut {NoStop}%
\bibitem [{\citenamefont {Castin}(2001)}]{castin:book}%
  \BibitemOpen
  \bibfield  {author} {\bibinfo {author} {\bibfnamefont {Y.}~\bibnamefont
  {Castin}},\ }in\ \href@noop {} {\emph {\bibinfo {booktitle} {Coherent Atomic
  Matter Waves}}},\ \bibinfo {editor} {edited by\ \bibinfo {editor}
  {\bibfnamefont {R.}~\bibnamefont {Caiser}}, \bibinfo {editor} {\bibfnamefont
  {C.}~\bibnamefont {Westbrook}}, \ and\ \bibinfo {editor} {\bibfnamefont
  {F.}~\bibnamefont {David}}}\ (\bibinfo  {publisher} {EDP Sciences and
  Springer-Verlag},\ \bibinfo {year} {2001})\BibitemShut {NoStop}%
\bibitem [{\citenamefont {Berry}(1989)}]{Berry1989}%
  \BibitemOpen
  \bibfield  {author} {\bibinfo {author} {\bibfnamefont {M.}~\bibnamefont
  {Berry}},\ }in\ \href@noop {} {\emph {\bibinfo {booktitle} {Geometric Phases
  in Physics}}},\ \bibinfo {editor} {edited by\ \bibinfo {editor}
  {\bibfnamefont {A.}~\bibnamefont {Shapere}}\ and\ \bibinfo {editor}
  {\bibfnamefont {F.}~\bibnamefont {Wilczek}}}\ (\bibinfo  {publisher} {World
  Scientific,},\ \bibinfo {year} {1989})\BibitemShut {NoStop}%
\bibitem [{\citenamefont {Rhim}\ \emph {et~al.}(2020)\citenamefont {Rhim},
  \citenamefont {Kim},\ and\ \citenamefont {Yang}}]{Rhim2020}%
  \BibitemOpen
  \bibfield  {author} {\bibinfo {author} {\bibfnamefont {J.-W.}\ \bibnamefont
  {Rhim}}, \bibinfo {author} {\bibfnamefont {K.}~\bibnamefont {Kim}}, \ and\
  \bibinfo {author} {\bibfnamefont {B.-J.}\ \bibnamefont {Yang}},\ }\href
  {\doibase 10.1038/s41586-020-2540-1} {\bibfield  {journal} {\bibinfo
  {journal} {Nature}\ }\textbf {\bibinfo {volume} {584}},\ \bibinfo {pages}
  {59} (\bibinfo {year} {2020})}\BibitemShut {NoStop}%
\bibitem [{\citenamefont {Julku}\ \emph
  {et~al.}(2016{\natexlab{b}})\citenamefont {Julku}, \citenamefont {Peotta},
  \citenamefont {Vanhala}, \citenamefont {Kim},\ and\ \citenamefont
  {T\"orm\"a}}]{julku:2016}%
  \BibitemOpen
  \bibfield  {author} {\bibinfo {author} {\bibfnamefont {A.}~\bibnamefont
  {Julku}}, \bibinfo {author} {\bibfnamefont {S.}~\bibnamefont {Peotta}},
  \bibinfo {author} {\bibfnamefont {T.}~\bibnamefont {Vanhala}}, \bibinfo
  {author} {\bibfnamefont {D.-H.}\ \bibnamefont {Kim}}, \ and\ \bibinfo
  {author} {\bibfnamefont {P.}~\bibnamefont {T\"orm\"a}},\ }\href {\doibase
  10.1103/PhysRevLett.117.045303} {\bibfield  {journal} {\bibinfo  {journal}
  {Phys. Rev. Lett.}\ }\textbf {\bibinfo {volume} {117}},\ \bibinfo {pages}
  {045303} (\bibinfo {year} {2016}{\natexlab{b}})}\BibitemShut {NoStop}%
\bibitem [{\citenamefont {T\"orm\"a}\ \emph {et~al.}(2018)\citenamefont
  {T\"orm\"a}, \citenamefont {Liang},\ and\ \citenamefont
  {Peotta}}]{Torma2018}%
  \BibitemOpen
  \bibfield  {author} {\bibinfo {author} {\bibfnamefont {P.}~\bibnamefont
  {T\"orm\"a}}, \bibinfo {author} {\bibfnamefont {L.}~\bibnamefont {Liang}}, \
  and\ \bibinfo {author} {\bibfnamefont {S.}~\bibnamefont {Peotta}},\ }\href
  {\doibase 10.1103/PhysRevB.98.220511} {\bibfield  {journal} {\bibinfo
  {journal} {Phys. Rev. B}\ }\textbf {\bibinfo {volume} {98}},\ \bibinfo
  {pages} {220511} (\bibinfo {year} {2018})}\BibitemShut {NoStop}%
\bibitem [{\citenamefont {Iskin}(2020)}]{Iskin2020}%
  \BibitemOpen
  \bibfield  {author} {\bibinfo {author} {\bibfnamefont {M.}~\bibnamefont
  {Iskin}},\ }\href {\doibase https://doi.org/10.1016/j.physb.2020.412260}
  {\bibfield  {journal} {\bibinfo  {journal} {Physica B: Condensed Matter}\
  }\textbf {\bibinfo {volume} {592}},\ \bibinfo {pages} {412260} (\bibinfo
  {year} {2020})}\BibitemShut {NoStop}%
\bibitem [{\citenamefont {Ozawa}\ and\ \citenamefont {Baym}(2012)}]{ozawa2012}%
  \BibitemOpen
  \bibfield  {author} {\bibinfo {author} {\bibfnamefont {T.}~\bibnamefont
  {Ozawa}}\ and\ \bibinfo {author} {\bibfnamefont {G.}~\bibnamefont {Baym}},\
  }\href {\doibase 10.1103/PhysRevLett.109.025301} {\bibfield  {journal}
  {\bibinfo  {journal} {Phys. Rev. Lett.}\ }\textbf {\bibinfo {volume} {109}},\
  \bibinfo {pages} {025301} (\bibinfo {year} {2012})}\BibitemShut {NoStop}%
\bibitem [{\citenamefont {Barnett}\ \emph {et~al.}(2012)\citenamefont
  {Barnett}, \citenamefont {Powell}, \citenamefont {Gra\ss{}}, \citenamefont
  {Lewenstein},\ and\ \citenamefont {Das~Sarma}}]{barnett2012}%
  \BibitemOpen
  \bibfield  {author} {\bibinfo {author} {\bibfnamefont {R.}~\bibnamefont
  {Barnett}}, \bibinfo {author} {\bibfnamefont {S.}~\bibnamefont {Powell}},
  \bibinfo {author} {\bibfnamefont {T.}~\bibnamefont {Gra\ss{}}}, \bibinfo
  {author} {\bibfnamefont {M.}~\bibnamefont {Lewenstein}}, \ and\ \bibinfo
  {author} {\bibfnamefont {S.}~\bibnamefont {Das~Sarma}},\ }\href {\doibase
  10.1103/PhysRevA.85.023615} {\bibfield  {journal} {\bibinfo  {journal} {Phys.
  Rev. A}\ }\textbf {\bibinfo {volume} {85}},\ \bibinfo {pages} {023615}
  (\bibinfo {year} {2012})}\BibitemShut {NoStop}%
\bibitem [{\citenamefont {T\"orm\"a}\ and\ \citenamefont
  {Sengstock}(2014)}]{Torma_book}%
  \BibitemOpen
  \bibfield  {author} {\bibinfo {author} {\bibfnamefont {P.}~\bibnamefont
  {T\"orm\"a}}\ and\ \bibinfo {author} {\bibfnamefont {K.}~\bibnamefont
  {Sengstock}},\ }\href@noop {} {\emph {\bibinfo {title} {Quantum Gas
  Experiments Exploring Many-Body States}}}\ (\bibinfo  {publisher} {Imperial
  College Press},\ \bibinfo {year} {2014})\BibitemShut {NoStop}%
\bibitem [{\citenamefont {Bloch}\ \emph {et~al.}(2008)\citenamefont {Bloch},
  \citenamefont {Dalibard},\ and\ \citenamefont {Zwerger}}]{bloch:2008}%
  \BibitemOpen
  \bibfield  {author} {\bibinfo {author} {\bibfnamefont {I.}~\bibnamefont
  {Bloch}}, \bibinfo {author} {\bibfnamefont {J.}~\bibnamefont {Dalibard}}, \
  and\ \bibinfo {author} {\bibfnamefont {W.}~\bibnamefont {Zwerger}},\ }\href
  {\doibase 10.1103/RevModPhys.80.885} {\bibfield  {journal} {\bibinfo
  {journal} {Rev. Mod. Phys.}\ }\textbf {\bibinfo {volume} {80}},\ \bibinfo
  {pages} {885} (\bibinfo {year} {2008})}\BibitemShut {NoStop}%
\bibitem [{\citenamefont {Leung}\ \emph {et~al.}(2020)\citenamefont {Leung},
  \citenamefont {Schwarz}, \citenamefont {Chang}, \citenamefont {Brown},
  \citenamefont {Unnikrishnan},\ and\ \citenamefont
  {Stamper-Kurn}}]{Leung2020}%
  \BibitemOpen
  \bibfield  {author} {\bibinfo {author} {\bibfnamefont {T.-H.}\ \bibnamefont
  {Leung}}, \bibinfo {author} {\bibfnamefont {M.~N.}\ \bibnamefont {Schwarz}},
  \bibinfo {author} {\bibfnamefont {S.-W.}\ \bibnamefont {Chang}}, \bibinfo
  {author} {\bibfnamefont {C.~D.}\ \bibnamefont {Brown}}, \bibinfo {author}
  {\bibfnamefont {G.}~\bibnamefont {Unnikrishnan}}, \ and\ \bibinfo {author}
  {\bibfnamefont {D.}~\bibnamefont {Stamper-Kurn}},\ }\href {\doibase
  10.1103/PhysRevLett.125.133001} {\bibfield  {journal} {\bibinfo  {journal}
  {Phys. Rev. Lett.}\ }\textbf {\bibinfo {volume} {125}},\ \bibinfo {pages}
  {133001} (\bibinfo {year} {2020})}\BibitemShut {NoStop}%
\bibitem [{\citenamefont {Caleffi}\ \emph {et~al.}(2020)\citenamefont
  {Caleffi}, \citenamefont {Capone}, \citenamefont {Menotti}, \citenamefont
  {Carusotto},\ and\ \citenamefont {Recati}}]{Caleffi2020}%
  \BibitemOpen
  \bibfield  {author} {\bibinfo {author} {\bibfnamefont {F.}~\bibnamefont
  {Caleffi}}, \bibinfo {author} {\bibfnamefont {M.}~\bibnamefont {Capone}},
  \bibinfo {author} {\bibfnamefont {C.}~\bibnamefont {Menotti}}, \bibinfo
  {author} {\bibfnamefont {I.}~\bibnamefont {Carusotto}}, \ and\ \bibinfo
  {author} {\bibfnamefont {A.}~\bibnamefont {Recati}},\ }\href {\doibase
  10.1103/PhysRevResearch.2.033276} {\bibfield  {journal} {\bibinfo  {journal}
  {Phys. Rev. Research}\ }\textbf {\bibinfo {volume} {2}},\ \bibinfo {pages}
  {033276} (\bibinfo {year} {2020})}\BibitemShut {NoStop}%
\end{thebibliography}

%

\end{document}


\section*{Supplementary Figures}

\begin{figure}[!htb]
 \centering
  \includegraphics[width=1.0\textwidth]{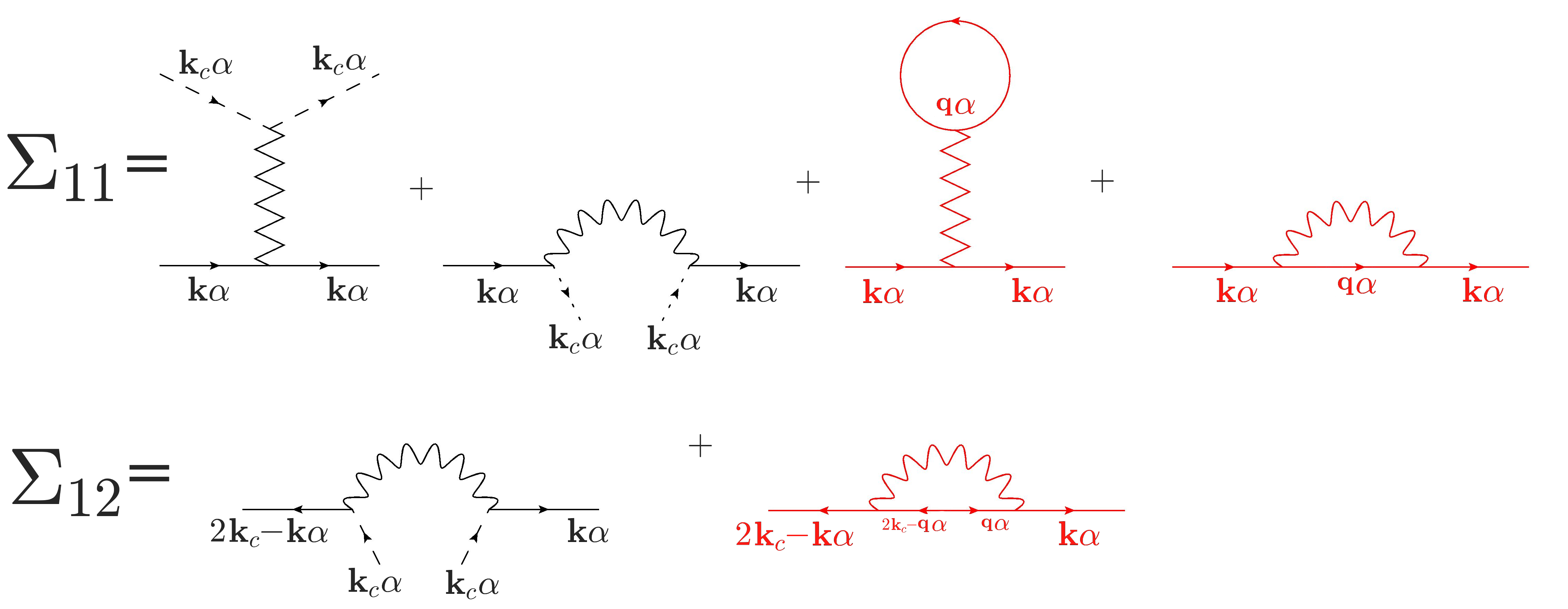}
   \caption{Self-energy diagrams for the Bogoliubov approximation (black diagrams) and Hartree-Fock-Bogoliubov (black and red diagrams). The solid (dashed) lines depict propagators of non-condensed (condensed) bosons, wiggly lines present the momentum-independent interaction vertices $U$ and $\alpha$ is the sublattice index. The upper (lower) row diagrams are for the block diagonal (off-diagonal or anomalous) self-energy $\Sigma_{11}$ ($\Sigma_{12}$). The internal momenta $\bq$ are integrated over. The momentum of the latter anomalous diagram (red diagram in the lower row) is conserved with respect of the condensate momentum $\bk_c$.}
  \label{Fig:S1}
\end{figure}

\clearpage

\begin{figure}[!htb]
 \centering
  \includegraphics[width=1.0\textwidth]{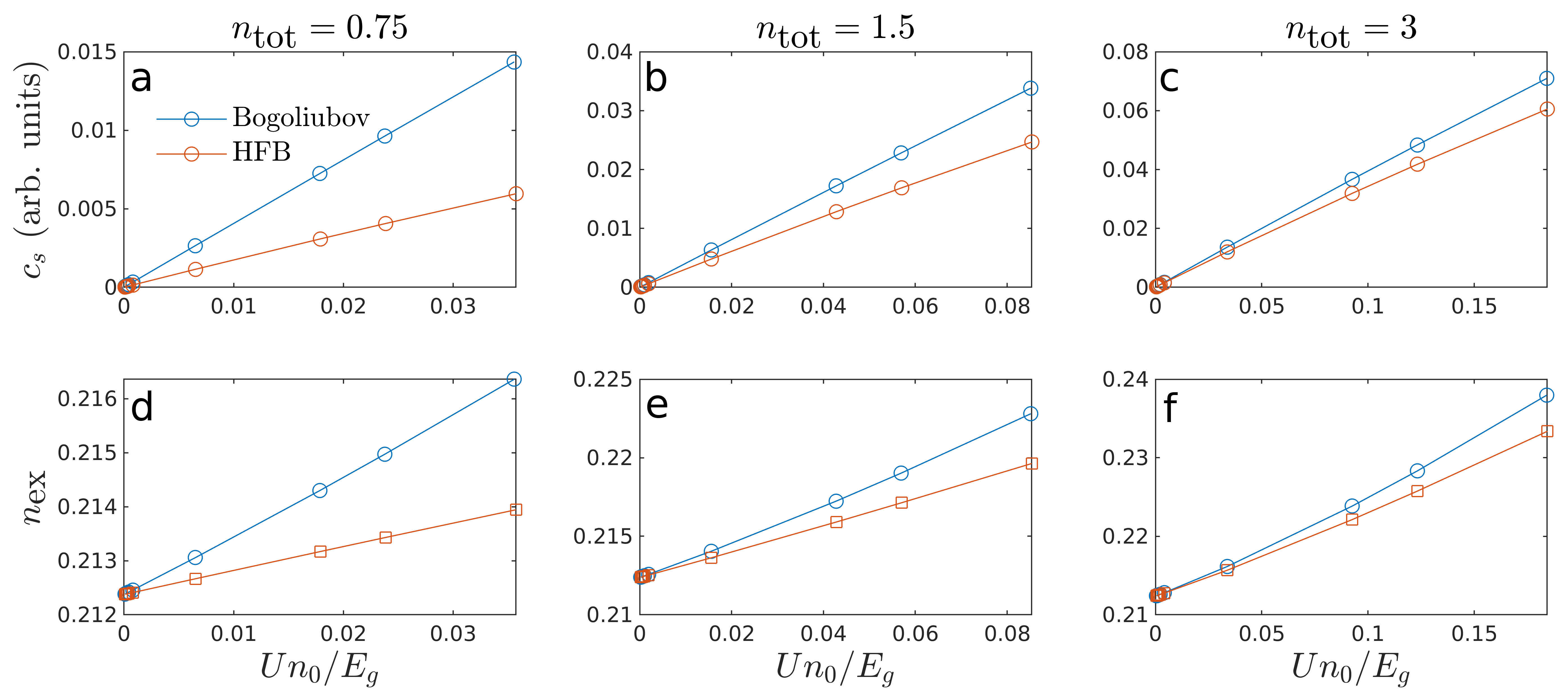}
   \caption{Comparison between the Bogoliubov and  Hartree-Fock-Bogoliubov (HFB) approximations. Panels \textbf{a}-\textbf{b} show speed of sound $c_s$ for the kagome flat band condensate as a function of interaction in case of three different values of the total density $n_{\textrm{tot}}$. Panels \textbf{d}-\textbf{f} show the corresponding results for the excitation density $n_{\textrm{ex}}$. We see that for smaller $U$ the agreement between the two methods become better: at the limit of $U\rightarrow 0$ both methods yield the same value for $\lim_{U\rightarrow 0}n_{\textrm{ex}}$.}
  \label{Fig:S2}
\end{figure}

\clearpage
\section*{Supplementary section S1: Second order coherence function}
In this section we provide the details on the local second order coherence function $g^{(2)}$. We start by considering the following function:
\begin{align}
G^{(2)} = \frac{1}{N} \sum_{i\alpha\beta} \langle c^\dag_{i\alpha} c_{i\alpha} c^\dag_{i \beta} c_{i \beta} \rangle = \frac{1}{N}\sum_{i} \langle \rho(\br_i)\rho(\br_i)\rangle,   
\end{align}
where $\rho(\br_i) = \sum_\alpha c_{i\alpha}^\dag c_{i\alpha}$ is the density within the $i$th unit cell so that $G^{(2)}$ depicts the average taken over all the unit cells of the function $\langle\rho(\br_i)\rho(\br_i)\rangle$. Now, by Fourier transforming, using $c_{\bk\alpha} = \sqrt{Nn_0}\langle \alpha | \phi_0 \rangle \delta_{\bk,\bk_c} + c_{\bk\neq \bk_c\alpha}$,  and discarding the linear and third-order terms in $c_{\bk\neq \bk_c\alpha}$ (as their expectation values vanish in the Bogoliubov appxroximation), one obtains, after tedious but straightforward algebra, the following:
\begin{align}
\label{g2_1}
&G^{(2)} =    n_0^2 + 2n_0 n_{ex} \nonumber \\
&+ \frac{n_0}{N}\sum_{\bk\alpha\beta}{}^{'} \Big[ e^{i(\bk-\bk_c)\cdot (\br_\alpha -\br_\beta)}\Big( \langle \phi_0 | \alpha \rangle  \langle \phi_0 | \beta \rangle \langle  c_{\bk\alpha}  c_{2\bk_c-\bk,\beta} \rangle +  \langle \phi_0 | \alpha \rangle  \langle \beta | \phi_0 \rangle  \langle  c_{\bk\alpha}  c^\dag_{\bk \beta} \rangle \Big)\Big]\nonumber \\
& + \frac{n_0}{N}\sum_{\bk\alpha\beta}{}^{'} \Big[ e^{-i(\bk-\bk_c)\cdot (\br_\alpha -\br_\beta)} \Big( \langle \alpha | \phi_0 \rangle \langle \phi_0 | \beta \rangle \langle  c^\dag_{\bk\alpha} c_{\bk\beta} \rangle + \langle \alpha | \phi_0 \rangle \langle \beta | \phi_0 \rangle \langle  c^\dag_{\bk\alpha} c^\dag_{2\bk_c-\bk \beta} \rangle \Big)\Big] \nonumber \\
& + \frac{1}{N^2}\sum_{\bk,\bk',\bq}{}^{'}\sum_{\alpha\beta} e^{i(\bk'-\bk)\cdot (\br_\alpha -\br_\beta)} \langle  c^\dag_{\bk\alpha}  c_{\bk'\alpha}  c^\dag_{\bq -\bk \beta}  c_{\bq -\bk' \beta}  \rangle,
\end{align}
where the primed sum indicates that all the bosonic operators inside the sum are for non-condensed states. Here $n_{ex}$ is the density of the non-condensed bosons, i.e. $n_{tot} = n_0 + n_{ex}$, where $n_{tot}$ is the total number of bosons per unit cell. Furthermore, $\br_\alpha$ is the spatial coordinate of the $\alpha$th orbital within an unit cell. For convenience, we now deploy a $U(1)$ gauge transformation of the form $\tilde{c}_{\bk\alpha} = \exp(i \bk \cdot \br_\alpha) c_{\bk\alpha}$. Explicitly, we rewrite the Bogoliubov Hamiltonian as:
\begin{align}
&H_B =  \frac{1}{2}\sum_{\bk \neq \bk_c} \Psi^\dag_\bk  \mathcal{H}_B(\bk)  \Psi_{\bk} = \frac{1}{2}\sum_{\bk \neq \bk_c} \tilde{ \Psi}^\dag_\bk \tilde{W}(\bk) \mathcal{H}_B(\bk) \tilde{W}^\dag(\bk) \tilde{\Psi}_\bk \equiv \frac{1}{2}\sum_{\bk \neq \bk_c} \tilde{ \Psi}^\dag_\bk \tilde{\mathcal{H}}_B(\bk)\tilde{\Psi}_\bk, \textrm{ where} \nonumber \\
&\tilde{W}(\bk) = \begin{bmatrix}
	\tilde{V}(\bk) &  0 \\
	0 & \tilde{V}^\dag_{2\bk_c-\bk},
\end{bmatrix}, \nonumber \\
&[\tilde{V}(\bk)]_{\alpha\beta} = \delta_{\alpha\beta} \exp[i\bk \cdot \br_{\alpha}].
\end{align}
Solving the Bogoliubov Hamiltonian in this new basis yields naturally the same Bogoliubov excitation energies as before but the Bloch states are transformed as $|\tilde{\psi}^s_m(\bk) \rangle = \tilde{W}(\bk) |\psi^s_m(\bk) \rangle$.

By working in the new basis, $G^{(2)}$ can be rewritten as
\begin{align}
\label{g2_2}
&G^{(2)} =    n_0^2 + 2n_0 n_{ex} + \frac{n_0}{N}\sum_{\bk\alpha\beta}{}^{'} \Big( \langle \tilde{\phi}_0 | \alpha \rangle  \langle \tilde{\phi}_0 | \beta \rangle \langle  \tilde{c}_{\bk\alpha}  \tilde{c}_{2\bk_c-\bk,\beta} \rangle +  \langle \tilde{\phi}_0 | \alpha \rangle  \langle \beta | \tilde{\phi}_0 \rangle  \langle  \tilde{c}_{\bk\alpha}  \tilde{c}^\dag_{\bk \beta} \rangle \Big)\nonumber \\
& + \frac{n_0}{N}\sum_{\bk\alpha\beta}{}^{'}\Big( \langle \alpha | \tilde{\phi}_0 \rangle \langle \tilde{\phi}_0 | \beta \rangle \langle  \tilde{c}^\dag_{\bk\alpha} \tilde{c}_{\bk\beta} \rangle + \langle \alpha | \tilde{\phi}_0 \rangle \langle \beta | \tilde{\phi}_0 \rangle \langle  \tilde{c}^\dag_{\bk\alpha} \tilde{c}^\dag_{2\bk_c-\bk \beta} \rangle \Big) + \frac{1}{N^2}\sum_{\bk,\bk',\bq}{}^{'}\sum_{\alpha\beta} \langle  \tilde{c}^\dag_{\bk\alpha}  \tilde{c}_{\bk'\alpha}  \tilde{c}^\dag_{\bq -\bk \beta}  \tilde{c}_{\bq -\bk' \beta}  \rangle.
\end{align}
One can now express the expectation values inside equation~\ref{g2_2} with the help of the Bogoliubov states. Furthermore, by recalling that at zero temperature one has for the Bose-Einstein distribution $n_{B}(x>0) = 0$ and $n_B(x<0)=-1$ and that the Bogoliubov states are taken to be non-interacting, one eventually finds, after lengthy but elementary algebra, the following:
\begin{align}
G^{(2)} =   G^{(2)}_1 + G^{(2)}_2 + G^{(2)}_3
\end{align}
with
\begin{align}
&  G^{(2)}_1 = n_0^2 + 2n_0 n_{ex} + n_{ex}^2 = n_{\textrm{tot}}^2 \nonumber \\
& G^{(2)}_2 = \frac{n_0}{N}\sum_{\bk,l}\langle \tilde{\Phi} | \tilde{\psi}^+_l(\bk)\rangle \langle \tilde{\psi}^+_l(\bk) | \tilde{\Phi} \rangle \nonumber \\
& G^{(2)}_3 = \frac{1}{2N^2}\sum_{\bk \bk'll'} \langle  \tilde{\psi}_l^-(\bk)| \tilde{\psi}_{l'}^+(\bk') \rangle \langle \tilde{\psi}_{l'}^+(\bk') | \tilde{\psi}_l^-(\bk) \rangle .
\end{align}
Here $|\tilde{\Phi}_0\rangle \equiv [|\tilde{\phi}_0 \rangle, |\tilde{\phi}_0^* \rangle]^T$.
By noting that $g^{(2)}(0,0) = \frac{G^{(2)} -n_{\textrm{tot}}}{n_{\textrm{tot}}^2}$, we get the expressions for $g^{(2)}$, $g^{(2)}_1$, $g^{(2)}_2$ and $g^{(2)}_3$:
\begin{align}
&g^{(2)}(0,0) = g^{(2)}_1 + g^{(2)}_2 + g^{(2)}_3 \nonumber \\
&g^{(2)}_1 = 1 - \frac{1}{n_{\textrm{tot}}} \\
&g^{(2)}_2 = \frac{n_0}{N n_{\textrm{tot}}^2}\sum_{\bk,l}{}^{'}\langle \tilde{\Phi}_0 | \tilde{\psi}^+_l(\bk)\rangle \langle \tilde{\psi}^+_l(\bk) | \tilde{\Phi}_0 \rangle \\
&g^{(2)}_3 =  \frac{1}{2(Nn_{\textrm{tot}})^2}\sum_{\bk \bk'll'}{}^{'} \langle  \tilde{\psi}_l^-(\bk)| \tilde{\psi}_{l'}^+(\bk') \rangle \langle \tilde{\psi}_{l'}^+(\bk') | \tilde{\psi}_l^-(\bk) \rangle. 
\end{align}

\section*{Supplementary section S2: Superfluid weight}

In this section we provide the details on the superfluid weight $D^s$. In a two-dimensional system $D^s$ is a $2\times2$ matrix and it is defined as the long-wavelength, zero frequency limit of the current-current linear response function $K_{\mu\nu}(\bq,\omega)$ \cite{Liang2017}, i.e.
\begin{align}
&D^s_{\mu\nu} = \lim_{\bq \rightarrow 0} \lim_{\omega \rightarrow 0} K_{\mu\nu}(\bq,\omega),
\end{align}
where
\begin{align}
\label{curLRF}
&K_{\mu\nu} = \langle T_{\mu\nu} \rangle - i \int^\infty_0 dt e^{i\omega t} \langle \big[ j^p_\mu(\bq,t) ,j^p_\mu(-\bq,0)\big] \rangle \equiv  \langle T_{\mu\nu} \rangle + \Pi_{\mu\nu}(\bq,\omega).
\end{align}
Here the diamagnetic and paramagnetic current operators read
\begin{align}
&T_{\mu\nu} = \sum_\bk c^\dag_{\bk} \partial_\mu \partial_\nu \mathcal{H}(\bk) c_{\bk} \nonumber \\
&j^p_\mu(\bq) = \sum_\bk c^\dag_{\bk+\bq}\partial_\mu \mathcal{H}(\bk + \bq/2) c_\bk.
\end{align}
We start by solving the paramagnetic term $\Pi(\bq,\omega)$ by deploying the standard method of computing the current Green's function in the Matsubara space and then at the end of the computation invoke the analytical continuation:
\begin{align}
&\Pi_{\mu\nu}(\bq,\omega) = \lim_{i\omega_n \rightarrow \omega + i\eta} \int_0^\beta d\tau e^{i\omega_n \tau}\Pi_{\mu\nu}(\bq,\tau) \nonumber \\ 
& \Pi_{\mu\nu}(\bq,\tau) = - \langle T_\tau j^p_\mu(\bq,\tau)j_\nu^p(-\bq,0) \rangle
\end{align}
where $\tau$ is the imaginary-time and $i\omega_n$ is the bosonic Matsubara frequency. We proceed by using $c_{\bk\alpha} = \sqrt{Nn_0}\langle \alpha | \phi_0 \rangle \delta_{\bk,\bk_c} +  c_{\bk\neq \bk_c\alpha}$ and discarding linear and third order fluctuation terms to find
\begin{align}
&  \langle T_\tau j^p_\mu(\bq,\tau)j_\nu^p(-\bq,0) \rangle = A + B(\bq,\tau) + C(\bq,\tau), \\
& A = [n_0^2N^2 j^0_\mu j^0_\nu  + n_0 N j^0_\mu \delta j_\nu + n_0 N j^0_\nu \delta j_\mu ]\delta_{\bq,0}, \nonumber \\
& j^0_\mu = \langle \psi_0 | \partial_\mu \mathcal{H}(\bk_c) | \psi_0 \rangle \nonumber \\
& \delta j_\mu = \sum_\bk{}^{'} \langle c^\dag_{\bk+\bq}\partial_\mu \mathcal{H}(\bk + \bq/2) c_\bk \rangle \nonumber \\
& B(\bq,\tau) = n_0 N \langle c^\dag_{\bk_c+\bq}(\tau)\partial_\mu \mathcal{H}(\bk_c + \bq/2) | \phi_0 \rangle \langle  c^\dag_{\bk_c-\bq} \partial_\nu \mathcal{H}(\bk_c - \bq/2) | \phi_0\rangle \nonumber \\
& + n_0 N \langle c^\dag_{\bk_c+\bq}(\tau)\partial_\mu \mathcal{H}(\bk_c + \bq/2) | \phi_0 \rangle \langle  \phi_0 | \partial_\nu \mathcal{H}(\bk_c + \bq/2) c_{\bk_c+\bq} \rangle \nonumber \\
& + n_0 N \langle \phi_0 | \partial_\mu \mathcal{H}(\bk_c - \bq/2) c_{\bk_c-\bq}(\tau) \rangle \langle c^\dag_{\bk_c-\bq} \partial_\nu \mathcal{H}(\bk_c - \bq/2) | \phi_0 \rangle \nonumber \\
& + n_0 N \langle \phi_0 | \partial_\mu \mathcal{H}(\bk_c - \bq/2) c_{\bk_c-\bq}(\tau) \rangle \langle \phi_0 | \partial_\nu \mathcal{H}(\bk_c + \bq/2) c_{\bk_c+\bq} \rangle \nonumber \\
& C(\bq,\tau) =  \sum_{\bk\bk'}{}^{'}\sum_{\alpha\beta\gamma\delta}\langle c^\dag_{\bk+\bq\alpha}(\tau) \partial_\mu \mathcal{H}_{\alpha\beta}(\bk+\bq/2) c_{\bk\beta}(\tau) c^\dag_{\bk'-\bq\gamma} \partial_\nu \mathcal{H}_{\gamma\delta}(\bk-\bq/2)c_{\bk'\delta} \rangle
\end{align}
We can discard the constant term $A$ as $\int^\beta_0 d\tau e^{i\omega_n \tau} =0$. By using the bosonic Green's function, defined in the Appendix B and the fact that $\partial_\mu \mathcal{H}_B(\bk)$ is block-diagonal, one finds, after lengthy but straightforward algebra, the following for $B(\bq,\tau)$:
\begin{align}
B(\bq,\tau) = -n_0 N \Tr\Big[ G(\bk_c-\bq,\tau)\sigma_z\partial_\nu \mathcal{H}_B(\bk_c-\bq/2)|\Phi_0 \rangle \langle \Phi_0 | \sigma_z \partial_\mu \mathcal{H}_B(\bk_c-\bq/2)\Big].    
\end{align}
By Fourier transforming  $B(\bq,\tau)$ and taking the zero-frequency and zero-momentum limits one finds
\begin{align}
\label{ds_B}
& \lim_{\bq\rightarrow 0}\lim_{\omega \rightarrow 0} \Pi^B_{\mu\nu}(\bq,\omega ) = \lim_{\bq\rightarrow 0} \lim_{i\omega_n \rightarrow 0} -\int_0^\beta d\tau e^{i\omega_n\tau} B(\bq,\tau) \nonumber \\ 
&= n_0 N \lim_{\bq\rightarrow 0} \lim_{i\omega_n \rightarrow 0} \Tr\Big[ G(\bk_c-\bq,i\omega_n)\sigma_z\partial_\nu \mathcal{H}_B(\bk_c-\bq/2)|\Phi_0 \rangle \langle \Phi_0 | \sigma_z \partial_\mu \mathcal{H}_B(\bk_c-\bq/2)\Big] \nonumber \\
& = n_0 N  \lim_{\bq\rightarrow 0} \lim_{i\omega_n \rightarrow 0} \sum_{ms}\frac{s}{i\omega_n - sE_m(\bk_c-s\bq)} \langle \Phi_0 | \sigma_z \partial_\mu\mathcal{H}_B(\bk_c -\bq/2) | \psi^s_m(\bk_c-\bq)\rangle \nonumber \\
& \times\langle \psi^s_m(\bk_c-\bq)|\sigma_z \partial_\nu \mathcal{H}_B(\bk_c -\bq/2)|\Phi_0\rangle \nonumber \\
&= -n_0 N \lim_{\bq\rightarrow 0}  \sum_{ms}\frac{\langle \Phi_0 | \sigma_z \partial_\mu\mathcal{H}_B(\bk_c -\bq/2) | \psi^s_m(\bk_c-\bq)\rangle \langle \psi^s_m(\bk_c-\bq)|\sigma_z \partial_\nu \mathcal{H}_B(\bk_c -\bq/2)|\Phi_0\rangle }{E_m(\bk_c-s\bq)}.
\end{align}
Here $s,s'$ take values one and minus one.
This is the superfluid contribution $D^s_{\mu\nu,2}$.

Next, we need to evaluate $C(\bk,\tau)$. To begin with, we invoke the Wick's theorem (valid within the Bogoliubov approximation) and keep the connected diagrams:
\begin{align}
\langle c^\dag_{\bk+\bq\alpha}(\tau) c_{\bk\beta}(\tau) c^\dag_{\bk'-\bq\gamma} c_{\bk'\delta} \rangle 
=& \langle c^\dag_{\bk+\bq\alpha}(\tau)c^\dag_{\bk'-\bq\gamma} \rangle \langle c_{\bk\beta}(\tau) c_{\bk'\delta} \rangle \delta_{\bk',2\bk_c-\bk} \nonumber \\
&+ \langle c^\dag_{\bk+\bq\alpha}(\tau) c_{\bk'\delta} \rangle \langle c_{\bk\beta}(\tau) c^\dag_{\bk'-\bq\gamma} \rangle \delta_{\bk',\bk+\bq}.  
\end{align}
By using once again the definition of the bosonic Green's function and rearranging terms in $C(\bk,\tau)$, one finds, after tedious algebra, the following:
\begin{align}
C(\bq,\tau) = \frac{1}{2} \sum_\bk{}^{'}  \Tr\Big[ \sigma_z\partial_\mu \mathcal{H}_B(\bk+\bq/2) G(\bk,\tau) \partial_\nu \sigma_z \mathcal{H}_B(\bk+\bq/2) G(\bk+\bq,-\tau) \Big]. 
\end{align}
By Fourier transforming  $C(\bq,\tau)$ and taking the zero-frequency and zero-momentum limits one has
\begin{align}
\label{ds_C}
&\lim_{\bq\rightarrow 0}\lim_{\omega \rightarrow 0}\Pi^C_{\mu\nu}(\bq,\omega ) = \lim_{\bq\rightarrow 0}\lim_{i\omega_n \rightarrow 0} -\int_0^\beta d\tau e^{i\omega_n\tau} C(\bq,\tau) \nonumber \\
&=-\lim_{\bq\rightarrow 0}\lim_{\omega_n \rightarrow 0}\frac{\beta}{2}\sum_{\bk\Omega_n}{}^{'}\Tr\Big[ \sigma_z\partial_\mu \mathcal{H}_B(\bk+\bq/2) G(\bk,i\Omega_n) \partial_\nu \sigma_z \mathcal{H}_B(\bk+\bq/2) G(\bk+\bq,i\Omega_n -i\omega_n)  \Big] \nonumber \\
&=\frac{1}{2}\sum_\bk{}^{'}\sum_{m,m',s,s'}ss'\frac{n_B[sE_m(\bk_c+s\tilde{\bk})] - n_B[s'E_{m'}(\bk_c+s'\tilde{\bk})]}{sE_m(\bk_c+s\tilde{\bk}) - s'E_{m'}(\bk_c+s'\tilde{\bk})} \nonumber \\
&\times\langle \psi_{m'}^{s'}(\bk)|\sigma_z\partial_\mu  \mathcal{H}_B(\bk) | \psi^s_m(\bk) \rangle \langle \psi^s_m(\bk) | \sigma_z\partial_\nu  \mathcal{H}_B(\bk) | \psi_{m'}^{s'}(\bk) \rangle.
\end{align}
where $\tilde{\bk} = \bk - \bk_c$. This term contributes to $D^s_{3}$.

We now evaluate the diamagnetic contribution $\langle K_{\mu\nu} \rangle$. By using $c_{\bk\alpha} = \sqrt{Nn_0}\langle \alpha | \phi_0 \rangle \delta_{\bk,\bk_c} + \delta c_{\bk\alpha}$ and discarding linear fluctuation terms (which vanish in the Bogoliubov approximation), one gets
\begin{align}
\langle K_{\mu\nu} \rangle \equiv K^0_{\mu\nu} + \delta K_{\mu\nu}.    
\end{align}
Here the first (second) term arises from the condensate (non-condensed particles). Explicitly:
\begin{align}
\label{ds_K0}
K^0_{\mu\nu} &= n_0 N \langle \phi_0 | \partial_\mu \partial_\nu \mathcal{H}(\bk_c) |\phi_0 \rangle \nonumber \\
&=n_0N\partial_\mu \partial_\nu \epsilon_1(\bk_c) + n_0N\sum_{n\neq 1}\Big\{[\epsilon_n(\bk_c) - \epsilon_0] \langle \partial_\mu \phi_0 | u_n(\bk_c)\rangle \langle u_n(\bk_c) | \partial_\nu \phi_0 \rangle + (\mu \leftrightarrow \nu) \Big\}.
\end{align}
This is the pure condensate superfluid contribution $D^s_1$.

The fluctuation term $\delta K_{\mu\nu}$ can be written as:
\begin{align}
\label{diam}
&\delta K_{\mu\nu} = \lim_{\tau \rightarrow 0} \sum_\bk{}^{'} \langle c^\dag_\bk(\tau) \partial_\mu \partial_\nu\mathcal{H}(\bk)c_\bk \rangle \nonumber \\
& = \lim_{\tau \rightarrow 0} \frac{1}{2}\sum_\bk{}^{'} \partial_\mu \partial_\nu \mathcal{H}_{\alpha\beta}(\bk)\Big[ \langle c^\dag_{\bk\alpha}(\tau) c_{\bk\beta} \rangle + \langle c_{\bk\alpha}(\tau) c^\dag_{\bk\beta}\rangle +  \delta_{\alpha\beta}\Big] \nonumber \\
&=-\lim_{\tau \rightarrow 0} \frac{1}{2}\sum_{\bk}{}^{'} \Tr\Big[ \partial_\mu \partial_\nu \mathcal{H}_B(\bk) G(\bk,\tau) \Big]  = \frac{1}{2\beta}\sum_{\bk,\Omega_n}{}^{'} \Tr\Big[ \partial_\mu \partial_\nu \mathcal{H}_B(\bk) G(\bk,i\Omega_n) \Big], 
\end{align}
where in the second last step the Kronecker Delta term vanishes due to the translational invariance. As $G^{-1}(\bk,i\Omega_n) = i\Omega_n\sigma_z -\mathcal{H}_B(\bk)$ and $\partial_\nu(GG^{-1}) =0$, one has $\partial_\nu G = G\partial_\nu \mathcal{H}_B G$. By using this expression, after partial integrating the last line of equation~\eqref{diam}, one obtains
\begin{align}
\label{ds_dK}
&\delta K_{\mu\nu} = \frac{1}{2\beta}\sum_{\bk,\Omega_n}{}^{'} \Tr[\partial_\mu \mathcal{H}_B(\bk) G(\bk,i\Omega_n)\partial_\nu \mathcal{H}_B(\bk) G(\bk,i\Omega_n)]  \nonumber \\
&=\frac{1}{2}\sum_{\bk}{}^{'}\sum_{m,m',s,s'}ss'\frac{n_B[sE_m(\bk_c+s\tilde{\bk})] - n_B[s'E_{m'}(\bk_c+s'\tilde{\bk})]}{s'E_{m'}(\bk_c+s'\tilde{\bk}) - sE_{m}(\bk_c+s\tilde{\bk})} \nonumber \\
&\times\langle \psi_{m'}^{s'}(\bk)|\partial_\mu  \mathcal{H}_B(\bk) | \psi^s_m(\bk) \rangle \langle \psi^s_m(\bk) | \partial_\nu  \mathcal{H}_B(\bk) | \psi_{m'}^{s'}(\bk) \rangle.
\end{align}
This gives a contribution to $D^s_3$.

By collecting all the terms, i.e.\ equations~\eqref{ds_B},\eqref{ds_C},\eqref{ds_K0} and \eqref{ds_dK}, we can finally write the total superfluid weight divided by system size $N$ as 
\begin{align}
&D^s_{\mu\nu} = D^s_{1,\mu\nu} +  D^s_{2,\mu\nu} + D^s_{3,\mu\nu}, \\
&D^s_{1,\mu\nu} = n_0\partial_\mu \partial_\nu \epsilon_1(\bk_c) + n_0\sum_{n\neq 1}\Big\{[\epsilon_n(\bk_c) - \epsilon_0] \langle \partial_\mu \phi_0 | u_n(\bk_c)\rangle \langle u_n(\bk_c) | \partial_\nu \phi_0 \rangle + (\mu \leftrightarrow \nu) \Big\}, \\
&D^s_{2,\mu\nu} = -n_0  \lim_{\bq\rightarrow 0}  \sum_{ms}\frac{\langle \Phi_0 | \sigma_z \partial_\mu\mathcal{H}_B(\bk_c -\bq/2) | \psi^s_m(\bk_c-\bq)\rangle \langle \psi^s_m(\bk_c-\bq)|\sigma_z \partial_\nu \mathcal{H}_B(\bk_c -\bq/2)|\Phi_0\rangle }{E_m(\bk_c-s\bq)}, \\
&D^s_{3,\mu\nu} = \frac{1}{2N} \sum_\bk{}^{'} \sum_{mm'ss'} ss'\frac{n_B[sE_m(\bk_c+s\tilde{\bk})] - n_B[s'E_{m'}(\bk_c+s'\tilde{\bk})] }{s'E_{m'}(\bk_c+s'\tilde{\bk}) - sE_{m}(\bk_c+s\tilde{\bk})} \times \nonumber \\
&\Big[\langle \psi^{s'}_{m'}(\bk)| \partial_\mu \mathcal{H}_B(\bk) | \psi^{s}_{m}(\bk) \rangle \langle \psi^{s}_{m}(\bk)| \partial_\nu \mathcal{H}_B(\bk) | \psi^{s'}_{m'}(\bk) \rangle \nonumber \\
&-\langle \psi^{s'}_{m'}(\bk)| \sigma_z \partial_\mu \mathcal{H}_B(\bk) | \psi^{s}_{m}(\bk) \rangle \langle \psi^{s}_{m}(\bk)| \sigma_z \partial_\nu \mathcal{H}_B(\bk) | \psi^{s'}_{m'}(\bk) \rangle\Big].
\end{align}

We based our derivation for $D^s$ on Eq.~\eqref{curLRF} where the form of the retarded current Green's function assumes the translational invariance. However, we know that in the flat band systems the Bose-condensation can take place in a non-zero momentum state, i.e. $\bk_c \neq 0$. When this is the case, the condensate wavefunction has the form $\psi_0(\br_{i\alpha}) = \exp(i\bk_c \cdot \br_{i\alpha}) \langle \alpha | \phi_0 \rangle$, meaning the translational invariance of the original lattice geometry is broken and the periodicity is dictated by the condensate momentum $\bk_c$ instead. For example in case of the kagome flat band condensate, $\bk_c = [4\pi/3,0]$ and $|\phi_0 \rangle = [-1,-1,1]$ which yields the condensate wavefunction $\psi_0(\br_{i\alpha})$ whose periodicity is three unit cells in the directions of both the lattice basis vectors $\textbf{a}_1$ and $\textbf{a}_2$. This means the system is translationally invariant when the new unit cell is chosen to consist of $3\times 3$ original unit cells. Thus, superfluid weight calculations for kagome lattice were carried out by utilizing this new unit cell scheme for which the number of sublattices is $M = 3\times 3 \times 3 = 27$. In the momentum space, this corresponds to a $9$ times smaller folded BZ with $27$ Bloch bands of which $9$ are degenerate flat bands with the condensate momentum folded to $\bk_c = 0$. The new supercell scheme is notably more complicated than the original $3$ band model and thus finding connections between the quantum distance and $D^s$, in the same way as with $n_{\textrm{ex}}$ or $c_s$, is highly non-trivial and thus out of scope of this work.

Following a similar procedure as in the case of Fermi-Hubbard systems, we can express the Bogoliubov states in the Bloch basis as 
\begin{align}
\label{psi_bloch}
|\psi^s_m(\bk) \rangle = \sum_{n} c^{ms}_{n}(\bk) |+\rangle \otimes |u_n(\bk) \rangle +  \sum_{n} d^{ms}_{n}(\bk) |-\rangle \otimes |u_n^*(\bk) \rangle,   
\end{align}
where $|u_n\bk)\rangle$ ($|u_n^*(\bk)\rangle$) are the Bloch functions of $\mathcal{H}(\bk)$ ($\mathcal{H}^*(\bk_c-\bk)$). Furthermore, we have $|+\rangle = [1,0]^T$ and $|-\rangle = [0,1]^T$ in the particle-hole basis. By inserting~\eqref{psi_bloch} to the expression of $D^s_3$, one finds at zero temperature the following:
\begin{align}
D^s_{3,\mu\nu} = \frac{1}{N}\sum_\bk{}^{'}\sum_{mm'}\frac{2}{E_m(\bk) + E_{m'}(\bk)}& \Re \Big[ \sum_{nn'} [c^{m+}_n(\bk)]^* c^{m'-}_{n'}(\bk) i\langle u_n(\bk) | \partial_\mu \mathcal{H} | u_{n'}(\bk) \rangle \nonumber \\
&\times\sum_{nn'} d^{m+}_{n}(\bk) [d^{m'-}_{n'}(\bk)]^* i\langle u^*_{n'}(\bk)|\partial_\nu \mathcal{H}^*(2\bk_c-\bk) |u^*_n(\bk) \rangle\Big].   
\end{align}
The current terms can be rewritten as
\begin{align}
i\langle u_n(\bk) &| \partial_\mu \mathcal{H}(\bk) | u_{n'}(\bk) \rangle = i\partial_\mu \epsilon_n(\bk)\delta_{nn'} + [\epsilon_{n'}(\bk) - \epsilon_n(\bk)]i\langle \partial_\mu u_n(\bk) | u_{n'}(\bk) \rangle.    
\end{align}
We see that there are two kinds of current terms: intraband terms that are proportional to the derivatives of the Bloch energies and interband terms that depend on the geometric properties of the Bloch functions in the form of the interband Berry connection functions. We can do the division $D^s_3 = D^s_{3,\textrm{conv}} + D^s_{3,\textrm{geom}}$, where the so-called conventional term $D^s_{3,\textrm{conv}}$ includes only the intraband current terms and the geometric contribution  $D^s_{3,\textrm{geom}}$ features the interband terms. For a flat band condensate we have $\partial_\mu\epsilon_1(\bk) = 0$ and therefore it is predominantly the geometric contribution that dictates superfluidity. In previous studies of fermionic superfluidity, the lower bound of the geometric contribution can be linked to various quantum geometric properties such as Chern number~\cite{Peotta2015}, Berry curvatures and quantum metric of the flat band~\cite{Liang2017}. It is thus likely that similar connections can be found for bosonic flat band superfluidity as the form of $D^s_3$ is similar to that of the fermionic $D^s$~\cite{Liang2017}, however the calculation is beyond the scope of the present article, and likely to be technically involved for cases with non-zero condensate momentum, as explained above.

In the main text we noted that it is the fluctuation term $D^s_3$ that dominates the superfluidity of the flat band condensation, in contrast to the usual case of a dispersive band condensate. We demonstrate this in the weak-coupling limit, i.e.\ we show that $\lim_{U\rightarrow 0} D^s_1 + D^s_2 = 0$ for a flat band condensate. We first note that $\partial_\mu \partial_\nu \epsilon_1(\bk_c) = 0$ for the flat band so that 
\begin{align}
\label{ds1_fb}
D^s_{1,\mu\nu} =  n_0\sum_{m\neq 1}\Big\{[\epsilon_m(\bk_c) - \epsilon_0] \langle \partial_\mu \phi_0 | u_m(\bk_c)\rangle \langle u_m(\bk_c) | \partial_\nu \phi_0 \rangle + (\mu \leftrightarrow \nu) \Big\}.   
\end{align}
Thus we have to show that this term is cancelled by $D^s_2$. Now, in the limit of $U\rightarrow 0$, one has $|\psi^+_m(\bk)\rangle \rightarrow [|u_m\rangle,0]^T$ and $|\psi^-_m(\bk)\rangle \rightarrow [0,|u^*_m\rangle]^T$ and that $E_m \rightarrow \epsilon_m -\epsilon_0$. From these limits, it follows that the $m=1$ contribution in $D^s_2$ vanishes and we are left with
\begin{align}
\lim_{U\rightarrow 0} D^s_{2,\mu\nu} &= -n_0 \sum_{m\neq 1}\Big[\frac{\langle u_m(\bk_c) | \partial_\mu \mathcal{H}(\bk_c) | \phi_0 \rangle \langle \phi_0 | \partial_\nu \mathcal{H}(\bk_c) | u_m(\bk_c) \rangle}{\epsilon_m(\bk_c) - \epsilon_0}  \nonumber \\
& + \frac{\langle u^*_m(\bk_c) | \partial_\mu \mathcal{H}^*(\bk_c) | \phi_0^* \rangle \langle \phi_0^* | \partial_\nu \mathcal{H}^*(\bk_c) | u^*_m(\bk_c) \rangle}{\epsilon_m(\bk_c) - \epsilon_0} \Big] \nonumber \\
& = -n_0  \sum_{m\neq 1}\Big\{\frac{\langle u_m(\bk_c) | \partial_\mu \mathcal{H}(\bk_c) | \phi_0 \rangle \langle \phi_0 | \partial_\nu \mathcal{H}(\bk_c) | u_m(\bk_c) \rangle}{\epsilon_m(\bk_c) - \epsilon_0} + (\mu \leftrightarrow \nu) \Big\}\nonumber \\
& = -n_0  \sum_{m\neq 1}\Big\{\frac{(\epsilon_m(\bk_c) - \epsilon_0)^2 \langle u_m(\bk_c) | \partial_\mu \phi_0 \rangle \langle \partial_\nu \phi_0 | u_m(\bk_c) \rangle}{\epsilon_m(\bk_c) - \epsilon_0} + (\mu \leftrightarrow \nu) \Big\}
\end{align}
which cancels $D^s_{1,\mu\nu}$ of Eq.~\eqref{ds1_fb}. Thus, at the weak coupling regime, $D^s_1 + D^s_2\sim 0$ for the flat band condensate. The numerical superfluid result, presented in Fig.~7 of the main text, however shows that $D^s_1 + D^s_2$ in case of the kagome flat band condensate remains very small also for notably large interaction strengths. Showing this analytically turns out to be difficult as computing $D^s$ in a translationally invariant form requires the utilization of the unit cell of $27$ sublattices, making it cumbersome to write down $D^s_2$ for general non-zero interaction $U$. We therefore leave a more general analysis of $D^s_1 + D^s_2$ at arbitrary $U$ for future studies. It suffices to say that our results indicate that superfluidity of flat band Bose-condensates is dominated by fluctuations, not by the condensed particles.




\title{Excitations of a  Bose-Einstein condensate and the quantum geometry of a flat band}

\newcommand{\affiliationAalto}{Department of Applied Physics, Aalto University, P.O.Box 15100, 00076 Aalto, Finland}
\newcommand{\affiliationAarhus}{Center for Complex Quantum Systems, Department of Physics and Astronomy, Aarhus University, Ny Munkegade 120, DK-8000 Aarhus C, Denmark}
\newcommand{\affiliationChina}{Shenzhen Institute for Quantum Science and Engineering and Department of Physics, Southern University of Science and Technology, Shenzhen 518055, China}

\author{Aleksi Julku}
\affiliation{\affiliationAalto}
\affiliation{\affiliationAarhus}
\author{Georg M.\ Bruun}
\affiliation{\affiliationAarhus}
\affiliation{\affiliationChina}
\author{P\"aivi T\"orm\"a}
\affiliation{\affiliationAalto}


\date{\today}

\begin{abstract}

The quantum geometry of Bloch states fundamentally affects a wide range of physical phenomena. The quantum Hall effect, for example, is governed by the Chern number, and flat band superconductivity by the distance between the Bloch states -- the quantum metric. While understanding quantum geometry phenomena in the context of fermions is well established, less is known about the role of quantum geometry in bosonic systems where particles can undergo Bose-Einstein condensation (BEC). In conventional single band or continuum systems, excitations of a weakly-interacting BEC are determined by the condensate density and the interparticle interaction energy. In contrast to this, we discover here fundamental connections between the properties of a weakly-interacting BEC and the underlying quantum geometry of a multiband lattice system. We show that, in the flat band limit, the defining physical quantities of BEC, namely the speed of sound and the quantum depletion, are dictated solely by the quantum geometry. We find that the speed of sound becomes proportional to the quantum metric of the condensed state. Furthermore, the quantum distance between the Bloch functions forces the quantum depletion and the quantum fluctuations of the density-density correlation to obtain finite values for infinitesimally small interactions. This is in striking contrast to dispersive bands where these quantities vanish with the interaction strength. Additionally, we show how in the flat band limit the supercurrent is carried by the quantum fluctuations and is determined by the Berry connections of the Bloch states. Our results reveal how non-trivial quantum geometry allows reaching strong quantum correlation regime of condensed bosons even with weak interactions. This is highly relevant for example for polariton and photon BECs where interparticle interactions are inherently small. Our predictions can be experimentally tested with flat band lattices already implemented in ultracold gases and various photonic platforms.  
        
\end{abstract}

\pacs{}

\maketitle

\section{Introduction}
In recent years, it has become clear that the quantum geometry of Bloch states is a fundamentally important property that complements the information given by energy band dispersions. If the unit cell hosts multiple orbitals (lattice sites, spins, etc.), the Bloch states become vectors in the orbital basis. Consequently, the quantum geometry of the band, namely the phase and amplitude distances of the Bloch states, may become non-trivial. In other words, they may differ drastically from the single band or continuum systems. This is quantified by the quantum geometric tensor~\cite{Provost:1980} whose imaginary part is the Berry curvature and real part is the quantum metric -- a measure of distance between two quantum states~\cite{Resta:2011}. 

Quantum geometry can play a central role in determining the properties of a given system. For instance, Berry curvature governs the anomalous transport of an electron wave-packet~\cite{Sundaram1999}, while its integral over the Brillouin zone (BZ) gives the Chern number, which, among other topological invariants, is central in explaining the quantum Hall effect and topological insulators~\cite{Klitzing1980,Thouless1982,Haldane1988,Kane2005,Hasan2010,Bernevig2013}. The importance of the quantum metric for phenomena such as superconductivity~\cite{Peotta2015,Julku2016,Liang2017,Wang2020}, orbital magnetic susceptibility~\cite{Montambaux:2016,Rhim2020} and light-matter coupling~\cite{Topp2021} has been understood only recently, and the interest in the quantum metric is rapidly growing~\cite{hu:2019,Julku:2020,xie:2019,Viewpoint2020}. Quantum geometric concepts have also been proposed for bosonic systems composed of light, bosonic atoms, or collective excitations~\cite{Haldane2008,Wang2009ObservationStates,Miyake:2013,Bloch2015_JM,Spielman2018_JM,Lu2014,Fainman2017,Khanikaev2017,ozawa_topological_2018,Gianfrate:2019}.

While prominent topological properties can already arise from single particle physics, interactions between particles lead to even more intriguing phenomena~\cite{Kane2005,Pesin2010,Neupert2011,Tang2011,Vanhala2016TopolModel,Sato2017,Rachel2018,Roy2014} such as topological superconductors and fractional Chern insulators. In the context of interacting systems, (nearly) dispersionless bands with diverging effective mass -- so-called flat bands~\cite{Leykam:2018} -- are particularly exciting for two reasons. First, quantum many-body interaction and correlation effects are expected to be strong when the kinetic energy scale is quenched. Second, the effects of the Bloch state quantum geometry are likely to dominate if the band dispersion itself is featureless. Superconductivity is an important example of this, as the critical temperature is predicted to be exponentially enhanced in a flat band~\cite{kopnin:2011}, and the stability of a supercurrent guaranteed by a non-zero quantum metric and the Chern number~\cite{Peotta2015,Julku2016,Liang2017}. Flat band systems, both for fermions and bosons, can be experimentally realized, for example in ultracold gases, photonic and polaritonic systems, and atomistic designer matter~\cite{Leykam:2018,Mukherjee2015,Taie:2015,Vicencio2015,kajiwara2016,Leung2020,Drost2017,Slot2017}. 
A well-known example of a nearly flat band system is given by twisted bilayer graphene, where the observed superconductivity~\cite{li:2010,cao:2018:1,yankowitz:2019} has indeed been proposed to be influenced by quantum geometry~\cite{hu:2019,Julku:2020,xie:2019,Viewpoint2020}.

While geometric properties of quantum states in fermionic systems are widely explored, the significance of quantum geometry in the context of bosonic systems remains less studied. It is therefore natural to ask \textit{how the quantum geometry affects the Bose-Einstein condensation (BEC) of interacting bosons, particularly in a flat band}. There are obvious outstanding puzzles. Since the states of the flat band are degenerate, which one is chosen for BEC? Since single particles are localized in a flat band, is superfluidity even possible? Due to the high degeneracy, is the condensate immediately fragmented? It has been theoretically shown that interactions between the bosons enable mass current~\cite{Vidal:2000,DiLiberto2019,Gligoric2019,danieli2020manybody,Mukherjee2018,Goblot2019} 
, even in a flat band, and thereby a BEC~\cite{Huber:2010}. Semiconductor polariton condensates have been experimentally studied in Lieb~\cite{Baboux2016,Whittaker2018,Harder2020a,Scafirimuto2021} and kagome~\cite{Harder2020b} lattices, showing fragmentation and localization. Theory work on the kagome lattice predicts that due to interactions, a certain lattice momentum is favorable for BEC, even when initially all momenta have degenerate energies~\cite{you:2012}. Thus far, studies of flat band BECs have mainly focused on mean-field properties, although Ref.~\cite{you:2012} analysed the stability of the condensate against quantum fluctuations using Bogoliubov theory. However, the density-density response and excitation density have not been calculated and, importantly, the relation between the quantum geometry and the excitations of a BEC has not been considered. 

Here, we investigate excitations of weakly interacting bosonic condensates and reveal fundamental connections to quantum geometry. In a flat band, we find condensate behavior radically different from that in a dispersive band. By applying multiband Bogoliubov theory, we study the speed of sound $c_s$, density-density correlations, superfluid weight $D^s$, and excitation density $n_{\textrm{ex}}$, i.e. the amount of particles depleted from the condensate due to interactions (so-called quantum depletion). 

In a flat band, $c_s$ is found to be determined by the quantum metric of the condensate state such that a finite quantum metric guarantees a finite $c_s$. Moreover, in contrast to the usual square root dependence~\cite{Fetter1971}, $c_s$ in a flat band is linearly proportional to the interaction, because the quantum metric provides an interaction-dependent effective mass.

We furthermore show that the excitation density $n_{\textrm{ex}}$ is found to behave in a striking way in a flat band: it has a finite value for both finite and infinitesimally small interactions. This is remarkable for two reasons. First, it is totally different from the conventional dispersive case where $n_{\textrm{ex}}$ vanishes with the interactions. Second, for finite interactions, one could intuitively expect the excitation fraction to diverge in a band with flat energy spectrum, as the excitations have no kinetic energy cost. However, it turns out that the excitation density for small interactions is determined by the quantum distance between the condensed state and the other states of the band: a finite distance between them curtails the excitation fraction from diverging. A salient point here is that the excitation density does not depend on the total density. Therefore, by decreasing the condensation density, one can increase the depletion of the condensate, even in the $U\rightarrow 0$ limit. In this way, the importance of quantum fluctuations and correlations can be significantly enhanced at the weak-coupling limit.

Density-density correlations are typically dominated by the macroscopic population of the condensate. We show that in a flat band the quantum correlations between excitations are also prominent. In fact, the quantum fluctuation contribution obtains a finite value, even in the $U\rightarrow 0$ limit, due to quantum geometric properties of the flat band, similar to $n_{\textrm{ex}}$. This means that in flat band BECs, quantum geometry can provide access to manifestly quantum (beyond mean-field) correlations, even for weak interactions.

Finally, the superfluid weight is usually given by the density of the condensed bosons and the band dispersion relation. In a flat band, we find that finite superfluidity arises extraordinarily due to the quantum fluctuations only and is determined by the Berry connections of the Bloch states. We prove this analytically in the non-interacting limit and show that for the kagome lattice the result is valid for a wide range of interaction values.  

\begin{figure}
 \centering
  \includegraphics[width=1.0\columnwidth]{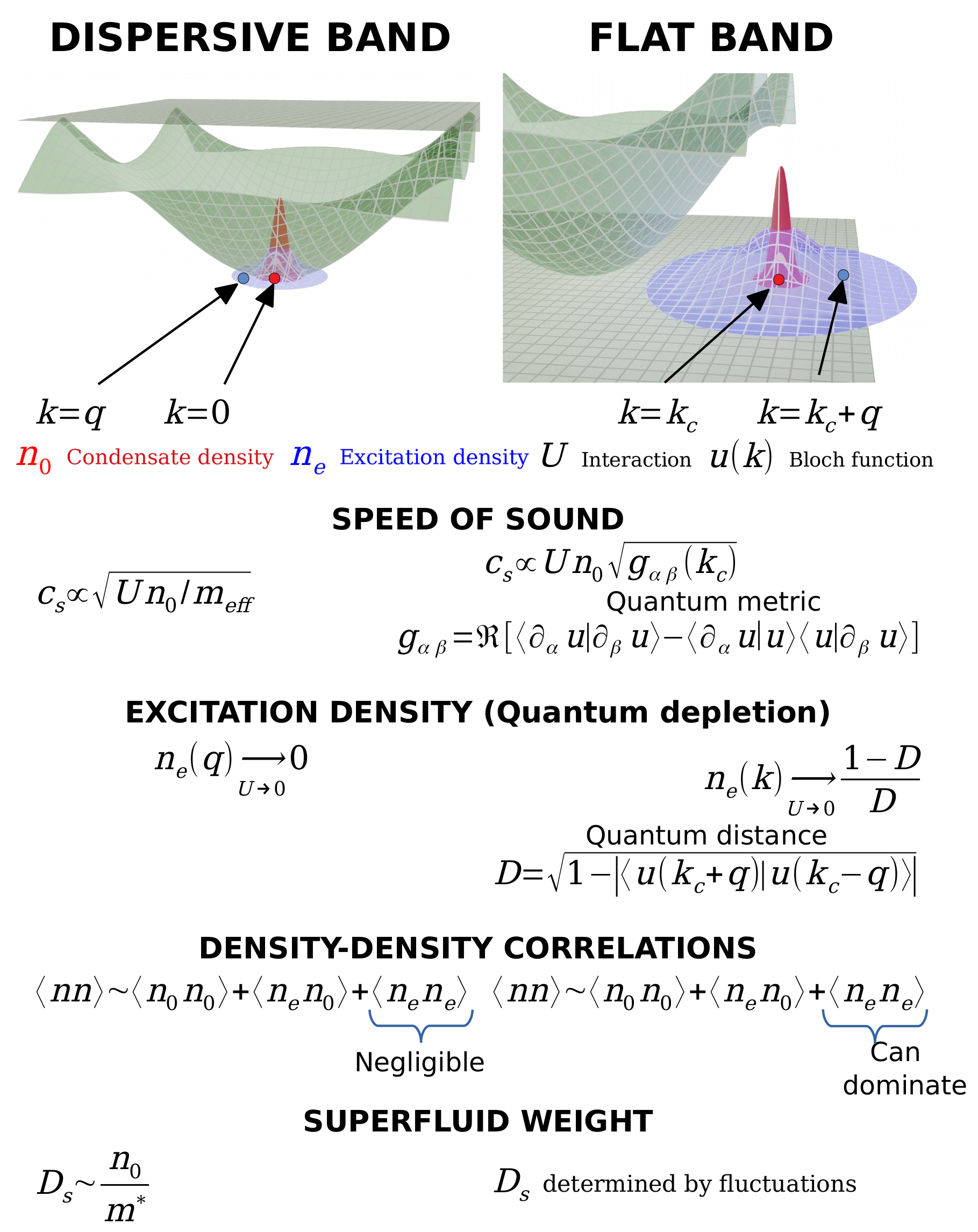}
  \caption{The connection between flat band Bose-Einstein condensation and quantum geometry. The right panel summarizes the main results of this article; for comparison, the left panel shows known results on dispersive band BEC.
   Here we show results for the case of a kagome flat band model, however the general formulas (see text) have essentially similar dependence on the geometric quantities.  
   }
  \label{Fig:summary}
\end{figure}

To summarize, in contrast to a dispersive band BEC, the excitation properties of a weakly-interacting flat band BEC depend on the underlying quantum geometry in the following way. 1) The speed of sound $c_s$ is given by the quantum metric and depends linearly (in contrast to the conventional square root dependence) on the interaction strength. 2) The excitation density can be non-zero even in the limit $U\rightarrow 0$ and is given by the quantum distance among the flat band states. 3) The quantum fluctuations can be significant even for $U\rightarrow 0$ and are determined by the quantum distance. 4) The superfluid weight is induced by the quantum fluctuations and has a quantum geometric origin. Our main results are  summarized in Fig.~\ref{Fig:summary}.

Our present work accompanies the joint study of Ref.~\cite{julku2021b} and together these two works establish fundamental new relations between the quantum geometry and BEC. We show that bosonic condensation and superfluidity can be stable in a flat band if there is a finite quantum metric and a quantum distance between the condensed state and non-condensed states of the band. Importantly, the fluctuations dominate over mean-field properties in several physical observables, making flat bands a promising platform for realizing strongly correlated bosonic systems, even at the weak interaction limit.

\section{Theoretical framework of BEC and Bogoliubov approximation in a flat band system}
\label{FlatBEC}
We consider a weakly interacting BEC in a multiband system described by the Bose-Hubbard Hamiltonian
\begin{align}
\label{BHH}
H = &\sum_{i\alpha,j\beta} c^\dag_{i\alpha} \mathcal{H}_{i\alpha,j\beta} c_{j\beta}  \nonumber \\
&-\mu\sum_{i\alpha} c^\dag_{i\alpha}c_{i\alpha} + \frac{U}{2}\sum_{i\alpha} c^\dag_{i\alpha}c_{i\alpha}( c^\dag_{i\alpha}c_{i\alpha}-1).
\end{align}
Here, $c_{i\alpha}$ is a bosonic annihilation operator for the $\alpha$th sublattice site within the $i$th unit cell; the matrix $\mathcal{H}$ contains the hopping coefficients between different sites; $\mu$ is the chemical potential; $U>0$ is the repulsive on-site interaction. The sublattice index ranges from $1$ to $M$, where $M$ is the number of lattice sites per unit cell. 
By assuming periodic boundary conditions and introducing the Fourier transforms
$c_{i\alpha} = \frac{1}{\sqrt{N}}\sum_\bk \exp(i \bk \cdot \br_{i\alpha}) c_{\bk\alpha}$, 
where $N$ is the number of unit cells, $\br_{i\alpha}$ is the location of the $\alpha$th lattice site in the $i$th unit cell, and $\bk$ is the momentum, one gets
\begin{align}
\label{BHH2}
&H = \sum_\bk\big( c^\dag_\bk \mathcal{H}(\bk)c_\bk -\mu c^\dag_\bk c_\bk \big)  \nonumber \\
&+\frac{U}{2N} \sum_\alpha \sum_{\bk,\bk',\bq} c^\dag_{\bk\alpha} c_{\bk-\bq\alpha}c^\dag_{\bk'\alpha} c_{\bk'+\bq\alpha}.   
\end{align}
Here, the one-particle Hamiltonian $\mathcal{H}(\bk)$ is a $M\times M$ matrix and $c_\bk$ is a $M\times 1$ vector such that $[c_\bk]_\alpha = c_{\bk\alpha}$. One can diagonalize  $\mathcal{H}(\bk)$ as $\mathcal{H}(\bk) |u_n(\bk) \rangle = \epsilon_n(\bk)|u_{n}(\bk) \rangle$, where $\epsilon_n(\bk)$ ($|u_{n\bk} \rangle$) are the eigenenergies (Bloch states) and $n$ is the band index so that $\epsilon_1(\bk) \leq \epsilon_2(\bk) \leq ... \leq \epsilon_M(\bk)$.

As we consider an equilibrium situation, the condensation takes place within the lowest Bloch band. Furthermore, we are mainly interested in condensation occurring within a flat or quasi-flat Bloch band; the latter is defined by $J<< Un_0$, where $J$ is the width of the band and $n_0$ is the number of condensed bosons per unit cell. Since the Bloch states of the flat band are degenerate in energy, the question arises: on which state does the condensation occur? To answer this, we use the approach of Ref.~\cite{you:2012}, i.e. utilize the mean-field (MF) approximation~\cite{Pitaevskii2003},
where we substitute operators in Eq.~\eqref{BHH2} by complex numbers, and the resulting MF energy $E_{\textrm{MF}}(\bk)$ is solved separately for each $\bk$ with fixed density $n_0$ (see Appendix A). The condensation takes place at the Bloch state $|\phi_0 \rangle \equiv| u_1(\bk_c)\rangle$ that minimizes $E_{\textrm{MF}}(\bk)$~\cite{you:2012}. The momentum (energy) of the condensed Bloch state is denoted as $\bk_c$ ($\epsilon_0\equiv\epsilon_1(\bk_c)$).

Therefore, even if the lowest Bloch band is strictly flat, the condensate can still occur at some specific Bloch state, as the repulsive on-site interaction favors Bloch states that distribute the particles among the sublattices as equally as possible. We herein assume uniform condensate density, i.e. $|\langle \alpha | \phi_0 \rangle|^2 = 1/M$ for all $\alpha$ with $\langle \alpha | \phi_0 \rangle$ being the projection of $|\phi_0 \rangle$ to the orbital $\alpha$. This is a rather general condition; examples will be presented below.

To analyse the excitations of the condensate, we express $c_{i\alpha}$ as
\begin{align}
c_{i\alpha} = \sqrt{n_0} \psi_0(\br_{i\alpha}) + \delta c_{i\alpha} \equiv c^0_\alpha + \delta c_{i\alpha}.   
\end{align}
Here, $\psi_0(\br_{i\alpha}) = \exp(i\bk_c \cdot \br_{i\alpha}) \langle \alpha | \phi_0 \rangle$ is the wavefunction of the condensate with 
the wave vector $\bk_c$ 
and $\delta c_{i \alpha}$ describes the fluctuations on top of the condensate. By Fourier transforming, one finds 
 $c_{\bk_c \alpha}=\sqrt{Nn_0} \langle \alpha | \phi_0 \rangle $ and $c_{\bk\neq\bk_c \alpha} = N^{-1/2}\sum_{i}e^{-i\bk\cdot \br_{i\alpha}}\delta c_{i\alpha}$.

We now treat the Hamiltonian within the multiband Bogoliubov approximation (for details, see Appendix B) by neglecting the interaction terms that are higher than quadratic order in the fluctuations $c_{\bk\alpha}$ and $c^\dag_{\bk\alpha}$ with $\bk\neq\bk_c$. One can then express Eq.~\eqref{BHH} as $H = E_c + H_B$, where $E_c$ is a constant giving the ground energy of the condensate. The excitations are described by the Bogoliubov Hamiltonian (See Appendix B for details)
\begin{align}
\label{ham}
H_B = \frac{1}{2}\sum_{\bk}{}^{'} \Psi^\dag_\bk  \mathcal{H}_B(\bk)  \Psi_{\bk},
\end{align}
where $\mathcal{H}_B(\bk)$ is a $2M\times2M$ matrix given by 
\begin{align}\label{BogMatrix}
&\mathcal{H}_B(\bk) = \begin{bmatrix}
	\mathcal{H}(\bk) -\mu_{\textrm{eff}} &  \Delta \\
	\Delta^* & \mathcal{H}^*(2\bk_c - \bk) -\mu_{\textrm{eff}},
\end{bmatrix}, \nonumber \\
& \Psi_\bk = [c_{\bk 1}, c_{\bk 2},...,c_{\bk M}, c^\dag_{2\bk_c -\bk 1},...,c^\dag_{2\bk_c -\bk M}]^T, \nonumber \\
&[\Delta]_{\alpha\beta} = \delta_{\alpha,\beta} U n_0 \langle \alpha | \phi_0 \rangle^2, \nonumber \\
&\mu_{\textrm{eff}} = (\epsilon_0 - \frac{Un_0}{M})\delta_{\alpha,\beta}.
\end{align}
The primed sum in Eq.~\eqref{ham} indicates that all the operators within the sum are for non-condensed states only, i.e. $\bk \neq \bk_c$ and $2\bk_c-\bk \neq \bk_c$.

To obtain the excitation energies of the condensate, one cannot directly diagonalize $\mathcal{H}_B$ since this would violate the bosonic commutation relations.  Instead, one needs to find the eigenstates of $L(\bk) \equiv \sigma_z \mathcal{H}_B(\bk)$, where $\sigma_z$ is the Pauli matrix~\cite{castin:book}. One obtains Bogoliubov bands of the energies $E_M(\bk) \geq ... E_2(\bk) \geq E_1(\bk) \geq 0 \geq -E_1(2\bk_c-\bk) \geq ... -E_M(2\bk_c-\bk)$. Here positive (negative) energies describe quasi-particle (-hole) excitations. The quasi-particle and -hole states are labelled as $|\psi^+_m (\bk) \rangle$ and $|\psi^-_m (\bk) \rangle$ such that
\begin{align}
\label{bogo_eig}
& L(\bk)|\psi^+_m(\bk) \rangle = E_m(\bk) |\psi^+_m(\bk) \rangle  \\
& L(\bk)|\psi^-_m(\bk) \rangle = -E_m(2\bk_c - \bk) |\psi^-_m(\bk). \rangle
\end{align}
The chemical potential in Eq.~\eqref{ham} is fixed such that the lowest quasi-particle energy band is gapless at $\bk_c$, i.e. $E_1(\bk \rightarrow \bk_c) = 0$. This is the usual Goldstone mode that emerges because the condensate wavefunction spontaneously picks up a complex phase, leading to the spontaneous gauge $U(1)$ symmetry breaking~\cite{Fetter1971,Pitaevskii2003}.

\section{Speed of sound at the weak-coupling regime} \label{Sound}
After setting up the Bogoliubov theory, we now proceed to derive our main results. We first compute the speed of sound $c_s$ for small $U$, which is given by the slope of the lowest Bogoliubov excitation band in the long-wavelength limit:
\begin{align}
c_s(\theta_\bq) = \lim_{|\bq|\rightarrow 0} \frac{E_1(\bk_c + \bq) - E_1(\bk_c)}{|\bq|} = \lim_{|\bq|\rightarrow 0} \frac{E_1(\bk_c + \bq)}{|\bq|}, 
\end{align}
with $|\bq|<<1$ (we use units where the lattice constant is unity). In general, $c_s$ depends on the direction of its momentum $\bq$, parametrised by the angle $\theta_\bq$ between $\bq$ and the $x$-axis. We assume that around $\bk_c$, the lowest Bloch band is isolated from the higher Bloch bands by an energy gap $E_g$ that is large compared to the interaction energy, i.e. $Un_0 \ll E_g$. In this case, the interaction term does not cause notable mixing between the lowest and dispersive bands. This is important as we are interested in the properties of flat band BECs. In the large interaction regime with $Un_0 \gg E_g$, the importance of a flat band would be obscured by the  dispersive bands. With $Un_0 \ll E_g$, we therefore discard the higher bands and project to the lowest band to obtain~(see Appendix C)
\begin{align}
\label{lk2}
L_p(\bk) =  \begin{bmatrix} \frac{\bq^2}{2m_{\textrm{eff}}} + \frac{Un_0}{M} & \frac{Un_0}{M}\alpha(\bq) \\
- \frac{Un_0}{M}\alpha^*(\bq)  & -\frac{\bq^2}{2m_{\textrm{eff}}} -\frac{Un_0}{M}
\end{bmatrix} 
\end{align}
for $\bq \rightarrow 0$. Here, the subscript $p$ indicates that this is the projected $2\times2$ matrix. We have furthermore defined the off-diagonal term as $\alpha(\bq) \equiv \frac{M}{Un_0}\langle u_1(\bk_c + \bq)|\Delta| u_1^*(\bk_c - \bq) \rangle$, which contains information about the quantum geometric properties of the Bloch states. It is this quantity that encodes quantum geometric effects to various physical properties of a BEC taking place in a multiband system, as can be strikingly seen in the flat band limit shown below. To contrast geometric effects to the behavior of a conventional BEC,
 we have retained a possible finite dispersion relation of the lowest band via the term $\bq^2/2m_{\textrm{eff}}$. 


It is easy to obtain $E_1(\bk_c+\bq)$ by diagonalizing Eq.~\eqref{lk2},
which yields 
\begin{align}
\label{e1}
E_1(\bk_c + \bq) = \sqrt{\Big(\frac{Un_0}{M}\Big)\frac{\bq^2}{m_{\textrm{eff}}} + \Big(\frac{Un_0}{M}\Big)^2\Big[1 - |\alpha(\bq)|^2\Big]}   
\end{align}
for small $|\bq|$. The two terms inside the square root in Eq.~\eqref{e1} have a completely different origin. The first is the usual term arising from the dispersion of the bosons in the considered Bloch band, which vanishes for a strictly flat band, i.e.\ when $m_{\textrm{eff}} \rightarrow \infty$. On the other hand, the second term involves overlaps of the Bloch states, and its connection to quantum geometry has not been considered in the context of Bose-Einstein condensation before. The quantity $\alpha(\bq)$reads
\begin{align}
\label{alphafunc}
\alpha(\bq) = M\sum_\alpha \langle  u_1(\bk_c + \bq) | \alpha \rangle \langle \alpha | \phi_0 \rangle \langle \phi^*_0 | \alpha \rangle \langle \alpha |  u_1^*(\bk_c - \bq) \rangle, 
\end{align}
 which gives an overlap between $|\phi_0\rangle$ and the states of the particle $| u_1(\bk_c + \bq) \rangle$ and the hole $| u_1^*(\bk_c - \bq) \rangle$.

Defining $\tilde{D}(\bq) \equiv \sqrt{1 - |\alpha(\bq)|^2}$, we can write Eq.~\eqref{e1} as 
\begin{align}
E_1(\bk_c + \bq)=\frac{Un_0}{M}\tilde{D}(\bq)
\end{align}
for a flat band with $m_\text{eff}\rightarrow\infty$. This shows that the Bogoliubov excitation energies for a flat band are determined by $\tilde{D}(\bq)$, which we call the "condensate quantum distance" between the condensed state and the neighbouring states, since $\tilde{D}(\bq) = 0$ when $\langle u_1(\bk_c + \bq) | \phi_0 \rangle = \langle u_1(\bk_c - \bq) | \phi_0 \rangle =1$. Indeed, $\tilde{D}(\bq)$ becomes identical to the Hilbert–Schmidt quantum distance $D(\bq)$~\cite{Rhim2020,Berry1989}
\begin{align}
\label{HSqd}
D(\bq) \equiv \sqrt{1- |\langle u_1(\bk_c + \bq) | u_1(\bk_c - \bq) \rangle|^2}.
\end{align}
when    $| u^*_1(\bk) \rangle = | u_1(\bk) \rangle$. 
By Taylor expanding up to second order in $\bq$, one has $D^2(\bq) = 4\sum_{\mu\nu}q_\mu q_\nu g^1_{\mu\nu}(\bk_c)$, 
where 
\begin{align}
g^n_{\mu\nu}(\bk) = \textrm{Re}\Big[\bra{\partial_\mu u_n(\bk)} \Big(1 - | u_n(\bk) \rangle \langle u_n(\bk)  |\Big) | \partial_\nu u_n(\bk)\rangle \Big]  
\end{align}
is the quantum metric $g^n_{\mu\nu}(\bk)$ of the $n$th Bloch band and $\partial_\mu \equiv \frac{\partial}{\partial q_\mu}$.
This implies that the speed of sound for the flat band, $c_{s,f.b.}$, reads 
\begin{align}
\label{fbcs2}
c_{s,f.b.}(\theta_\bq) = \frac{2Un_0}{M}\sqrt{ \hat{\textbf{e}}^T_\bq g^1(\bk_c) \hat{\textbf{e}}_\bq}, \textrm{ with } | u^*_1(\bk) \rangle = | u_1(\bk) \rangle    
\end{align}
is purely determined by the quantum metric. Here, $ \hat{\textbf{e}}_\bq = \bq/|\bq|$, $\tan \theta_\bq = q_y/q_x$ and $[g^1]_{\mu\nu} = g^1_{\mu\nu}$. 

From Eq.~\eqref{fbcs2} we see that finite quantum metric yields non-zero $c_s$ and thus allows superfluidity even in case of a flat band BEC. This result can be contrasted to fermionic flat band systems where superconductivity was shown to exists for non-zero Chern numbers or momentum space integrals of of the quantum metric~\cite{Peotta2015,Liang2017,Julku2016}. The difference to the present bosonic BEC case is that only the quantum metric of the condensed Bloch state $|\phi_0\rangle$ matters, not the geometric quantities integrated over the whole BZ. Furthermore, the speed of sound of a spin-orbit coupled Fermi gas was studied in Ref.~\cite{Iskin2020}: $c_s$ was shown to partially depend on the quantum geometric terms but their significance was hampered due to the presence of more dominant non-geometric contributions. In our case, quantum geometry manifests as the sole contribution for the weakly-coupled flat band BEC.

In the general case of $| u^*_1(\bk) \rangle \neq | u_1(\bk) \rangle$, we use the condition $|\langle \alpha | \phi_0 \rangle|^2 = 1/M$ and expand  $| u_1(\bk_c + \bq) \rangle$ and  $| u_1^*(\bk_c - \bq) \rangle$ up to second order in $\bq$. This gives 
\begin{align}
\label{d1}
&\tilde{D}^2(\bq) = \sum_{\mu\nu} q_\mu q_\nu \Big\{ 2g^1_{\mu\nu}(\bk_c) +2\textrm{Re}[ \langle \partial_\mu \phi_0 |\phi_0 \rangle \langle \phi_0|\partial_\nu \phi_0\rangle ] \nonumber \\
&+ 2M \textrm{Re}\Big[\sum_\alpha \langle \partial_\mu \phi_0 | \alpha \rangle \langle \alpha | \phi_0 \rangle \langle \phi_0^* | \alpha \rangle \langle \alpha | \partial_\nu \phi_0^* \rangle\Big] \Big\} \nonumber \\
&\equiv 4\sum_{\mu\nu}q_\mu q_\nu \tilde{g}_{\mu\nu}(\bk_c)
\end{align}
for $\bq \rightarrow 0$. We have introduced a generalised metric $\tilde{g}_{\mu\nu}$, which replaces $g^1_{\mu\nu}$ in Eq.~\eqref{fbcs2} for the speed of sound when 
$| u^*_1(\bk) \rangle \neq | u_1(\bk) \rangle$. Consequently, the discussion following Eq.~\eqref{fbcs2} remains valid for a general flat band with the replacement $g_{\mu\nu}\rightarrow\tilde g_{\mu\nu}$. 
With Eq.~\eqref{d1}, Eq.~\eqref{e1} and $c_{s,f.b.}$ can be recast as
\begin{align}
&E_1(\bk_c+\bq) = \sqrt{\Big(\frac{Un_0}{M m_{\textrm{eff}}} \Big)\bq^2 + \Big(\frac{2Un_0}{M} \Big)^2 \sum_{\mu\nu}q_\mu q_\nu \tilde{g}_{\mu\nu}(\bk_c)} \label{e2}  \\
& c_{c,f.b.} = \frac{2Un_0}{M}\sqrt{ \hat{\textbf{e}}^T_\bq \tilde{g}(\bk_c) \hat{\textbf{e}}_\bq}.
\end{align}
When the geometric contribution is zero, i.e. $\tilde{g}_{\mu\nu} =0$, the speed of sound for a dispersive band reduces to the 
usual form $c_s = \sqrt{Un_0/(Mm_{\textrm{eff}})}$, i.e. $c_s \propto \sqrt{U}$~\cite{Pitaevskii2003}. This should be compared to the flat band limit $m_{\textrm{eff}} \rightarrow \infty$, where the first term in Eq.~\eqref{e2} vanishes and $c_s \propto U$. The linear vs.\ square root dependence can be used to distinguish the sound velocities of geometric and conventional origin in an experiment where $U$ can be tuned. Equivalently, 
one can define an interaction-dependent effective mass for the flat band $\tilde{m}_{eff} = M/(4Un_0 \hat{\textbf{e}}^T_\bq \tilde{g} \hat{\textbf{e}}_\bq)$ so that the speed of sound is formally the same as for a dispersive band, $c_s = \sqrt{Un_0/(M\tilde{m}_{eff})}$. Different forms of $c_s$ for dispersive and flat band condensates are summarized in Fig.~\ref{Fig:summary}.

\begin{figure*}
  \centering
    \includegraphics[width=1.0\textwidth]{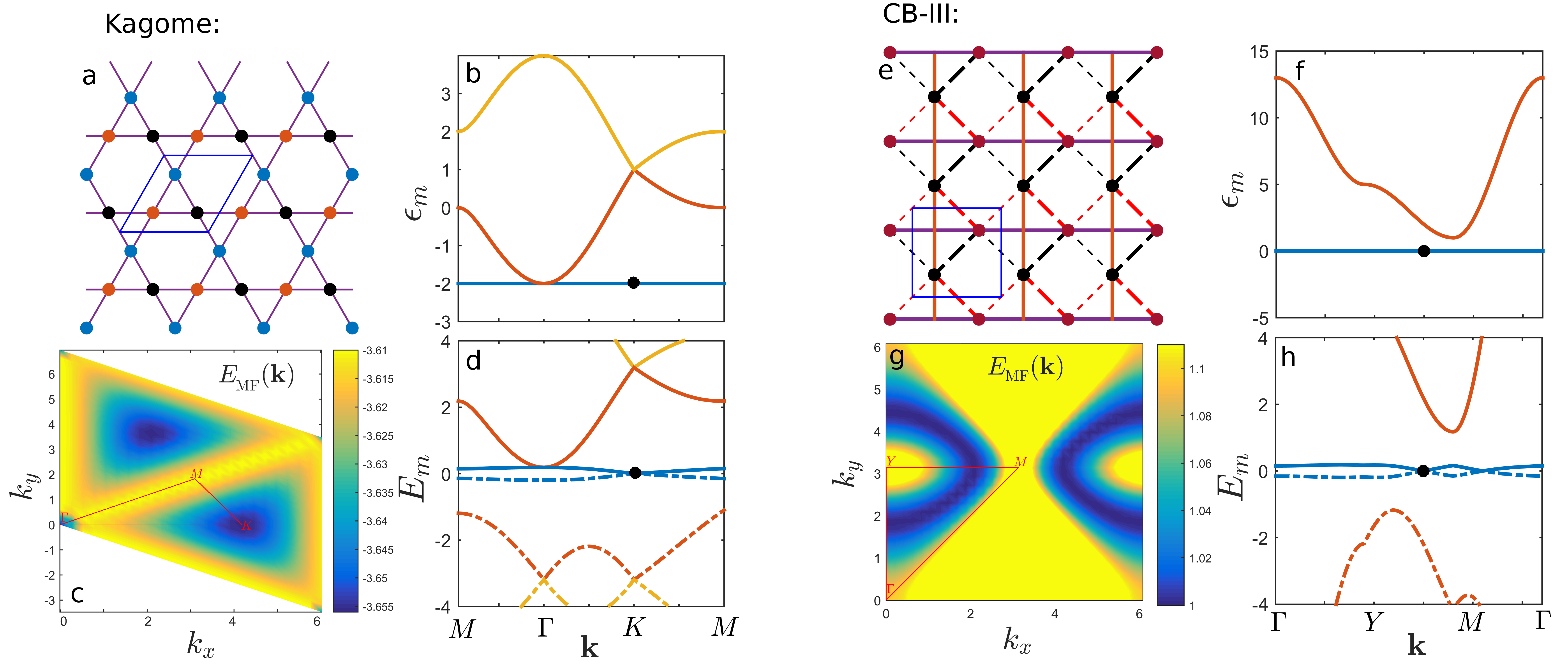}
    \caption{Flat band models considered in this work. (a) Kagome lattice geometry. The unit cell is shown as a blue parallelogram. Purple lines depict NN hopping terms of strength $t$. (b) Bloch bands of the kagome lattice with $t=1$ along high-symmetry points. (c) Mean-field energy $E_{\textbf{MF}}(\bk)$ across the  BZ for the kagome lattice. It is minimized at the $\Gamma$- and Dirac-points but fluctuation analysis~\cite{you:2012} shows that the Dirac-point condensate is the most favourable. The red lines depict the path along which the dispersions in panels (b) and (d) are plotted. (d) Bogoliubov spectrum for the kagome lattice along high-symmetry lines with $\bk_c = [4\pi/3,0]$ [marked as a black dot in (b) and (d)] and $Un_0/Mt=0.2$. Quasiparticle (-hole) modes are depicted with solid (dashed) lines. There is a gapless Goldstone mode at $\bk_c$. (e) Checkerboard-III (CB-III) geometry. The blue square represents the unit cell. Solid purple (orange) lines depict kinetic hopping terms of strength $t=1$ ($2t$), whereas bolded black (red) dashed lines are hoppings of strength $-2t$ ($-t$). Thin black and red dashed lines depict hoppings of strength $-2t(1-\delta)$ and $-t(1-\delta)$, where a finite staggering parameter $\delta \ll 1$ can be used to render the flat Bloch band slightly dispersive. In the computations we use $\delta = 10^{-5}$. The sublattice depicted as black (red) dots has the on-site energy of $5t$ ($2t$). (f) Bloch band structure for CB-III. (g) Mean-field energy $E_{\textbf{MF}}(\bk)$ for the CB-III model with $\delta = 0$. The red lines depict the path along which the dispersions in panels (f) and (h) are plotted. There is a continuous subset of Bloch states that minimize $E_\bk$ equally well. By using a finite $\delta$, one can introduce stable condensation to the momentum $\bk =[2\pi/3,\pi]$. (h) Bogoliubov spectrum for the CB-III model along high-symmetry lines with $\bk_c = [2\pi/\pi,\pi]$ [marked as a black dot in (b) and (d)] and $\delta = 10^{-5}$ at $Un_0/Mt=0.2$. Quasiparticle (-hole) modes are depicted with solid (dashed) lines. A gapless Goldstone mode exists at $\bk_c$. A small gap opens at the $M$-point, despite being barely visible in the plot.}
   \label{Fig:lattices}
\end{figure*}

\begin{figure}
  \centering
    \includegraphics[width=1.0\columnwidth]{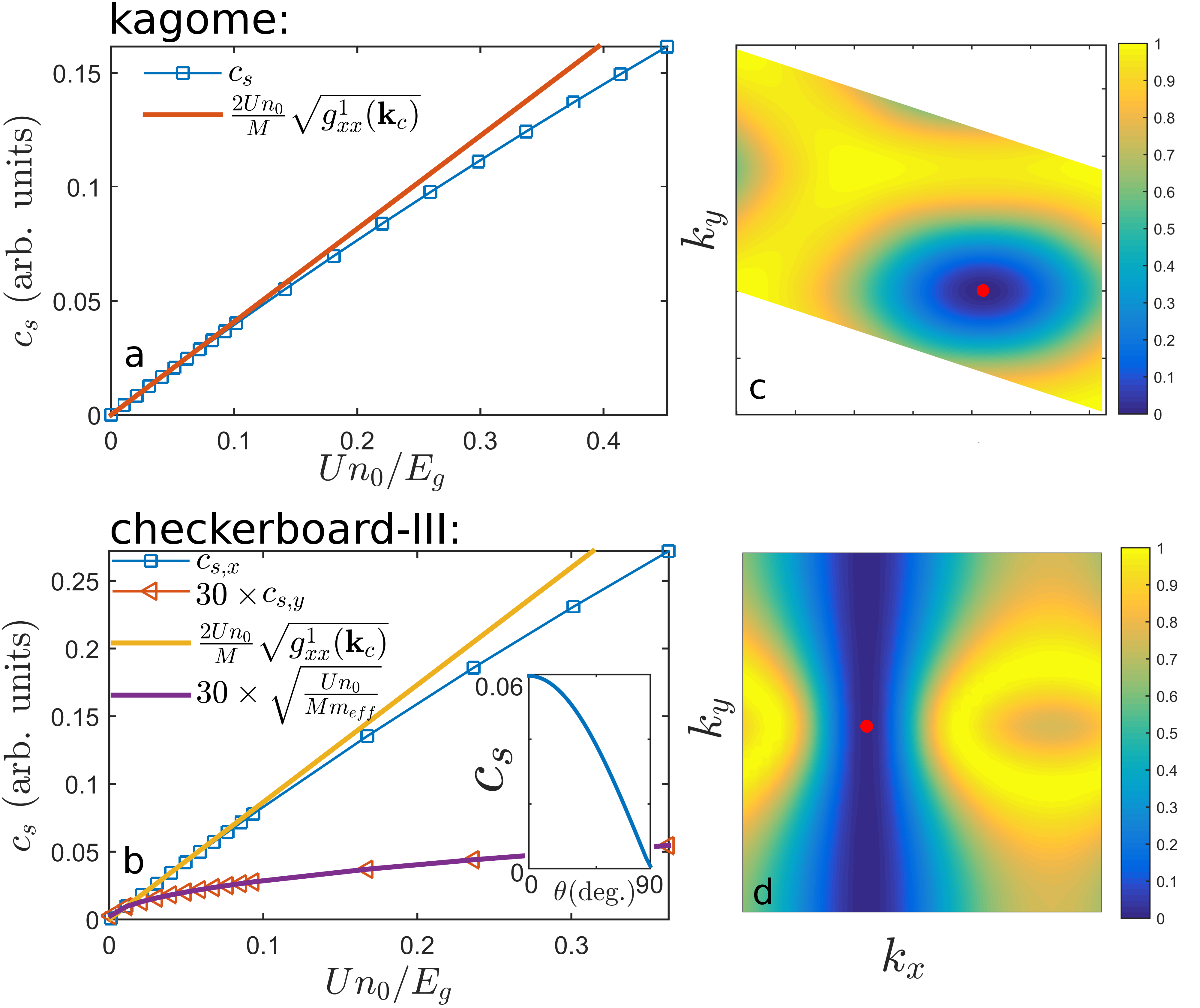}
    \caption{Speed of sound in flat band condensates. (a)-(b) The speed of sound $c_s$ for kagome and CB-III flat band condensates, respectively, as a function of $U$. Total density was chosen to be $n_{\textrm{tot}} = M$, i.e.\ one particle per lattice site. We also show the weak-coupling result of Eq.~\eqref{fbcs2}. The quasi-flat band of CB-III has a small but non-zero bandwidth $J \sim 10^{-5} Un_0$ due to a finite $\delta$. For the kagome model, $c_s$ is determined by the quantum metric of the condensed state, which in this case is isotropic. In CB-III, $c_s$ in the $x$-direction ($c_{s,x}$) depends on the quantum metric and in the $y$-direction ($c_{s,y}$) on the effective mass $m_{\textrm{eff}}$ of the quasi-flat band. The energy scaling $E_g$ is taken to be the energy gap between the flat band and higher bands at $\bk_c$. To make it more visible, data for $c_{s,y}$ is scaled by a factor of 30. Inset of (b) shows the angle dependence of $c_s$ for CB-III at $Un_0/E_g=0.07$. (c)-(d) The condensate quantum distance $\tilde{D}(\bq)$ for the kagome and CB-III lattices as a function of $\bk = \bk_c + \bq$. For the kagome (CB-III) lattice, $\tilde{D}$ is computed with respect to $\bk_c = [4\pi/3,0]$ ($\bk_c = [2\pi/3,\pi]$), marked as a red dot. For CB-III, $\tilde{D}(\bq)$ = 0 in the $k_y$-direction, consistent with the fact that for $\delta=0$ the condensation at $\bk_c =[2\pi/3,\pi]$ is unstable.
    }
   \label{Fig:sos}
\end{figure}

\subsection{Speed of sound in kagome and checkerboard lattices} \label{KagomeCheckerboard}
We now consider $c_s$ in two specific flat band models. The kagome lattice [see Fig.~\ref{Fig:lattices}(a)] consists of three sublattices, and one of the three Bloch bands is strictly flat [Fig.~\ref{Fig:lattices}(b)]~\cite{you:2012}. When the nearest-neighbour (NN) hopping is positive $t>0$, the flat band has the lowest energy and Bose condensation can take place within it. By minimizing the mean-field energy $E_{\textrm{MF}}(\bk)$, one finds~\cite{you:2012} that condensation at the $\Gamma$-point, i.e.\ $\bk=0$, or at one of the Dirac points, e.g.\ $\bk_c = \bk_K \equiv [4\pi/3,0]$, is favoured [see Fig.~\ref{Fig:lattices}(c)] since the particle density distributes uniformly among the sublattices for these Bloch states, minimizing the repulsive on-site interaction. A fluctuation analysis \cite{you:2012} shows the condensation at $\bk_K$ having a slightly smaller zero-point energy, being then more favourable than the $\Gamma$-condensate. We thus take $\bk_c = \bk_K$. Since the upper two dispersive bands are far away from the flat band at $\bk_K$, the $\bk_K$-condensate realises flat band condensation to a good approximation.

In Fig.~\ref{Fig:lattices}(d) we show a typical Bogoliubov spectrum for the kagome flat band condensate at $\bk_c = \bk_K$. A gapless Goldstone mode exists whose dispersion around $\bk_c$ is linear. By extracting the slope of the Goldstone mode, we obtain $c_s$ which is plotted in Fig.~\ref{Fig:sos}(a) as a function of interaction $U$. Furthermore, because $| u^*_1(\bk_c - \bq) \rangle = | u_1(\bk_c - \bq) \rangle$, Eq.~\eqref{fbcs2} holds for $U\rightarrow 0$. Thus, alongside the numerical result, we also plot in Fig.~\ref{Fig:sos}(a) the weak-coupling result of Eq. \eqref{fbcs2}. The agreement for small interactions is excellent. Moreover, $c_s$ is isotropic, consistent with the fact that $\tilde{D}(\bq)$ around $\bk_c$ is rotationally symmetric, see Fig.~\ref{Fig:sos}(c).

As another example, we consider the checkerboard-III (CB-III) geometry~\cite{Rhim2019} [Fig.~\ref{Fig:lattices}(e)] that consists of two sublattices and features a strictly flat band, see Fig.~\ref{Fig:lattices}(f). By minimizing the MF energy $E_{\textrm{MF}}(\bk)$, one finds that there exists a continuous subset of flat band Bloch states that minimize the condensation energy [Fig.~\ref{Fig:lattices}(g)]. One of these states is at $\bk = [2\pi/3,\pi]$, and computing $\tilde{D}(\bq)$ for this state gives $\tilde{D}(\bq) = 0$ at  $\bq || \textbf{e}_y$ as shown in Fig.~\ref{Fig:sos}(d). This means that $c_s=0$ in the $y-$direction, 
implying an unstable BEC.
This is consistent with $E_{\textrm{MF}}(\bk)$ being minimized for a continuous subset of flat band states, 
see Fig.\ \ref{Fig:lattices}(g), which makes condensation in a single Bloch state ambiguous. 

Our results demonstrate an important aspect of  flat band condensation. While finite interactions are needed, they are not sufficient to obtain a stable flat band BEC. An additional condition is to have a suitable quantum geometry for the flat band. Indeed, in the CB-III geometry, a flat band BEC cannot be obtained as the quantum distance is zero between some of the flat band states. Another example of an unstable flatband BEC is a single band model for which $\tilde{D}(\bq)=0$ for all $\bq$. In case of the kagome lattice, we have shown above that the flat band condensate can be stable due to a non-zero quantum distance between the Bloch states. The connection between the stability of a flat band BEC and the quantum distance is  elaborated further in the next section.

The checkerboard-III lattice condensation can be made stable by introducing a small perturbation to the NN hopping terms [Fig.~\ref{Fig:lattices}(e)] that renders the lowest Bloch band slightly dispersive. By computing $c_s$ with Eq.~\eqref{e2}, one finds that $c_s$ in the $y$-direction scales as $\sqrt{U}$ due to finite dispersion
. In the $x$-direction, however, $c_s$ scales linearly in $U$ and is therefore determined by the geometric term $\tilde{D}(\bq)$. 
Although the condition $|u_1^*\rangle = |u_1\rangle$ is not strictly met, $c_s$ in the $x$-direction is still mostly dictated by the quantum metric, see Fig.~\ref{Fig:sos}(b). By computing $c_s$ for all the directions $\theta_\bq$, we see that $c_s$ decreases monotonically between the directions $\theta_\bq = 0$ and $\theta_\bq = \pi/2$ [see inset of Fig.~\ref{Fig:sos}(b)], reflecting the anisotropy of $\tilde{D}(\bq)$ shown in Fig.~\ref{Fig:sos}(d).

\section{Excitation fraction} \label{Frag}


In addition to the speed of sound, quantum geometry can manifest also via other physical quantities. We demonstrate this here for the excitation density $n_{ex}$, i.e. the number of non-condensed bosons per unit cell, in the limit of $U\rightarrow 0$. In case of the usual dispersive band condensate, one has $\lim_{U\rightarrow 0}n_{ex} = 0$ at zero-temperature ~\cite{Pitaevskii2003,Fetter1971}. However, for a strictly flat band, the limit $U\rightarrow 0$ implies that the lowest Bogoliubov excitation band becomes flat as the off-diagonal terms of $\mathcal{H}_B$ in Eq.~\eqref{BogMatrix} vanish, i.e. $\Delta \rightarrow 0$.  This implies that the BEC at a single flat band state becomes unstable (see also Appendix D) as the lowest Bogoliubov band excitations have vanishing energy cost. Therefore, it is not intuitively clear what kind of asymptotic behavior $ \lim_{U\rightarrow 0}n_{ex}$ of a flat band condensate features. In this section we show that, remarkably, $n_{\textrm{ex}}$ can be {\it non-zero} and {\it finite} for vanishing interaction strength, and that its value is dictated by quantum geometry.

\begin{figure}
  \centering
    \includegraphics[width=1.0\columnwidth]{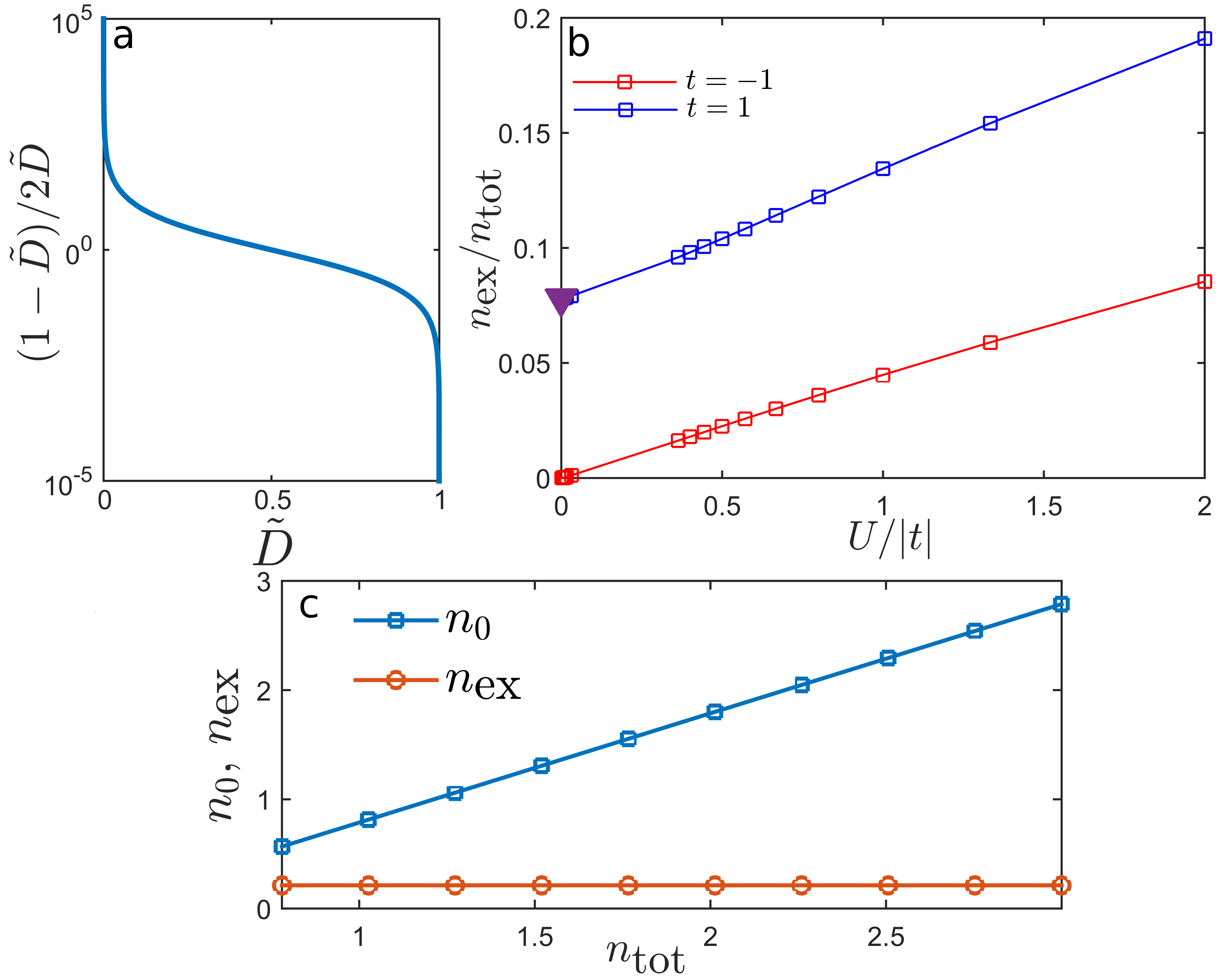}
    \caption{Excitation density. (a) Functional form of Eq.~\eqref{nex_ana}. For $\tilde{D} \rightarrow 0$, one has $(1-\tilde{D})/2\tilde{D} \rightarrow \infty$, which implies instability of the BEC and breakdown of the Bogoliubov theory. (b) Excitation fraction $n_{\textrm{ex}}/n_{\textrm{tot}}$ of the two kagome lattice condensation schemes for $n_{\textrm{tot}}=3$ as a function of $U$. For the flat band condensate ($t=1$), $n_{\textrm{ex}}$ remains finite even in the limit of $U\rightarrow 0$. Purple triangle depicts the analytical result of Eq.~\eqref{nex_ana}. (c) Densities $n_0$ and $n_{\textrm{ex}}$ as a function of $n_{\textrm{tot}}$ for kagome flat band condensation at $\bk_c = [4\pi/3,0]$ and $U = |t|/1800$. Excitation density $n_{\textrm{ex}}$ remains constant, as it is determined by only the quantum geometry.}
   \label{Fig:Kagome_n_ex}
\end{figure}

The expression for $n_{\textrm{ex}}$ can be written as (see Appendix D for details)
\begin{align}
\label{nex}
n_{ex} &= \frac{1}{N}\sum_{\bk m}{}^{'} \langle c^\dag_{\bk m} c_{\bk m} \rangle = \frac{1}{2N} \sum_{\bk m}{}^{'}[-1 + \langle \psi^-_m(\bk) |  \psi^-_m(\bk)  \rangle]\nonumber \\
&= \frac{1}{N}\sum_{\bk}{}^{'}n_{ex}(\bk),
\end{align}
where $c^\dag_{\bk m}$ creates a boson in the Bloch band $m$ with momentum $\bk$ and energy $\epsilon_m(\bk)$, obtained by diagonalising $\mathcal{H}(\bk)$ in~\eqref{BogMatrix}. 

It is instructive to consider first a single band case for which the one-particle energy spectrum reads $\epsilon(\bk) \propto \bk^2/m_{eff}$ corresponding to a continuum system or the long-wavelength limit of a square lattice. The corresponding Bogoliubov energy is then $E_1(\bk) = \sqrt{\tilde{\epsilon}(\bk)[ \tilde{\epsilon}(\bk) + 2Un_0]}$ with $\tilde{\epsilon}(\bk) \equiv \epsilon(\bk) - \epsilon_0$, and  one has $n_{ex}(\bk) = [\tilde{\epsilon}(\bk)+Un_0]/E_1(\bk) -1$~\cite{Fetter1971}. We see that $n_{ex}$ diverges  in the flat band limit $1/m_{eff} = 0$. Thus, a flat band BEC within a single band system is always unstable. This is consistent with the results of the previous sections since for a single band lattice system one always has $\alpha(\bk) =1$. Consequently, one needs to incorporate quantum geometric effects to reach a stable BEC, i.e.~consider a multiband system.

We now focus on a multiband lattice system, assuming the condensation to take place within a strictly flat band and neglecting higher bands, which is justified for a small $U$. We therefore deploy Eq.~\eqref{lk2} with $1/m_{\textrm{eff}} =0$. Solving  $L_p(\bk)|\psi^-_1(\bk)\rangle = -E_1(\bk)|\psi^-_1(\bk)\rangle$ and taking into account the proper normalization requirements [Eq.~\eqref{normalization} in Appendix B] gives
\begin{align}
\label{nex_ana}
\lim_{U\rightarrow 0}\langle c^\dag_{\bk 1} c_{\bk 1} \rangle =  \frac{1-\tilde{D}(\bq)}{2\tilde{D}(\bq)}, 
\end{align}
where $\bq = \bk - \bk_c$. This expression relates a simple way the density of non-condensed bosons to the condensate quantum distance $\tilde{D}(\bq)$. It is valid for infinitesimally small but finite $U$; in complete absence of interactions ($U=0$), a BEC in a strictly flat band cannot exist since all momentum states remain degenerate, as discussed in Section~\ref{FlatBEC}.

 The functional form of Eq.~\eqref{nex_ana} is depicted in Fig.~\ref{Fig:Kagome_n_ex}(a). If $\tilde{D}(\bq) = 0$, $n_{ex}(\bk)$ diverges, implying the breakdown of the Bogoliubov theory. This is intuitively easy to comprehend as a consequence of a perfect overlap between $|\phi_0\rangle$ and non-condensed Bloch states, which leads the particles to "spill" out of the condensate. 
 It is also in agreement with the results presented in the previous section where $\tilde{D}(\bq)=0$ indicated unstable flat band condensation.

To illustrate the role of the condensate quantum distance, we now consider $n_{\textrm{ex}}$ in case of the flat band BEC of the kagome lattice. For comparison, we also consider the dispersive band BEC by carryring out the calculations for negative NN hopping term, i.e. $t=-1$. With this choice, one of two dispersive Bloch bands is the lowest one in which the bosons condense at $\bk_c=0$.

In Fig.~\ref{Fig:Kagome_n_ex}(b) we plot $n_{\textrm{ex}}/n_{\textrm{tot}}$ (with $n_{\textrm{tot}}$ being the total density) as a function of interaction for the flat and dispersive band condensates. In the case of the dispersive band BEC (red curves), we see that 
$n_{\textrm{ex}}$ vanishes when the interaction goes to zero as usual. In contrast, for the flat band condensate $n_{\textrm{ex}}$ is non-zero when $U\rightarrow 0$, indeed approaching the value obtained by integrating Eq.~\eqref{nex_ana} over the BZ (purple triangle). The different behavior of $n_{\textrm{ex}}$ for dispersive and flat band condensates is summarized in Fig.~\ref{Fig:summary}.

\begin{figure}
  \centering
    \includegraphics[width=1.0\columnwidth]{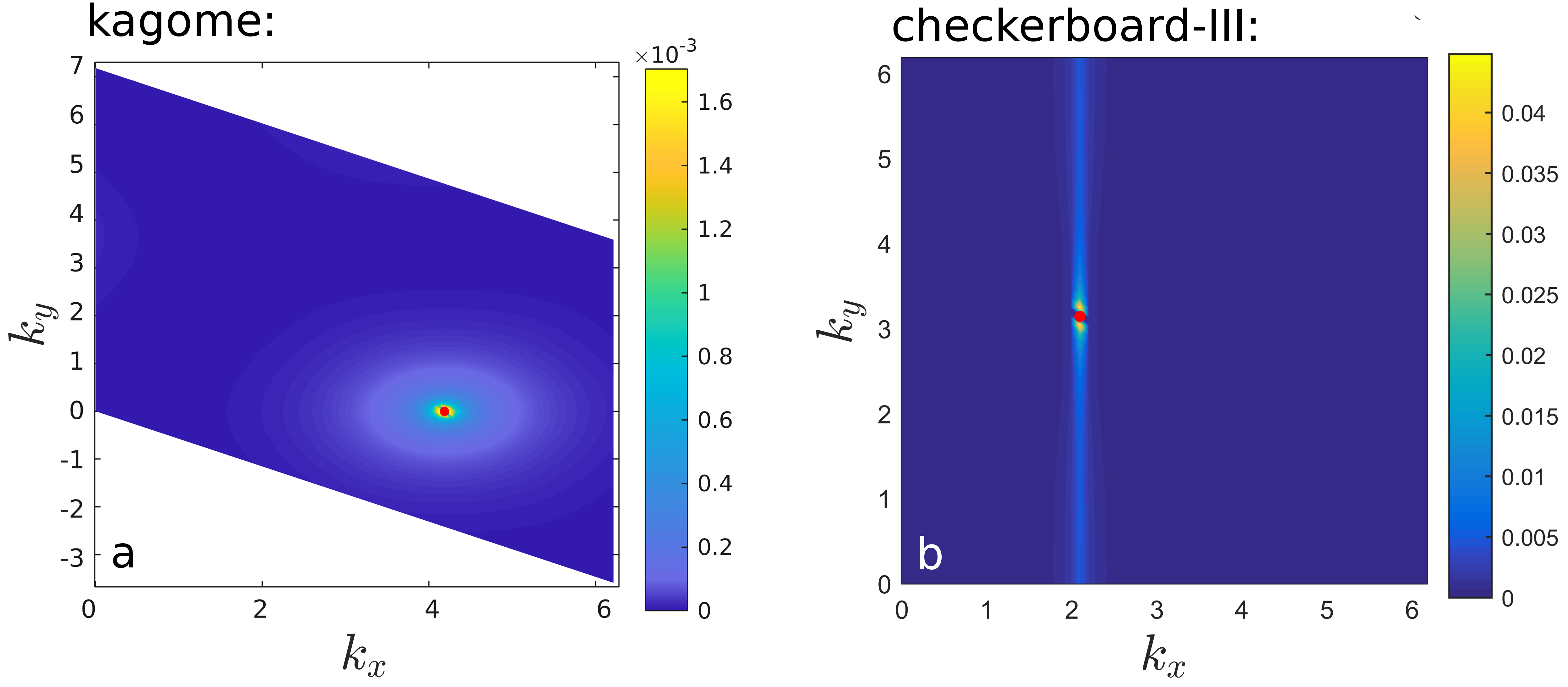}
    \caption{Momentum dependence of the excitation density. (a)-(b) Momentum dependence of $n_{\textrm{ex}}(\bk)$ for kagome and CB-III lattices, respectively, with $n_{\textrm{tot}}=M$. The interaction in the kagome (CB-III) model is chosen to be $Un_0/E_g= 5.13\times 10^{-4}$ ($Un_0/E_g= 0.011$); in the case of CB-III, we use $\delta = 10^{-5}$ to avoid a diverging $n_{\textrm{ex}}$. Remarkably, the excitation fraction follows the features of $\tilde{D}$ shown in Figs.~\ref{Fig:sos}(c)-(d), as predicted by Eq.~\eqref{nex_ana}.}
   \label{Fig:nex}
\end{figure}

In Fig.~\ref{Fig:nex}(a), we present $n_{\textrm{ex}}(\bk)$ of the kagome flat band BEC across the BZ for a small $U$. As predicted by Eq.~\eqref{nex_ana}, the momentum dependence of $\tilde{D}(\bk)$, shown in Fig.~\ref{Fig:sos}(c), is imprinted to the momentum distribution of $n_{\textrm{ex}}(\bk)$. Similar conclusions can be reached by considering the CB-III model [Fig.~\ref{Fig:nex}(b)], where $n_{\textrm{ex}}(\bk)$ is non-zero only near the momenta for which $\tilde{D}(\bk)$ vanishes [see Fig.~\ref{Fig:sos}(d)].

The kagome flat band condensate respects the condition $| u^*_1(\bk) \rangle = | u_1(\bk) \rangle$ and therefore $\tilde{D}(\bq)$ reduces to the usual Hilbert-Schmidt quantum distance $D(\bq)$~\eqref{HSqd}. Thus, in the kagome lattice, the Hilbert-Schmidt distance of the flat band states directly determines $n_{\textrm{ex}}$. The Hilbert-Schmidt distance has been previously connected to Landau level spreading in non-interacting flat band models~\cite{Rhim2020}, and our result is one of the first to reveal how the quantum distance affects physically relevant quantities in an \textit{interacting many-body quantum system}.

Remarkably, $\lim_{U\rightarrow 0}n_{\textrm{ex}}$ of the flat band BEC in Eq.~\eqref{nex_ana} 
does not depend on $n_{\textrm{tot}}$ but is determined by the quantum distance. Therefore, one can, by decreasing $n_{\textrm{tot}}$, increase the excitation fraction $n_{\textrm{ex}}/n_{\textrm{tot}}$ and the role of the interaction and quantum fluctuation effects even in the non-interacting limit $U\rightarrow 0$. We show this in Fig.~\ref{Fig:Kagome_n_ex}(c) where  $n_0$ and $n_{\textrm{ex}}$ are depicted as as a function of $n_{\textrm{tot}}$ for small interaction of $Un_0/E_g= 5.13\times 10^{-4}$ in the case of the kagome model. The excitation density $n_{\textrm{ex}}$ remains constant as a function of $n_{\textrm{tot}}$, in contrast to $n_0$ which decreases when the total density becomes smaller.

Note that in Fig.~\ref{Fig:Kagome_n_ex}(c) we consider the regime $n_{\textrm{ex}} \sim n_0$ and thus one might anticipate the usual Bogoliubov theory to break down~\cite{Shi1998}. We thus also considered the Hartree-Fock-Bogoliubov (HFB) approximation, which, unlike the simple Bogoliubov theory, takes into account the first order self-energy diagrams not containing any condensate propagators (see Appendix E and Fig. S1 in Supplementary Material (SM)~\cite{suppis}). Even for rather large $n_{\textrm{ex}}/n_{\textrm{tot}}$, the Bogoliubov and HFB approximations yield the same results in the limit of $U\rightarrow 0$. This agreement between HFB and Bogoliubov theory confirms the accuracy of our approach, even when $n_{ex}$ becomes rather large. 

The lowest total density used in Fig.~\ref{Fig:Kagome_n_ex}(c) is approximately  $n_{\textrm{tot}}\sim 0.75$. By utilizing the HFB approach for even smaller total densities, one notes that it is unable to yield self-consistent solutions, implying that our assumption about the existence of a homogeneous BEC might become invalid. It was shown in Ref.~\cite{Huber:2010} using mean-field theory that for 
densities $n_{tot}\lesssim 0.72$, a charge density wave (CDW) co-exists, alongside the condensate at $\bk_c = [4\pi/3,0]$. Therefore, the breakdown of our HFB method around this density may occur because of a co-existing competing order which is not taken into account in the HFB theory. In Ref.~\cite{Huber:2010}, the co-existence of the CDW order was confirmed by exact diagonalization computations. 

The dilute limit considered in Ref.~\cite{Huber:2010} sets important limits for the applicability of our Bogoliubov and HFB approach, that assume the existence of a BEC. Namely, it can easily be shown  analytically~\cite{Huber:2010} that for densities $n_{tot} \leq 1/3$, the ground state consists of spatially non-overlapping compact localized states (CLS), leading to CDW order. This is a unique feature of flat band systems, since CLS states can be constructed as superpositions of flat band states only~\cite{Bergman:2008,Rhim2019}. Below the critical density $n_{tot,c} = 1/3$, the particles are distributed among spatially non-overlapping CLSs. However, for $n_{tot} \geq n_{tot,c}$ it is not (due to the repulsive interaction) energetically favourable to accumulate particles only to the non-overlapping CLSs and added bosons hop between interstitial sites of the CDW order, forming a condensate at $\bk_c = [4\pi/4,0]$. In Ref.~\cite{Huber:2010}, the dilute densities of $n_{tot} \in [1/3,~\sim0.72]$ were studied showing that both the CDW and the BEC phases can exist or co-exist such that the order parameter of CDW (BEC) decreases (increases) as a function of $n_{tot}$. In this work, however, we are interested in BEC phenomena of flat band systems rather than different co-existing many-body phases, that possibly take place within a flat band system. Accordingly, we do not consider the dilute density limit of Ref.~\cite{Huber:2010} but instead higher densities. While our Bogoliubov and HFB theories ignore a possible co-existing CDW phase, it is expected that our method should capture the essential physical features of the BEC phase itself. Based on Ref.~\cite{Huber:2010}, the BEC is expected to be favorable as compared to the CWD order at higher densities. Our calculations are therefore reliable for the kagome lattice system  at the densities we consider.

\section{Second order correlations} \label{g2}
We further unravel the significance of the fluctuations by considering the second order coherence $g^{(2)}(\mathbf{r},\tau)$ where $\mathbf{r}$ is the separation of two spatial positions and $\tau$ of two time instances. Specifically, we focus on the correlations within a unit cell:
\begin{align}
\label{g2}
g^{(2)}(0,0) &= \frac{\frac{1}{N}\sum_{i\alpha\beta}\langle c_{i\alpha}^\dag c_{i\beta}^\dag c_{i\alpha} c_{i\beta} \rangle}{n_{\textrm{tot}}^2} \nonumber \\
&= \frac{\frac{1}{N}\sum_i\langle n_i n_i \rangle - n_{\textrm{tot}} }{n_{\textrm{tot}}^2},
\end{align}
where the second form follows from the bosonic commutation rules. Here $n_i = \sum_\alpha c^\dag_{i\alpha}c_{i\alpha}$ is the density operator of the $i$th unit cell.  In other words, we are interested in the sum of second order coherences within a unit cell, averaged over all the unit cells of the lattice. 

\begin{figure}
  \centering
    \includegraphics[width=1.0\columnwidth]{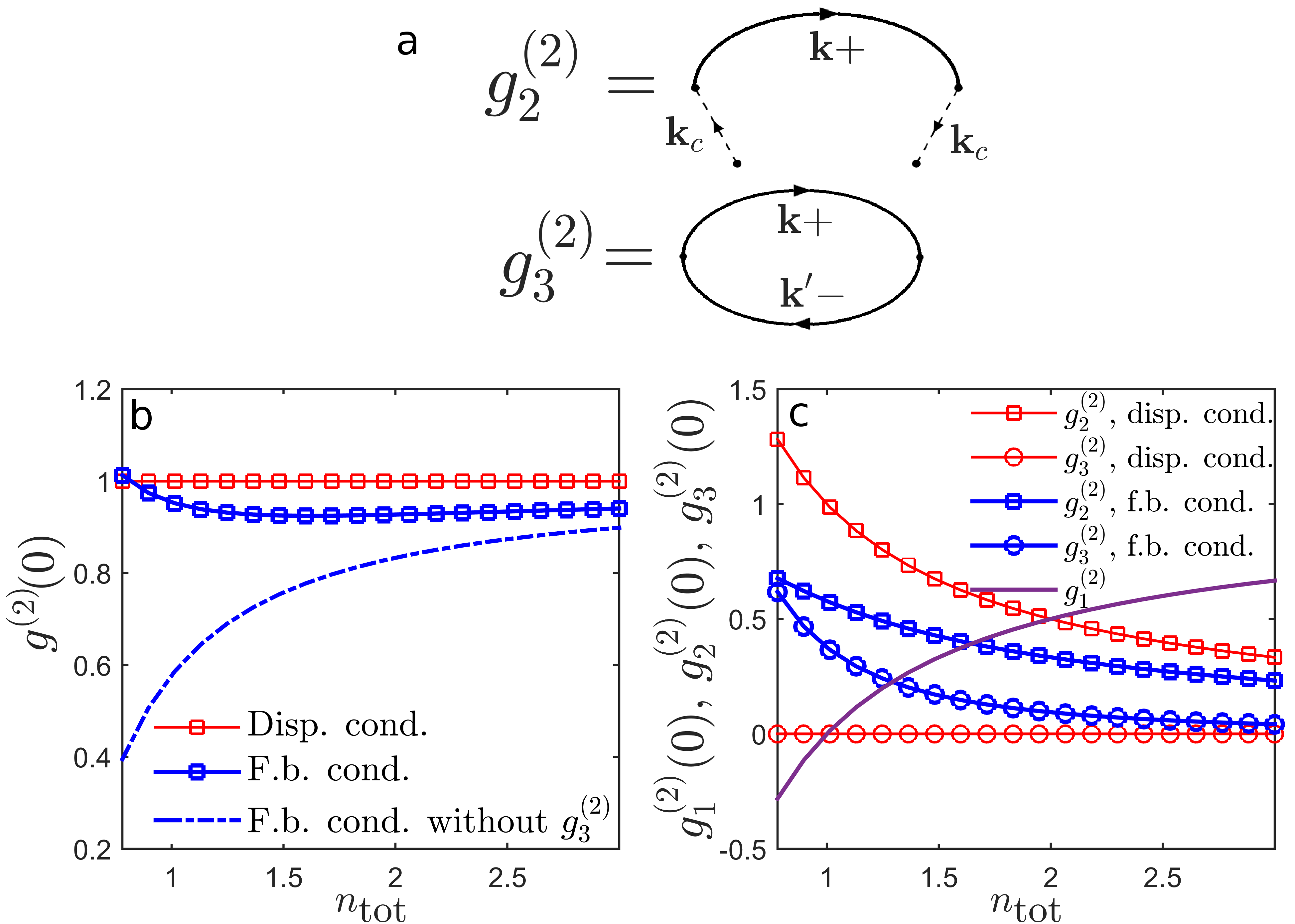}
    \caption{Second order local correlator. (a) Diagrammatic presentation of $g^{(2)}_2$ and $g^{(2)}_3$. The solid (dashed) lines depict the Bogoliubov (condensate) propagators. (b) The second order correlation function $g^{(2)}$ for the kagome lattice in the case of dispersive (red curve) and flat band (blue curve) condensates. The dashed blue line gives $ g^{(2)}_1 +  g^{(2)}_2$ for the flat band condensate. (c) Terms $g^{(2)}_2$ and $g^{(2)}_3$ for the dispersive band condensate (red curves) and flat band condensate (blue curves) as a function of $n_{\textrm{tot}}$. The term $g^{(2)}_1$ is also shown (purple curve) and is the same for both of the cases, as it depends only on $n_{\textrm{tot}}$. The interaction is set to $U=|t|/1800$.}
   \label{Fig:g2}
\end{figure}

By treating Eq.~\eqref{g2} at the Bogoliubov level, one finds (see SM~\cite{suppis} for details) that the resulting $g^{(2)}$ consists of three terms, i.e. $g^{(2)} = \sum_{i=1}^3 g^{(2)}_i$. The first term is the MF contribution, reading $g^{(2)}_1 = 1 - 1/n_{\textrm{tot}}$, whereas $g^{(2)}_2$ involves the overlap functions between the Bogoliubov excitations and the condensed Bloch state. Finally, $g^{(2)}_3$ describes the contribution arising solely from the quantum fluctuations. The expressions for $g^{(2)}_2$ and $g^{(2)}_3$ are provided in SM~\cite{suppis} and their Feynman diagrams are depicted in Fig.~\ref{Fig:g2}(a).

We find that as $n_{\textrm{ex}}$, $g^{(2)}_3$ of the flat band condensate at $U\rightarrow0$ is also governed by quantum geometry only. Namely, by considering again only the flat band degrees of freedom, one finds
\begin{align}
\lim_{U\rightarrow 0} g^{(2)}_3 &= 
\frac{1}{8N^2 n_{\textrm{tot}}^2} \left(\sum_\bk{}^{'} \frac{\alpha(\bk)}{\tilde{D}(\bk)} \sum_\bk{}^{'} \frac{\alpha^*(\bk)}{\tilde{D}(\bk)} \right. \nonumber \\
& + \left. \sum_\bk{}^{'} \frac{1 - \tilde{D}(\bk)}{\tilde{D}(\bk)} \sum_\bk{}^{'} \frac{1 + \tilde{D}(\bk)}{\tilde{D}(\bk)}  \right)  
\end{align}
This expression diverges (vanishes) when $\tilde{D}\rightarrow 0$ ($\tilde{D}\rightarrow 1$) and is finite for $0< \tilde{D} <1$. Quantum geometry therefore guarantees finite quantum fluctuations even in the non-interacting limit of $U\rightarrow 0$. 

The fluctuation term $g^{(2)}_3$ can be made significant by tuning down the total density $n_{\textrm{tot}}$. This is demonstrated for the kagome lattice in Fig.~\ref{Fig:g2}(c), where $g^{(2)}_2$ and $g^{(2)}_3$ are shown as a function of $n_{\textrm{tot}}$ for flat and dispersive band condensates for a small interaction of $U=|t|/1800$. We see $g^{(2)}_3$ being negligible for the dispersive band condensate, whereas quantum geometry guarantees finite $g^{(2)}_3$ for the flat band condensate. For decreasing $n_{\textrm{tot}}$, the fluctuation term $g^{(2)}_3$ increases and eventually becomes comparable to $g^{(2)}_2$. The full coherence function in Fig.~\ref{Fig:g2}(b) shows essentially the coherent BEC value $g^{(2)}=1$ in the dispersive case, as expected. Intriguingly, the flat band condensate shows antibunching behavior, despite the minute value of the interaction $U$, which crosses over to the bunching regime at small densities. This non-monotonic behavior as a function of the density could be used experimentally to probe the effect of quantum geometry in case of a flat band condensate. The fundamental difference between the second order coherence function for dispersive and flat band BECs is outlined in Fig.~\ref{Fig:summary}.

\section{Superfluid weight} \label{SFweight}
Finite $c_s$ guarantees the possibility of superfluidity, but to understand the phenomenon more deeply, we consider the superfluid weight tensor $D^s_{\mu\nu}$ (also referred to as superfluid density or superfluid stiffness in the literature). The superfluid weight is the long-wavelength, zero frequency limit of the current-current linear response $K_{\mu\nu}(\bq,\omega)$~\cite{Liang2017}, i.e.
\begin{align}
\label{ds}
&D^s_{\mu\nu} = \lim_{\bq \rightarrow 0} \lim_{\omega \rightarrow 0} K_{\mu\nu}(\bq,\omega).
\end{align}
We calculate this by deploying the Matsubara Green's function formalism in the framework of the Bogoliubov theory (see SM~\cite{suppis}) at zero temperature. The resulting $D^s$ can be divided into three contributions, i.e. $D^s_{\mu\nu} = \sum_{i=1}^3 D^s_{,i\mu\nu}$. The first term is the pure condensate contribution and reads $D^s_{1,\mu\nu} = n_0\langle \phi_0 |\partial_\mu \partial_\nu \mathcal{H}(\bk_c) | \phi_0 \rangle$. This generalizes the usual mean-field result of a single-band system $D^s = n_0/m_{\textrm{eff}}$, where the effective mass $m_{\textrm{eff}}$ is replaced by $\langle \phi_0 |\partial_\mu \partial_\nu \mathcal{H}(\bk_c) | \phi_0 \rangle$ accounting for the multiband nature of the system. The second term $D^s_2$ constitutes the mixed contribution of the condensate wave function and quantum fluctuations, and the third term $D^s_3$ arises solely from the quantum fluctuations, i.e. from the Bogoliubov excitations. The expressions for $D^s_2$ and $D^s_3$ are provided in SM~\cite{suppis}.

\begin{figure}
  \centering
    \includegraphics[width=1.0\columnwidth]{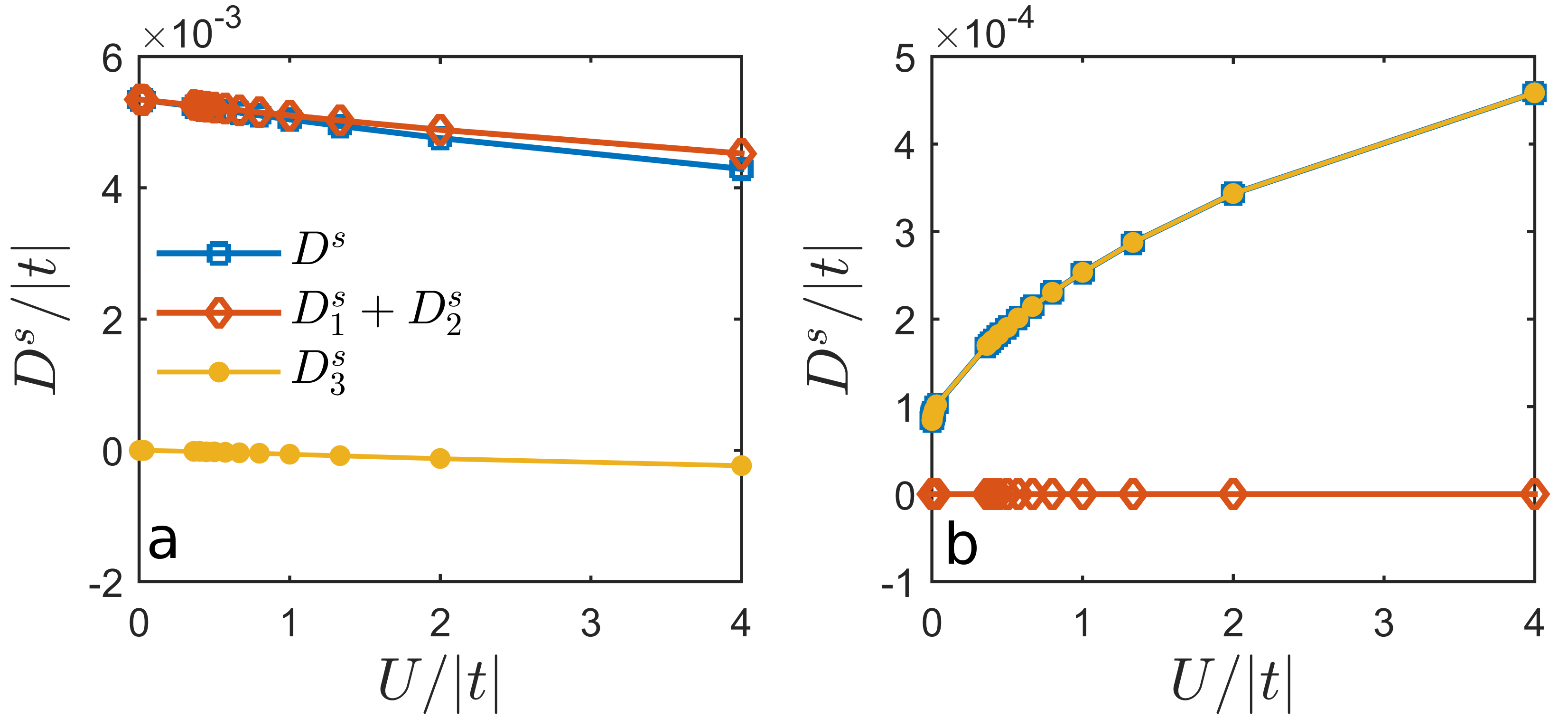}
    \caption{Zero temperature superfluid weight $D^s$ and its different components for the kagome lattice as a function of $U$. (a) Dispersive band condensate. (b) Flat band condensate. The particle density is set to $n_{\textrm{tot}}=3$. Superfluidity of the flat band is governed by the fluctuation term $D^s_3$ alone.}
   \label{Fig:SFW}
\end{figure}

For a single band square lattice, one has $D^s_{1,\mu\mu} = n_0 \partial^2_\mu \epsilon_1(\bk_c)$, $D^s_2 = 0$ and $D^s_3 <0$. Thus, the condensation contribution $D^s_1$ is determined by the inverse effective mass of bosons at $\bk_c$ and quantum fluctuations decrease the supercurrent via negative $D^s_3$. However, for a flat band condensate, the story is very different. We demonstrate this in Fig.~\ref{Fig:SFW} by depicting $D^s$ and its components as a function of $U$ in the case of the kagome model for both the dispersive [Fig.~\ref{Fig:SFW}(a)] and flat band condensates [Fig.~\ref{Fig:SFW}(b)]. For the dispersive band condensate, $D^s$ is mainly given by $D^s_1 + D^s_2$ and the fluctuation term $D^s_3$ gives a negative contribution. 
In stark contrast, the flat band condensate results depicted in Fig.~\ref{Fig:SFW}(b) reveal that $D^s$ is solely determined by the fluctuation term $D^s_3$. Therefore, instead of inhibiting the superfluidity, \textit{quantum fluctuations are actually responsible for finite superfluidity of the flat band condensate}. 

Flat band superfluidity is governed by $D^s_3$ because $D^s_2$ cancels the condensate contribution $D^s_1$. This can be proven in the weak-coupling limit  $U\rightarrow 0$ (shown in SM~\cite{suppis}). For the kagome lattice, the cancellation also takes place for finite interactions as shown in Fig.~\ref{Fig:SFW}(b). A very similar result was obtained recently in Ref.~\cite{yang2021}, where condensation was studied in a two-dimensional spin-orbit coupled system featuring a one-particle dispersion with a continuous subset of degenerate ground states on a circle of equal momentum amplitudes. Also in that case, finite superfluidity in the direction where the effective mass diverges was shown to arise from the quantum fluctuations only.

Our result regarding quantum fluctuation-mediated flat band superfluidity is intuitively easy to understand. Namely, due to destructive interference among the flat band states, bosons condensed to a single flat band Bloch state are spatially localized to a subset of lattice sites and thus have vanishing group velocity. In contrast, for a dispersive band BEC, it is the condensate that mainly carries the supercurrent as the effective mass of a dispersive band BEC is finite. In case of a flat band BEC, instead, it is the fluctuations, arising due to interactions, that are needed for a finite supercurrent. This fundamental difference regarding the origin of superfluidity in disperisve and flat band condensates is  summarized in Fig.~\ref{Fig:summary}.

The fluctuation term $D^s_3$ is similar to the superfluid weight of fermionic systems~\cite{Peotta2015,Liang2017}, except for factors accounting for fermionic statistics, as expected from the formal connection between bosonic Bogoliubov and fermionic BCS approximations. Like fermionic $D^s$, $D^s_3$ can also be split into intra- ("conventional") and interband ("geometric") terms, i.e. $D^s_3 = D^s_{3,\textrm{conv}} + D^s_{3,\textrm{geom}}$. Specifically, $D^s_3$ contains the current terms of the form (see SM~\cite{suppis})
\begin{align}
\langle u_n(\bk) &| \partial_\mu \mathcal{H}(\bk) | u_m(\bk) \rangle = \partial_\mu \epsilon_n(\bk)\delta_{mn} + \nonumber \\ 
&[\epsilon_m(\bk) - \epsilon_n(\bk)]\langle \partial_\mu u_n(\bk) | u_m(\bk) \rangle.    
\end{align}
The intraband term $D^s_{3,\textrm{conv}}$ contains only the diagonal current terms ($m=n$) proportional to the derivatives of the single-particle dispersions $\partial_\mu \epsilon_n(\bk)$, whereas the geometric contribution $D^s_{3,\textrm{geom}}$ includes the interband current terms ($m\neq n$) proportional to the interband Berry connection $\langle \partial_\mu u_n(\bk) | u_m(\bk) \rangle$. For a flat band $D^s_{3,\textrm{conv}}$ vanishes due to momentum independent dispersion, and $D^s_1$ and $D^s_2$ cancel each other, therefore it is the geometric part of the fluctuation term, $D^s_{3,\textrm{geom}}$, that is responsible for the superfluidity of the flat band condensate.

\section{Conclusion} \label{Conclusions}
We have deployed multiband Bogoliubov theory to study how quantum geometric properties of the Bloch states affect various physical quantities of a weakly-interacting BEC, with the emphasis being in lattice models that feature flat bands. We showed that in case of a flat band BEC, the speed of sound $c_s$ is dictated by the quantum metric and the excitation density $n_{\textrm{ex}}$ remains finite at $U\rightarrow 0$ due to finite quantum distance between the condensed state and other Bloch states of the band. Similarly, we demonstrated that also the quantum fluctuation part of the density-density correlations remains finite in the limit of $U\rightarrow 0$ and is given by the same quantum distance.

Our results demonstrate the unusually prominent role of fluctuations in a flat band BEC. The weak-coupling limit of the excitation fraction and the quantum fluctuation contribution of the density-density correlator do not depend on the condensate density, and therefore their relative importance can be enhanced by reducing the total density. Even the extreme situation where the superfluid weight is given by the quantum fluctuations alone can be achieved in the small interaction limit. 

The fact that interactions, quantum fluctuations, and correlation effects may become enhanced has been a key motivation for studies of flat bands. Our work shows that indeed this promise can be fulfilled in the case of bosonic condensates, and that these effects can be controlled by the underlying non-trivial quantum geometry, that is, a non-zero quantum metric, and distances between the Bloch states. A non-trivial quantum geometry guarantees the stability of the condensate via a non-diverging excitation fraction, and finite speed of sound. Furthermore, it allows to realize situations where quantum fluctuations and correlations dominate the system behavior.

Our work, alongside the accompanying study in Ref.~\cite{julku2021b}, establishes important connections between the quantum geometry and physical quantities of a BEC, showing how the geometric properties of quantum states can be utilized to reach the strongly correlated regime. This can have important consequences especially in systems where the interactions between particles are inherently small compared to kinetic tunneling energies and quantum correlations and depletion are weak, e.g. in photonic or polaritonic platforms. It could provide a key step towards realizing strongly correlated photons, important for both fundamental research and information processing. In long-lived, quasi-equilibrium photon and polariton systems our results should directly apply, while extending them to the strongly driven-dissipative case is an interesting future research task.


\begin{acknowledgments}
\textit{Acknowledgements}
We thank Long Liang and Menderes Iskin for useful discussions. A.J.~and P.T.~acknowledge support by the Academy of Finland under project numbers 303351, 307419 and 327293. A.J.~acknowledges financial support from the Jenny and Antti Wihuri Foundation. This research was supported in part by the National Science Foundation under Grant No.~PHY-1748958. We thank Antti Paraoanu for producing the illustrations for Fig.~1.
\end{acknowledgments}

\appendix
\section{Mean-field approximation}

To solve the Bloch momentum state in which the Bose-condensate takes place, we utilize the mean-field (MF) approximation where we substitute operators in Eq. \eqref{BHH2} by complex numbers, i.e. $c_\bk \rightarrow \psi_\bk \equiv [\psi_{\bk 1},...\psi_{\bk M}]^T$ and minimize the resulting MF energy
\begin{align}
\label{gpe} 
E_{\textrm{MF}}(\bk) = \psi_\bk^\dag \mathcal{H}(\bk) \psi_\bk + \frac{U}{2}\sum_\alpha |\psi_{\bk,\alpha}|^4
\end{align}
for each $\bk$ separately with respect to the constraint $\sum_{\alpha}|\psi_\alpha|^2 = n_0$~\cite{you:2012}. The constraint ensures that the condensate density $n_0$ is the same for all $\bk$. The condensation is then chosen to take place at the Bloch momentum $\bk_c$ and state $|\phi_0\rangle$ for which the MF energy $E_{\textrm{MF}}(\bk)$ is minimized, i.e. 
\begin{align}
E_{\textrm{MF}}(\bk_c) &= \textrm{min}[E_{\textrm{MF}}(\bk)] \nonumber \\
&= \langle \phi_0 | \mathcal{H}(\bk_c) | \phi_0 \rangle + \frac{U}{2}\sum_\alpha |\langle \alpha | \phi_0 \rangle|^4    
\end{align}
with $\langle \alpha | \phi_0 \rangle$ being the projection of $|\phi_0 \rangle$ to orbital $\alpha$. Thus, even though the flat band is strictly flat, the interaction term can favor a subset of Bloch states for which the condensate density is distributed as uniformly as possible between the sublattices to minimize the repulsive on-site interaction. For example, in the case of the kagome lattice, the condensed state at $\bk_c = [4\pi/3,0]$ reads $\phi_0 = [-1,-1,1]/\sqrt{3}$ whose sublattice density $|\langle \alpha | \phi_0 \rangle |^2 = 1/3$ is the same for all three sublattices. Similarly, in the case of the checkerboard-III flat band, the condensed state at $\bk_c = [2\pi/3,\pi]$ is $\phi_0 = [i,1]$, yielding uniform sublattice density $|\langle \alpha | \phi_0 \rangle |^2 = 1/2$.

\section{Details on Bogoliubov approximation}
Here we outline the the well-known Bogoliubov theory~\cite{Fetter1971,Shi1998} for a multiband system. To this end, one defines the bosonic Green's function for the non-condensed bosons in the imaginary-time domain as follows:
\begin{align}
\label{greenf}
&G(\bk,\tau) \equiv -\langle T_\tau   \begin{bmatrix}  c_{\bk}(\tau) \\  c_{2\bk_c-\bk}^\dag(\tau) \end{bmatrix}   
  \begin{bmatrix}  c_{\bk}^\dag(0) &  c_{2\bk_c-\bk}(0) \end{bmatrix}  \rangle = \nonumber \\ 
& = \begin{bmatrix}
   -\langle T_\tau  c_{\bk}(\tau)  c_{\bk}^\dag(0) \rangle & -\langle T_\tau  c_{\bk}(\tau)  c_{2\bk_c-\bk}(0) \rangle  \\   -\langle T_\tau  c_{2\bk_c-\bk}^\dag(\tau)  c_{\bk}^\dag(0) \rangle &  -\langle T_\tau  c_{2\bk_c-\bk}^\dag(\tau)  c_{2\bk_c-\bk}(0) \rangle
  \end{bmatrix} \nonumber \\
  &\equiv \begin{bmatrix}
   G_{11}(\bk,\tau) & G_{12}(\bk,\tau) \\ 
   G_{21}(\bk,\tau) & G_{22}(\bk,\tau) \\ 
  \end{bmatrix}.
\end{align}
Here, $T_\tau$ is the imaginary time ordering operator, $\tau$ is imaginary time, and $\bk \neq \bk_c$ as we consider the non-condensed bosons. Note that in our multiband case, each block $G_{ij}(\bk,\tau)$ in~\eqref{greenf} is a $M\times M$ matrix. In the bosonic Matsubara frequency space, one has the Dyson equation~\cite{Fetter1971,Shi1998}:
\begin{align}
\label{dyson}
&G^{-1}(\bk,i\omega_n) = G_{0}^{-1}(\bk,i\omega_n) -  \Sigma(\bk,i\omega_n) \Leftrightarrow \nonumber \\
& \begin{bmatrix}
   G_{11}(\bk,i\omega) & G_{12}(\bk,i\omega) \\ 
   G_{21}(\bk,i\omega) & G_{22}(\bk,i\omega) \\ 
  \end{bmatrix}^{-1} = \nonumber \\
&  \begin{bmatrix}
   G_{0,11}(\bk,i\omega) & 0 \\ 
   0 & G_{0,22}(\bk,i\omega) \\ 
  \end{bmatrix} -
  \begin{bmatrix}
   \Sigma_{11}(\bk,i\omega) & \Sigma_{12}(\bk,i\omega) \\ 
   \Sigma_{21}(\bk,i\omega) & \Sigma_{22}(\bk,i\omega) \\ 
  \end{bmatrix}.
\end{align}
Here, $\omega_n$ is a bosonic Matsubara frequency and $\Sigma$ is the self-energy arising from the finite interaction $U\neq 0$ and $G_{0}$ is the non-interacting Green's function which reads 
\begin{align}
G_{0}^{-1}(\bk,i\omega_n) = i\omega_n\sigma_z - \begin{bmatrix}
  \mathcal{H}(\bk) -\mu & 0 \\ 
   0& \mathcal{H}^*(\bk) - \mu \\ 
  \end{bmatrix}.    
\end{align}

Here, $\sigma_z$ is the Pauli matrix in the particle-hole basis. In general, the self-energy $\Sigma(\bk,i\omega_n)$ is evaluated by using the diagrammatic Beliaev theory~\cite{Shi1998,Fetter1971}. The Bogoliubov theory is the first order approximation of the Belieav approach~\cite{Shi1998}, containing the self-energy diagrams illustrated in black color in Supplementary Fig. S1~\cite{suppis}. Specifically, the Bogoliubov theory includes only the first order diagrams that contain the condensate propagators. In the case of a momentum-independent contact interaction, it is straightforward to evaluate both the diagonal and off-diagonal self-energy blocks $\Sigma_{11}$ and $\Sigma_{12}$:
\begin{align}
&[\Sigma_{11}(\bk)]_{\alpha\beta} = \delta_{\alpha\beta}2Un_0 |\langle \alpha | \phi_0 \rangle|^2 \\
&[\Sigma_{12}(\bk)]_{\alpha\beta} = \delta_{\alpha\beta} Un_0 \langle \alpha | \phi_0 \rangle^2.
\end{align}
We assume uniform condensate such that $|\langle \alpha | \psi_0 \rangle|^2 = 1/M$. The full Green's function is then obtained from the Dyson equation~\eqref{dyson}:
\begin{align}
\label{gf}
&G^{-1}(\bk,i\omega_n) = i\omega_n \sigma_z \nonumber \\
&-\begin{bmatrix}
\mathcal{H}(\bk) - \mu + 2\frac{Un_0}{M} & \Delta \\ \Delta^* & \mathcal{H}^*(2\bk_c-\bk) - \mu + 2\frac{Un_0}{M}
\end{bmatrix}.
\end{align}
Here, $[\Delta]_{\alpha\beta} = \delta_{\alpha,\beta} U n_0 \langle \alpha | \phi_0 \rangle^2$. By rewriting $\bk = \bk_c + \bq$, we see that the upper (lower) diagonal block in~\eqref{gf} represents particles (holes) travelling with momentum $\bq$ ($-\bq$) with respect to the momentum $\bk_c$ of the condensate. 

The poles of $G(\bk,i\omega_n)$ give the excitation spectrum of the Bose condensed system~\cite{Fetter1971}. As we are dealing with the equilibrium condensate, the lowest excitation band has to be gapless at $\bk_c$, i.e. a pole must exist at $\omega =0$ for $\bk = \bk_c$~\cite{Shi1998}. For our multiband system, we find that this condition forces $\mu$ to satisfy the following expression:
\begin{align}
\label{HP}
&\epsilon_0 -\mu + \frac{2Un_0}{M} - \frac{U^2n_0}{M^2}\Big( \epsilon_0 -\mu + \frac{2Un_0}{M} \Big)^{-1} =0 \nonumber \\
&\Leftrightarrow \mu = \epsilon_0 + \frac{Un_0}{M}.
\end{align}
This is the multiband generalization of the usual Hugenholtz-Pines relation~\cite{Shi1998,Fetter1971}. We should note that for obtaining Eq.~\eqref{HP}, the uniform condensation assumption $|\langle \alpha | \phi_0 \rangle|^2 = 1/M$ is crucial. If this was not the case, solving $\mu$ would most likely require numerical evaluation.

By combining the expression of $\mu$ to Eq.~\eqref{gf}, one gets
\begin{align}
&G^{-1}(\bk,i\omega_n) = i\omega_n \sigma_z - \begin{bmatrix}
\mathcal{H}(\bk) - \mu_{\textrm{eff}} & \Delta \\ \Delta^* & \mathcal{H}^*(2\bk_c-\bk) - \mu_{\textrm{eff}}
\end{bmatrix},
\end{align}
where $\mu_{\textrm{eff}} = \epsilon_0 - \frac{Un_0}{M}$. The Bogoliubov approximation yields a quadratic Hamiltonian, and it is easy to show that $G^{-1}(\bk,i\omega_n) = i\omega_n \sigma_z -\mathcal{H}_B$ so that $\mathcal{H}_B(\bk) = -G^{-1}(\bk,0)$. This is the Bogoliubov Hamiltonian expressed in Eq.~\eqref{ham}.

In our numerical Bogoliubov calculations, we fixed the total density $n_{\textrm{tot}}$ and chose an initial ansatz for $n_0$. This was then substituted to $L(\bk) = \sigma_z \mathcal{H}_B(\bk)$ and the Bogoliubov states $|\psi^\pm_m(\bk) \rangle$ were obtained by carrying out the diagonalization of Eq.~\eqref{bogo_eig}. To ensure the Bogoliubov states follow the bosonic commutation rules, we demand the standard normalization condition for the Bogoliubov states to hold~\cite{castin:book}:
\begin{align}
\label{normalization}
& _1\langle \psi^\pm_m(\bk) | \psi^\pm_m(\bk) \rangle_1 -\textrm{ } _2\langle \psi^\pm_m(\bk) | \psi^\pm_m(\bk) \rangle_2 = \pm 1, \textrm{ with} \nonumber\\
& |  \psi^\pm_m(\bk) \rangle \equiv \Big[|  \psi^\pm_m(\bk) \rangle_1, |  \psi^\pm_m(\bk) \rangle_2\Big]^T. 
\end{align}
Based on the obtained Bogoliubov states, $n_{\textrm{ex}}$ was then calculated with Eq.~\eqref{nex}. In this way, a new value for $n_0 = n_{\textrm{tot}} - n_{\textrm{ex}}$ was acquired and then substituted back to $L(\bk)$. This iteration procedure was continued until a self-consistent solution for $n_0$ was found.

In general, Bogoliubov theory is based on the assumption that the system is Bose-condensed and the remaining quantum fluctuation effects can be described by the quadratic Bogoliubov Hamiltonian~\eqref{ham}. This approximation is based on the assumption that the local on-site interaction energy $U$ times condensate density per unit cell $n_0$, i.e. $Un_0$, is small compared to the kinetic hopping terms $\mathcal{H}_{i\alpha j\beta}$ of the Hamiltonian of Eq.~\eqref{BHH}. In our calculations this is always the case. Furthermore, conventionally the Bogoliubov theory is considered hold only for cases with $n_{\textrm{ex}} \ll n_{\textrm{tot}}$. However, as we discuss at the end of Sec.~\ref{Frag} and Appendix~\ref{AppE}, our method should be still valid even for cases when $n_{\textrm{ex}}$ is of the same order of magnitude than $n_{\textrm{tot}}$ as long as the aforementioned condition on $Un_0 \ll \mathcal{H}_{i\alpha j \beta}$ holds. In this work we present results for two-dimensional systems, but extension to three dimensions is straightforward as the Bogoliubov theory works well there.

\section{Projection to the lowest band}
In obtaining our main results, we used the projected $L_p(\bk)$ of Eq.~\eqref{lk2} in which only the lowest Bloch band degrees of freedom were retained and the effect of other Bloch bands were discarded. To obtain \eqref{lk2}, we must first transform the full $L(\bk) =\sigma_z\mathcal{H}_B$ to the Bloch basis that diagonalizes the kinetic energy Hamiltonian $\mathcal{H}(\bk)$. Specifically, we first write $\mathcal{H}(\bk) = U(\bk) D(\bk) U^\dag(\bk)$, where $D(\bk)=\text{diag}[\epsilon_1(\bk),\epsilon_2(\bk),\ldots,\epsilon_M(\bk)]$ and the columns of $U(\bk)$ contain the corresponding Bloch states. We then define the unitary transformation $\mathcal{U}(\bk)$ as
\begin{align}
\mathcal{U}(\bk) = \begin{bmatrix}
 U(\bk) & 0 \\ 0 & U^*(2\bk_c -\bk).
\end{bmatrix}    
\end{align}
By transforming $L(\bk) \rightarrow \mathcal{U}^\dagger(\bk)L(\bk)\mathcal{U}(\bk)$ and moreover writing the momentum as $\bk = \bk_c + \bq$, we have in the Bloch basis the following:
\begin{widetext}
\begin{align}
\label{Lk}
L(\bk) =  \begin{bmatrix}
	D(\bk_c + \bq) -\mu_{\textrm{eff}} & U^\dag(\bk_c + \bq)\Delta U^*(\bk_c - \bq) \\
	-U^T(\bk_c - \bq)\Delta^* U(\bk_c + \bq) & -D(\bk_c - \bq) +\mu_{\textrm{eff}} 
\end{bmatrix}.
\end{align}
\end{widetext}
By only retaining the lowest Bloch band and projecting out all of the other Bloch band degrees of freedom in Eq.~\eqref{Lk}, one obtains $L_p(\bk)$ of Eq.~\eqref{lk2}.
\clearpage
\section{Density of non-condensed particles}
The expression~\eqref{nex} for the density of non-condensed particles can be derived as follows:
\begin{align}
\label{nex_m}
&n_{\textrm{ex}} = \frac{1}{N}\sum_{\bk\alpha}{}^{'} \langle  c^\dag_{\bk\alpha}  c_{\bk\alpha} \rangle \nonumber \\
&=  \frac{1}{2N}\sum_{\bk \alpha}{}^{'}\big[ \langle c^\dag_{\bk\alpha}  c_{\bk\alpha} \rangle + \langle c_{\bk\alpha}  c^\dag_{\bk\alpha} \rangle -1 \big] \nonumber \\
&= -\frac{\lim_{\tau \rightarrow 0}}{2N}\sum_{\bk}{}^{'} \Tr[G(\bk,\tau) +1] \nonumber \\
&= -\frac{1}{2N}\sum_\bk{}^{'}\Big\{ M+ \frac{1}{\beta}\sum_{i\omega_n} \Tr[G(\bk,i\omega_n)]\Big\}.
\end{align}
Here, $\beta = 1/(k_B T)$ with $k_B$ is the Boltzmann constant. We can proceed by noting that for the quadratic Bogoliubov Hamiltonian, one can write the Green's function with the aid of the Bogoliubov states and energies as 
\begin{align}
\label{greenf2}
G(\bk,i\omega_n) = \sum_{m,s}s\frac{| \psi^s_m(\bk) \rangle \langle \psi^s_m(\bk) |}{i\omega_n - sE_m(\bk_c + s\bq)}.    
\end{align}
Here, $\bq = \bk-\bk_c$ and $| \psi^s_m(\bk) \rangle$ are the Bogoliubov states fulfilling the equations \eqref{bogo_eig} and \eqref{normalization}. The validity of Eq.~\eqref{greenf2} can be confirmed by multiplying it with $G^{-1}(\bk,i\omega_n) = i\omega_n \sigma_z -\mathcal{H}_B$. By substituting Eq.~\eqref{greenf2} to~\eqref{nex_m} and carrying out the trace, one finds at $T=0$
\begin{align}
\label{nex_m2}
&n_{\textrm{ex}} = \frac{(-1)}{2N} \sum_{\bk,m}{}^{'}\Big\{ 1 + \frac{1}{\beta} \sum_{s,i\omega_n}\frac{s\langle  \psi^s_m(\bk) |  \psi^s_m(\bk) \rangle}{i\omega_n - sE_m(\bk_c + s\bq)} \Big\} \nonumber \\
&= \frac{1}{2N}\sum_{\bk,m}{}^{'}\Big[-1 + \sum_s s \langle  \psi^s_m(\bk) |  \psi^s_m(\bk) \rangle n_B(sE_m(\bk_c + s\bq))   \Big] \nonumber \\
& = \frac{1}{2N}\sum_{\bk,m}{}^{'}\Big[-1 +  \langle  \psi^-_m(\bk) |  \psi^-_m(\bk) \rangle \Big],
\end{align}
where $n_B(x) = 1/(e^{\beta x}-1)$ is the Bose-Einstein distribution, and in the last step we have used the fact that at $T=0$ we have $n_B(x>0) = 0$ and $n_B(x<0) = -1$.  

It is instructive to consider $n_{\textrm{ex}}$ in a lattice geometry which respects the time-reversal symmetry, i.e. $\mathcal{H}(\bk) = \mathcal{H}^*(-\bk)$, with zero-momentum condensate $\bk_c =0$ and a scalar-valued $\Delta$. This special case covers, for example, the usual square lattice, honeycomb lattice and the kagome lattice with a dispersive band chosen as the lowest Bloch band (i.e. with negative NN hopping, $t<0$). We then find
\begin{align}
\label{n_ex_trs}
n_{\textrm{ex}} = \frac{1}{2N}\sum_{\bk,m}{}^{'}\Big[-1 + |u^m(\bk)|^2 + |v^m(\bk)|^2\Big] \end{align}
where the coherence factors $u^m(\bk)$ and $v^m(\bk)$ are
\begin{align}
& u^m(\bk) = \frac{1}{\sqrt{2}}\sqrt{ \frac{\tilde{\epsilon}_m(\bk) + U n_0/M}{ E_m(\bk) } +1}  \nonumber \\
& v^m(\bk) = -\frac{1}{\sqrt{2}}\sqrt{ \frac{\tilde{\epsilon}_m(\bk) + U n_0/M}{ E_m(\bk) } -1}  
\end{align}
with the Bogoliubov energies $E_m(\bk) = \sqrt{\tilde{\epsilon}_m(\bk)[ \tilde{\epsilon}_m(\bk) + 2Un_0/M]}$ and $\tilde{\epsilon}_m(\bk) \equiv \epsilon_m(\bk) - \epsilon_0$. Note that each Bogoliubov band $m$ depends only on the $m$th Bloch band. When the lowest Bloch band is not strictly flat ($\tilde{\epsilon}_1 \neq 0$), one has $\lim_{U\rightarrow 0}n_{\textrm{ex}} = 0$ as $u^m(\bk) \rightarrow 1$ and $v^m(\bk) \rightarrow 0$. This is consistent with the result shown in Fig.~\ref{Fig:Kagome_n_ex} where $\lim_{U\rightarrow 0}n_{\textrm{ex}} = 0$ for the dispersive band condensate of the kagome lattice. On the other hand, in the case of a flat band condensate, one finds with Eq.~\eqref{n_ex_trs} that $\lim_{U\rightarrow 0}n_{\textrm{ex}} \rightarrow \infty$, meaning that one can never achieve a stable flat band BEC with the aforementioned conditions. Indeed, in the case of the kagome flat band condensate, finite condensate momentum $\bk_c = [4\pi,0]$ breaks the $\bk_c=0$ condition, implying that Eq.~\eqref{n_ex_trs} is no longer valid and one can then find stable BEC within the flat band. 

\section{Hartree-Fock-Bogoliubov approximation}\label{AppE}
In our calculations we can have $n_{\textrm{ex}} \sim n_0$, and therefore it is not evidently clear whether the Bogoliubov theory is an appropriate method, since Bogoliubov theory usually assumes that $n_0 \gg n_{\textrm{ex}}$~\cite{Shi1998}. It turns out that in the $U\rightarrow 0$ limit (which we are mostly interested in), the Bogoliubov approximation should be a valid choice for our calculations even at the regime of $n_{\textrm{ex}} \sim n_0$. We show this by considering the Hartree-Fock-Bogoliubov (HFB) theory, which is an extension to the Bogoliubov theory. Specifically, HFB counts for all the first order self-energy diagrams shown in Fig. S1 of SM~\cite{suppis}. In addition to the Bogoliubov diagrams, one also accounts the diagrams that do not contain the condensate propagators~\cite{Shi1998}. By evaluating the diagrams, one finds
\begin{align}
& [\Sigma_{11}(\bk)]_{\alpha\beta} = \delta_{\alpha\beta}2U( n_0|\langle \alpha | \phi_0 \rangle|^2 + n_{\textrm{ex},\alpha}) \nonumber \\
& \equiv \delta_{\alpha\beta}2Un_{\textrm{tot},\alpha} \\
& [\Sigma_{12}(\bk)]_{\alpha\beta} = \delta_{\alpha\beta}Un_0 \langle \alpha | \phi_0 \rangle^2 + \frac{\delta_{\alpha\beta}U}{N}\sum_\bq{}^{'} \langle c_{\bq\alpha} c_{2\bk_c-\bq \alpha} \rangle \nonumber \\
&\equiv \delta_{\alpha\beta}\Big(Un_0 \langle \alpha | \phi_0 \rangle^2 + \Lambda_\alpha \Big).
\end{align}
Here, $n_{\textrm{ex},\alpha}$ ($n_{\textrm{tot},\alpha}$) is the excitation (total) density of the $\alpha$th sublattice and solving $\Lambda_\alpha$ in $\Sigma_{12}$ requires a self-consistent fixed point iteration scheme. It is well-known that the self-consistent HFB method has problems with yielding  the gapless excitation modes at $\bk_c=0$~\cite{Shi1998,Yukalov2011}. Here we circumvent this inconsistency by explicitly demanding the existence of the gapless Goldstone mode at $\bk_c$. By assuming uniform density $n_{\textrm{tot},\alpha} = n_{\textrm{tot}}/M$ and furthermore $\Lambda_\alpha = \frac{U}{M}\Lambda$, one finds $\mu = \epsilon_0 +\frac{2U n_{\textrm{tot}}}{M} - \frac{U}{M}|n_0 + \Lambda|$. We therefore obtain the following quadratic Hamiltonian for the HFB approximation: 
\begin{align}
\label{HFB}
&H_{HFB}(\bk) = \begin{bmatrix}
\mathcal{H}_\bk - \tilde{\mu}_{\textrm{eff}} & \tilde{\Delta} \\ \tilde{\Delta}^* & \mathcal{H}^*_\bk - \tilde{\mu}_{\textrm{eff}}\end{bmatrix}  \\
& \tilde{\Delta}_{\alpha\beta} = [\Delta]_{\alpha\beta} + \frac{\delta_{\alpha\beta}U\Lambda}{M} \nonumber \\
& \tilde{\mu}_{\textrm{eff}} = \epsilon_0 - \frac{U}{M}|n_0 + \Lambda| \nonumber
\end{align}
We see that for $U\rightarrow 0$, the HFB approximation reduces to the Bogoliubov theory. This is due to the fact that the interaction is taken to be momentum-independent contact interaction: if this was not the case, $\Sigma_{11}$ and thus the diagonal blocks of $H_{\textrm{HFB}}$ would acquire a more complicated momentum-dependent form. Therefore, the Bogoliubov theory gives the same results as HFB theory at the limit of $U\rightarrow0$ even when the excitation density $n_{\textrm{ex}}$ is comparable to the condensate density $n_0$, i.e. $n_{\textrm{ex}}\sim n_0$.

Figure S2 in SM~\cite{suppis} compares both Bogoliubov and HFB methods in the case of the kagome flat band condensate. In panels a-b we depict $c_s$ and in d-f $n_{\textrm{ex}}$ as a function of interaction for three different total densities. One can see that at the limit of $U\rightarrow 0$, the $n_{\textrm{ex}}$ of the HFB method approaches the Bogoliubov result, consistent with our discussion in the preceding paragraph. We have confirmed that the HFB results of Fig. S2 are self-consistent and that their excitation spectra feature the gapless Goldstone mode at $\bk_c$. We further note that utilizing the HFB method at total densities even smaller than those depicted in Fig. S2 does not yield self-consistent results, indicating that either the condensate becomes unstable or the HFB method breaks down in the small density limit.


%